\documentclass[final,10pt,onecolumn,a4paper,twoside]{thesis_mm}

\usepackage{afterpage}
\usepackage{epsfig}
\usepackage{graphicx}
\usepackage{graphics}
\usepackage{float}
\usepackage{floatfig}
\usepackage{makeidx}
\usepackage{multicol}
\usepackage{setspace}
\usepackage[sort&compress,sectionbib]{natbib}
\bibpunct{(}{)}{;}{a}{}{,}
\usepackage{palatino}
\usepackage{amssymb}
\usepackage{amsmath}
\usepackage{rotating}
\usepackage{layout}

\newcommand{\kmskpc}{km\, s$^{-1}$\, kpc$^{-1}$}

\newcommand{\los}{line-of-sight}

\newcommand{\vlos}{$v_{\mathrm{los}}$}

\newcommand{\um}{$\mu$m}
\newcommand{\kms}{km\thinspace s$^{-1}$}

\newcommand{\Al}{{\it A$_\lambda$}}
\newcommand{\ks}{{\it K$_{\rm S}$}}
\newcommand{\Aks}{{\it A$_{\rm K_{\rm S}}$}}
\newcommand{\Ak}{{\it A$_{\rm K}$}}
\newcommand{\Av}{{\it A$_{\rm V}$}}
\newcommand{\sig}{$\sigma_{\rm A_{\rm K_{\rm S}}}$}

\newcommand{\be}{\begin{equation}}
\newcommand{\ee}{\end{equation}}
\newcommand{\ba}{\begin{eqnarray}}
\newcommand{\ea}{\end{eqnarray}}

\makeindex

\def\exps{ps}

\begin{document}


{\clearpage\thispagestyle{empty}}
\begin{center}
\ \\ {\LARGE Late-type Giants in the Inner Galaxy} \ \\
\vspace{1.5 cm}
\ \\
PROEFSCHRIFT \\
\ \\
\vspace{1.5 cm}
\ \\
ter verkrijging van \\
de graad van Doctor aan de Universiteit Leiden, \\
op gezag van de Rector Magnificus Dr. D.D. Breimer, \\
hoogleraar in de faculteit der Wiskunde en \\
Natuurwetenschappen en die der Geneeskunde, \\
volgens besluit van het College voor Promoties \\
te verdedigen op woensdag 30 juni 2004 \\
klokke 16.15 uur \\
\ \\
\vspace{1.5 cm}
\ \\
door \\
\ \\
\vspace{2.25 cm}
\ \\
Maria Messineo \\
geboren te Petralia Soprana (Itali\"e)\\
in 1970 \\
\end{center}
{\clearpage\thispagestyle{empty}}
\noindent
\ \ Promotiecomissie
\ \\
\bigskip
\ \\
\begin{tabular}{ll}
Promotor : & Prof. dr. H.~J. Habing\\
Referent : & Dr. J. Lub \\ & \\
Overige leden :
& Prof. dr. W.~B. Burton      \\ 
& Dr. M.R. Cioni (European Southern Observatory, Garching bei M\"unchen)\\ 
& Prof. dr. K. Kuijken\\ 
& Prof. dr. K.~M. Menten (Max-Planck-Institut fuer Radioastronomie, Bonn)\\ 
& Prof. dr. A. Omont  ( Institut d'Astrophysique de Paris)\\ 
& Prof. dr. P.~T. de Zeeuw\\ 
\end{tabular}

{\clearpage\thispagestyle{empty}}
\ \\
\vspace{10cm}
\ \\
\begin{flushright}
{A Lucia e Leonardo}
\end{flushright}
{\clearpage\thispagestyle{empty}}
\ \\
\vspace{13cm}
\ \\
\noindent
 Front cover: ``I Girasoli'' by Renato Guttuso.  Permission to
 reproduce this painting was kindly granted by Dr. Fabio
 Carapezza Guttuso and authorised by SIAE 2004 (Italian Society of
 Authors and Publishers).
\\Back cover: ISOGAL image taken  at 7 \um\ (LW5 filter) of a field
 of 20\arcmin$\times$20\arcmin  centred on  $(l,b)=-0.27^\circ,-0.6^\circ$,
 field FC--00027--00006. \ \\ \normalsize

\frontmatter
\tableofcontents
\mainmatter
\chapter{Introduction}
The Milky Way is the cornerstone of our understanding of galaxies.
The structure and kinematics of its gas and stars can be studied in
unique detail due to their relative proximity. However, being located
well within the Galactic disk and thereby observing the Milky Way in
non--linear projection makes it difficult to properly map its
large-scale morphology.

One of the interesting findings has been that observations of
molecular line emission (CO, HI) and stellar motions show signatures
of a Galactic Bar in the inner Galaxy. However, its characteristics
such as elongation, thickness and viewing angle are still poorly
constrained. One of the main obstacles has been the strong obscuration
by interstellar dust toward the inner Galaxy, which makes optical
studies of the stellar population in that region almost impossible.
The extinction is less severe at near- and mid-infrared wavelengths.
To characterise the structure and formation history of the Milky Way,
several infrared surveys were conducted during the past decade:
ISOGAL, MSX, DENIS, 2MASS.  These data contain a wealth of information
on the structure of the stellar populations that has yet to be fully
analysed. Having entered a golden age for Galactic astronomy, soon
even more detailed imaging and spectroscopy will be provided by the
Spitzer Space Telescope, while the GAIA satellite will provide
unprecedented astrometry.

\begin{figure}[h]
\begin{center}
\resizebox{0.7\hsize}{!}{\includegraphics{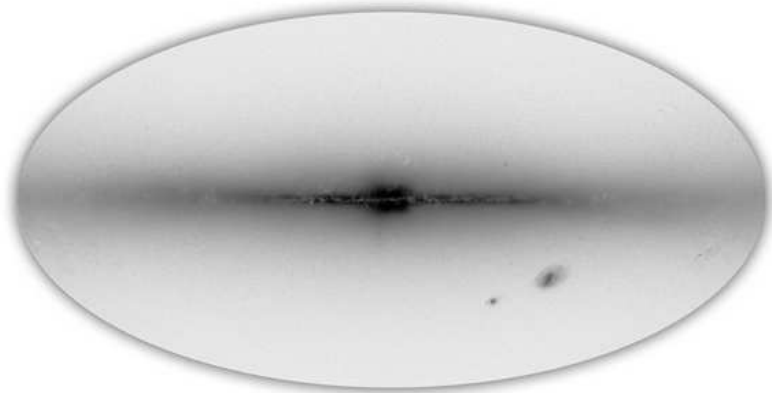}}
\end{center} \hfill
\caption{\label{fig:dirbe123_2p6dec_s.ps} This infrared image taken by
the 2MASS satellite shows the plane of our Milky Way Galaxy as a thin
disk. Dust obscuration makes the optical appearance of the Milky Way
much more patchy.}
\end{figure}

My thesis research has focused on the structure and stellar population
of the inner 4 kpc of the Milky Way. I have analysed data from recent
infrared surveys and obtained SiO radio maser line observations of
late-type giants to study the star formation history and the
gravitational potential of the inner Galaxy. With ages ranging from
less than 1 to 15 Gyr, the infrared-luminous late-type giant stars are
representative of the bulk of the Galactic stellar population, and
hence trace its star formation history. Their spatial abundance
variation maps the stellar mass distribution, and thereby probes the
Galactic gravitational potential.  The reddening of their spectral
energy distribution can be used to map the interstellar extinction.
Their envelopes often emit strong molecular masers (OH, SiO) that can
be detected throughout the Galaxy, and through the precise measurement
of the maser line velocity they reveal the stars' line-of-sight
velocities.  Therefore they are ideal tracers of the Galactic
kinematics and gravitational potential.

\section{Late-type giants}

In this section I shall briefly discuss the life cycle of stars,
with emphasis on the red giant and asymptotic giant branch
phases.

\begin{figure}[h]
\begin{center}
\resizebox{0.6\hsize}{!}{\includegraphics{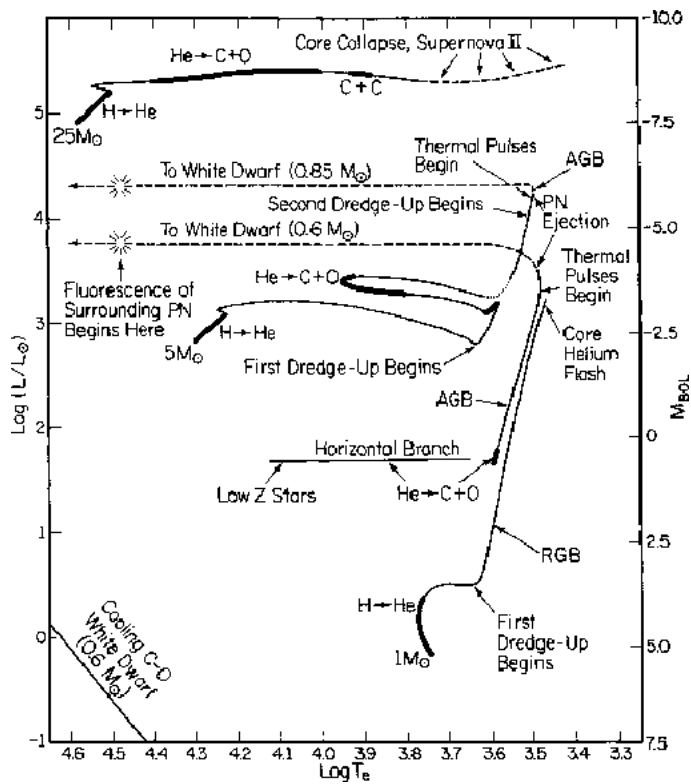}}
\end{center} \hfill
\caption{\label{fig:iben.ps} Stellar evolution tracks of  1, 5, and 25
M$_\odot$ stars in the H-R diagram \citep{iben85}.
 }
\end{figure}

A low- to intermediate-mass star ($M_*<8\, M_\odot$) spends 80 to 90
percent of its life on the so called main-sequence phase.  This phase
ends when a large fraction of the star's hydrogen has been converted
to helium. Then the stellar core contracts and heats until hydrogen
fusion starts in a shell surrounding the core. This causes the stellar
envelope to expand to about 50 to 100 solar diameters, while the
surface temperature decreases.  Stars in this phase are called red
giant branch (RGB) stars.

When the core temperature is high enough, helium nuclei fuse into
carbon and oxygen. For stellar masses less than 2.3 solar masses (low
mass stars), the core is degenerate and core helium burning begins
abruptly in a so called core helium flash. In the Hertzsprung-Russell
(HR) diagram this event marks the tip of the red giant branch.  For
higher mass stars helium burning begins more gradually.  The core
helium-burning phase lasts between 10 and 25 percent of a star's
main-sequence lifetime.

When the core helium is exhausted, the core contracts, the envelope
expands and the stellar surface temperature decreases.  The star is
now powered by hydrogen and helium burning in shells surrounding the
core, which consists of carbon and oxygen nuclei with a degenerate
distribution of electrons. A star in this phase is called an
asymptotic giant branch (AGB) star. This name originates from the fact
that in the HR diagram for low mass stars the AGB branch approaches
the RGB sequence asymptotically.  They can be as large as several
hundred solar radii and have a relatively cool surface temperature of
about 3000 K.

At the beginning of the AGB phase, helium shell burning prevails over
shell hydrogen burning, so the C-O core grows steadily in mass,
approaching the hydrogen shell (E-AGB).  When the mass of helium
between the core and the hydrogen shell drops below a critical value,
the helium shell exhibits oscillations that eventually develop into
the first helium shell flash and the thermally pulsating (TP-AGB)
phase begins.  A dredge-up (the third dredge-up) may take place during
this phase, bringing carbon to the surface.

Mass-loss reduces the envelope mass until the residual envelope is
ejected in a short superwind phase.  The strength of the wind controls
the decrease of the stellar mass (as the star climbs the AGB in the HR
diagram), which also affects the evolution of its surface composition.
Mass-loss may occur also in RGB stars close to the RGB tip, but with
much lower intensity than in AGB stars.

The third dredge-up is fundamental to explain the conversion of a
fraction of oxygen-rich AGB stars into carbon-rich AGB stars (for
which [C/O]$>1$) and predicts that the latter form only above a
specific minimum luminosity.  Carbon stars are virtually absent in the
Galactic bulge, whereas they are numerous in the Magellanic Clouds,
suggesting that the lower metallicity there provides for a more
efficient dredge-up.

AGB stars produce roughly one third of the carbon in the Galaxy,
almost the same amount as supernovae and Wolf-Rayet stars.  By
returning dust and gas to the interstellar medium, RGB and AGB stars
pave the way for the formation of future generations of stars and
planets.

Due to their low surface temperatures, late-type giants (RGB and AGB)
are bright at infrared wavelengths. Facing a high interstellar
extinction toward the central regions of the Milky Way that obscures
stars at visible wavelengths, red giants are the best targets for
studies of the stellar populations, dynamics, and star formation
history in the inner Galaxy.

\subsection{AGB star and variability}
An important property of AGB stars with direct applications to
Galactic structure studies is their luminosity variability.  The radial
pulsations of AGB stars are confined to the large convective envelopes
and should not be confused with the thermal pulse that originates in
the  helium burning shell. The latter leads to a longer-term
variability.

Variable AGB stars are named in several different ways  based on 
the light curve properties and periods: large amplitude variables
(LAV), long period variables (LPV), Mira variables, semiregular (SR)
and irregular variables.

By definition, Mira stars show pulsations with large amplitudes at
visual wavelengths (more than 2.5 mag) and vary relatively regularly
with typical periods of 200 to 600 days.  Semiregular variables show
smaller amplitudes (less than 2.5 mag) and they have a definite
periodicity. Since they are obscured in the visual, this
classification cannot be applied to inner Galactic variable AGB stars.
Therefore, for inner Galactic variable stars `LAVs' and `LPVs' refer
to variations in the $K$-band.  Mira stars are usually LAVs with
$K$-band variation amplitudes larger than 0.3 mag.

Another important class of AGB variable stars are OH/IR stars, which are
dust-enshrouded infrared variable stars. They are discovered in the infrared
and show 1612 MHz OH maser emission.  Their periods are typically longer than
600 days and can exceed 1500 days.

Recent observations from the MACHO, EROS and OGLE surveys initiated a
discussion on the pulsation modes of long period variables and on the
period-luminosity relations.  Such period-luminosity relations are
important for Galactic structure studies as they yield estimates for a
star's distance. The new data reveals four parallel period-luminosity
sequences (A-D). However, the classical period-luminosity relation
discovered by \citet{feast89}, which is based on visual observations
of Mira stars, still holds and coincides with the C sequence. Large
amplitude variables with a single periodicity, like probably most of
our SiO targets, populate this sequence.

\subsection{Circumstellar maser emission}
MASER stands for Microwave Amplification by Stimulated Emission of
Radiation. In 1964 Charles Townes, Nicolay Gennadiyevich Basov and
Aleksandr Mikhailovich Prochorov received the Nobel Prize
for their discovery of the maser phenomenon. Now, forty years later, we
know of thousands of astronomical masers, ``radio radiation detected
in some lines of certain astronomical molecules, attributed to the
natural occurrence of the maser phenomenon'' (Elitzur).

Maser radiation is caused by a population inversion in the energy
levels of atoms or molecules.  The non-equilibrium inversion is caused
by different pumping mechanisms, in astronomical objects usually
infrared radiation and collisions.

An observed line can be identified as a maser line on the basis of its
unusually narrow line-width, or when line ratios indicate deviations
from thermal equilibrium.

Various molecules can show maser emission. Astronomical masers are
found around late-type stars (circumstellar masers), and in the cores
of dense molecular clouds (interstellar masers).  A comprehensive
review of astronomical masers was given by \citet{elitzur92} and
\citet{reid88}.  In this thesis we study circumstellar maser emission.

The circumstellar envelopes of oxygen-rich late-type stars can exhibit
maser emission from SiO, H$_2$O, and OH molecules
\citep{habing96}. Masers occur in distinct regions at various
distances from the central star. SiO masers at 43 and 86 GHz originate
from near the stellar photosphere, within the dust formation zone
\citep{reid97}. Water masers originate further out, at distances of up
to $10^{15}$ cm from the central star, while OH masers are found in
the cooler outer regions of the stellar envelope, about ten times
further out.

The presence or absence of particular maser lines in a circumstellar
envelope appears to depend on the opacity at 9.7\um: a higher
mass-loss rate leads to a more opaque dust shell, which shields
molecules better against photodissociation by interstellar UV
radiation.

SiO masers arise from rotational transitions in excited vibrational
states. These levels can be highly populated only near the star where
the excitation rates are high.  SiO maser emission has been detected
in different transitions towards oxygen-rich AGB stars (i.e.\ Mira
variables, semi-regular variables, OH/IR stars) and supergiants. The
relative intensity of different SiO maser lines varies among different
sources, indicating that the SiO maser pumping mechanism depends on
the mass-loss rate. Maser pumping is dominantly radiative, as
suggested by the observed correlation between the maser intensity
and the stellar infrared luminosity. However, collisional pumping
cannot be ruled out.  A maser line can show the stellar line of sight
velocity with an accuracy of a few \kms.

H$_2$O (22 GHz) and OH (1612 MHz, 1667 MHz and 1665 MHz) masers
originate from transitions in the ground vibrational state. H$_2$O
maser spectra are irregular and variable. Therefore they are not
useful for an accurate determination of stellar line of sight
velocities.  The 1612 MHz OH maser line is pumped by radiation from
the circumstellar dust, which excites the 35 \um\ OH line and has a
typical double-peaked profile. The stellar velocity lies between the
two peaks and the distance between the two peaks yields a measure of
the expansion velocity of the circumstellar envelope.

\citet{lewis89}  analysed colours of and masers from IRAS stars,
suggesting a chronological sequence of increasing
mass-loss rate: from SiO, via H$_2$O to OH masers. This sequence 
links AGB stars via the Mira and OH/IR stages with Planetary
Nebulae. However, parameters other than mass-loss, such as stellar
abundance, probably also play an important role \citep{habing96}.

\section{The Milky Way galaxy}

\begin{figure}[h]
\begin{center}
\resizebox{0.6\hsize}{!}{\includegraphics{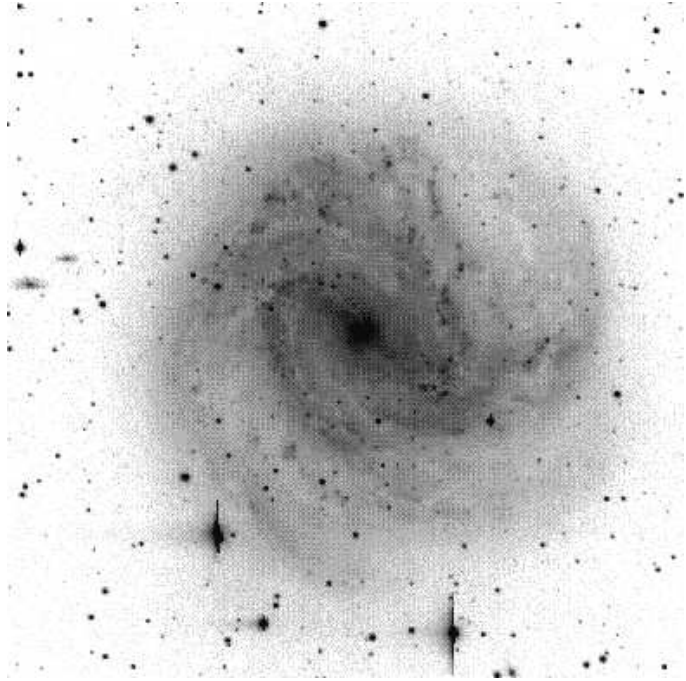}}
\end{center} \hfill
\caption{\label{fig:m83.ps} The Southern Pinwheel galaxy, M83, was
classified as intermediate between normal and barred spiral galaxies
by G. de Vaucouleurs. It has both a pronounced disk component
exhibiting a spiral structure, and a prominent nuclear region, which
is part of a notable bulge/halo component.  The Milky Way might look
similar to M83.}
\end{figure}

Our home Galaxy, the Milky Way, is a large disk galaxy. It is likely
to be of Hubble type SBbc, with its main components being the bulge,
the disk, and the halo.

The Milky Way today is the result of star formation, gas flow, and
mergers integrated over time.  The different Galactic components were
not formed by independent events, and their formation history is
largely unknown. The possible connection between the star formation
history and the formation of Galactic structures is equally unknown.

\vskip 0.3cm
\noindent {\em Halo}\\
\noindent
The halo is composed of a dark matter and a stellar halo.  The dark halo
is of yet unknown nature and dominates the total Galactic mass, as
suggested by dynamical studies of satellite galaxies. The stellar
halo, a roughly spherical distribution of stars whose chemical
composition, kinematics, and evolutionary history are quite different
from stars in the disk, contains the most metal-poor and possibly some
of the oldest stars in the Galaxy. It retains important information on
the Galactic accretion history.  The recent discovery of stellar
streams in the halo \citep[e.g.][]{helmi99,ibata94} supports the
hierarchical clustering and merging scenario of galaxy formation.

\vskip 0.3cm
\noindent {\em Disk}\\
\noindent
The disk is usually divided into two components, the thin and thick
disks.  The thick disk \citep{gilmore83} is older that 10 Gyr.  Its
metallicity ranges from -1.7 to -0.5 [Fe/H], it has a scale height of
0.7-1.5 kpc, a scale length of 2-3.5 kpc and a vertical velocity
dispersion of 40 \kms. The thin disk has a scale-height of about 250
pc and contains stars of all ages.  The thick disk was probably formed
from the thin disk during a merger event that heated the disk
\citep{gilmore02}.

\vskip 0.3cm
\noindent {\em Bulge}\\
\noindent
There is some confusion in the use of the term ``bulge''.  In the
literature it is often used to indicate everything in the inner few
kiloparsec of the Galaxy, i.e., the bar and the nuclear disk.  \citet{wyse97}
prefer to define a ``bulge'' as a ``centrally concentrated stellar
distribution with an amorphous, smooth appearance. This excludes
gas, dust and recent star formation by definition, ascribing all such
phenomena in the central parts of the Galaxy to the central disk, not
to the bulge with which it cohabits.''

The ``bulge'' is dominated by an old stellar component (10 Gyr).  Its
abundance distribution is broad, with a mean of [Fe/H]$\sim -0.25$ dex
\citep{mcwilliam94}. It has a scale height of about 0.4 kpc, and a
radial velocity dispersion of about 100 \kms.

There is growing evidence of a non-axisymmetric mass distribution in
the inner Galaxy. This is found from the near-infrared light
distribution, source counts, gas and stellar kinematics, and
microlensing studies. However, it is not clear yet whether a
distinction should be made between the triaxial Galactic bulge and the
bar in the disk -- a bar is defined as a thin, elongated structure in
the plane.

In face-on galaxies, it is not unusual to observe both a central bulge
and a bar (i.e.\ NGC1433). In general, bars may be populated by both
old and young stars.  In addition, barred galaxies often show a ring
around the bar.  From studies of edge-on galaxies there is indication
that peanut-shaped or box-shaped (rather than spheroidal) bulges may
be associated with bars. The Milky Way provides the closest example of
a box-shaped bulge, and therefore it is a unique laboratory to
investigate the structure and kinematics of a boxy bulge and its
relation with the disk. A bar and a triaxial bulge could both be
present, distinct, and coexisting.

\begin{figure}[h]
\begin{center}
\resizebox{0.6\hsize}{!}{\rotatebox{0}{\includegraphics{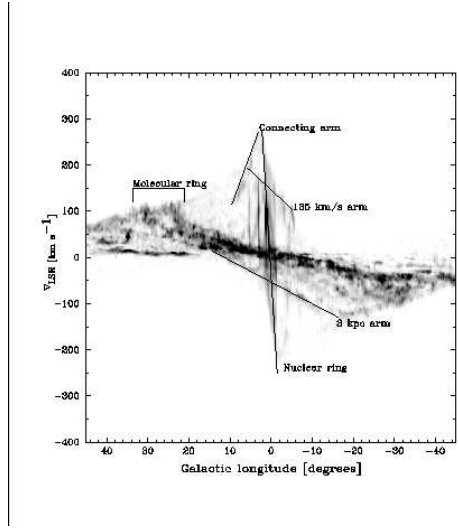}}}\caption{
\label{fig:gas.ps}  CO ($l,v$) diagram from \citet{dame01}.
Gas features are labelled as in Fig.\ 1 of \citet{fux99}.}
\end{center} \hfill
\end{figure}

The existence of a distinct disk-like dense molecular cloud complex in
the central few hundred pc of our Galaxy, the Central Molecular Zone
(CMZ), was established in the early 1970s.  Observations of ongoing
star formation and the presence of ionizing stars suggest that this is
a component different from the bulge.  In the longitude-velocity
diagram (Fig. \ref{fig:gas.ps}) it generates a remarkable feature
called the 180 pc-Nuclear Ring. It can be understood as a gaseous
shock region at the transition between the innermost non intersecting 
X1 orbits and the X2 orbits \citep{binney91}.  The total mass of the CMZ
(including the central stellar cluster) amounts to $(1.4\pm0.6) 10^9
M_\odot$, of which 99\% is stellar mass, and 1\% gaseous mass. Its
stellar luminosity amounts to $(2.5\pm1) 10^9 L_\odot$, 5\% of the
total luminosity of the Galactic disk and bulge taken together 
\citep{mezger02}.

The centre of our Galaxy contains a massive black hole.  The advent of
adaptive optics has permitted high spatial resolution imaging studies
of the Galactic centre. A dense cluster of stars surrounds Sgr A, and
proper motions of these stars were recently obtained, showing them to
have high velocities of up to 5000 \kms. Thereby the mass of the central
black hole has been estimated to be ($3.3\pm0.7)\times 10^6$ M$_\odot$
\citep[e.g.][]{schodel03,ghez00}.  Massive star formation is still going
on in the central parsec of the Galaxy.

\subsection{Stellar line of sight velocity surveys and the importance of maser surveys}
Though the AGB phase is very short ($\sim 10^6$ yr) and therefore AGB
stars are rare among stars, they are representative of all low and
intermediate mass stars, i.e.\ of the bulk of the Galactic population.
They are evolved stars and therefore dynamically relaxed and their
kinematics traces the global Galactic gravitational potential.
Thermally pulsing AGB stars are surrounded by a dense envelope of
dust and molecular gas. They are bright at infrared wavelengths and
can be detected even throughout highly obscured regions.  Furthermore,
the OH and SiO maser emission from their envelopes can be detected
throughout the Galaxy, providing stellar line-of-sight velocities to
within a few km s$^{-1}$.  AGB stars thus permit a study of the
Galactic kinematics, structure and mass-distribution.

This is especially useful in the inner regions of the Galaxy where the
identification of other tracers like Planetary Nebulae is extremely
difficult. Observations of H$\alpha$ and [OIII] emission lines which
easily reveal velocities of Planetary Nebulae are hampered by high
interstellar extinction.  A dynamical study of planetary nebulae
($-5$\degr$<b<-10$\degr) was performed by \citet{beaulieu00}, who found that
the spatial distribution of planetary nebulae agrees very well with
the COBE light distribution. However, no conclusive results were found
comparing the stellar kinematics properties with models of a barred
Galaxy.  The poor statistics was the main problem.

Performing radio maser surveys is the most efficient way to obtain
line of sight velocities in the inner Galaxy.  Two extensive blind
surveys have been made at 1612 MHz searching for OH/IR stars in the
Galactic plane ($|l|<45$\degr, $|b|<4$\degr), one in the South using
the ATCA, and another in the North using the VLA
\citep{sevenster97a,sevenster97b,sevenster01}, yielding a sample of
766 compact OH-masing sources.

Searches at 43 or 86 GHz for SiO maser emission are also
successful. SiO maser lines have the advantage to be found more
frequently than 1612 MHz OH maser and the disadvantage that they can
only be searched in targeted surveys, since the cost of an unbiased
search is too high. Several 43 GHz SiO maser surveys of IRAS point
sources have been conducted by Japanese groups using the Nobeyama
telescope \citep[e.g.][]{izumiura99,deguchi00a,deguchi00b}. However,
those surveys are not complete at low latitudes, since there IRAS
suffers from confusion.

Up to day more than 1000 maser stars are known in the inner Galaxy.  A
kinematical analysis of Sevenster's sample of OH/IR stars
\citep{sevenster99,debattista02} shows clear signs of a barred
potential. However, the number of line of sight velocities is still
too small to allow us an unambiguous determination of the parameters
of the bar. Furthermore, although most of the Galactic mass is in
stars, a stellar longitude-velocity, $(l - v)$, diagram alone is not
sufficient to constrain a model of Galactic dynamics, mainly due to
the dispersion velocity of stars, which smooths the various features.

Improved statistics together with additional information on the
distance distribution of masing stars will notably improve the
understanding of the Galactic $(l - v)$ diagram. Low latitude stars
are of particular importance since their motion may contain a
signature of in-plane Galactic components, e.g. the nuclear ring, and
they may show better the effect of a thin bar. New targeted maser surveys
in the Galactic plane are now possible using ISOGAL and MSX sources,
it is this simple idea from which the work presented in this thesis
originates.

\section{Outline of this thesis}
To increase the number of measured line of sight velocities in the
inner Galaxy ($30^\circ < l < -30 ^\circ$, mostly at $|b|<1$\degr), we
began a survey of 86 GHz ($v = 1, J = 2 \rightarrow 1$) SiO maser
emission.  In {\bf Chapter 2} we present the survey that was conducted
with the IRAM 30-m telescope. Stars were selected from the ISOGAL and
MSX catalogues to have colours of Mira-like stars. SiO maser emission
was detected in 271 sources (a detection rate of 61\%), doubling the
number of maser derived line-of-sight velocities toward the inner
Galaxy.  I observed and detected the first line on August 26th, 2000:
it was an unforgettable moment of joy!

The collection of near- and mid-infrared measurements of SiO targets
allow us to study their energy distribution and determine their
luminosity and mass-loss.  {\bf Chapter 3} describes a compilation of
DENIS, 2MASS, ISOGAL, MSX and IRAS 1--25\um\ photometry of the 441
late-type stars which we searched for 86 GHz SiO maser emission.  The
comparison between DENIS and 2MASS $J$ and $K_{\rm S}$ magnitudes
shows that most of the sources are variable stars.  MSX colours and
the IRAS $[12]-[25]$ colour are consistent with those of Mira type
stars with a dust silicate feature at 9.7 $\mu$m in emission,
indicating only a moderate mass-loss rate.

Towards the inner Galaxy the visual extinction can exceed 30
magnitudes, and even at infrared wavelengths the extinction is
significant.  In {\bf Chapter 4} we carry out the analysis of 2MASS
colour magnitude diagrams of several fields in the plane at longitudes
$l$ between 0 and 30$^\circ$ in order to obtain extinction estimates
for all SiO targets.  With this analysis we are also able to put new
constraints on the near-infrared extinction power-law.

The luminosity of our SiO targets is derived in {\bf Chapter 5} and
compared to  that of a sample of OH/IR stars.  We computed stellar
bolometric magnitudes by direct integration under the observed energy
distribution.  Assuming a distance of 8 kpc for all stars within
5\degr\ from the Galactic centre we find the luminosity distribution
to peak at M$_{\mathrm {bol}}= -5$ mag, which coincides with the peak
shown by OH/IR stars in the Galactic centre.  We found that the main
difference between SiO targets and OH/IR stars is mass loss, which is
higher in OH/IR stars.  This fact offers several advantages. In
contrast to OH/IR stars, SiO target stars are readily detectable in the
near-infrared and therefore ideal for follow-up studies to better
characterise the central star.

Considerations on the kinematics of  SiO targets and future work
plans are reported in {\bf Chapter6}.

Finally in the last Chapter ({\bf Chapter 7}) I briefly describe the
ISOGAL survey, which is a 7 and 15 \um\ survey of $\sim$16 deg$^2$
towards selected fields along the Galactic plane, mostly toward the
Galactic centre.  In collaboration with A. Omont (P.I.) and the ISOGAL
team, I worked on the finalisation of the ISOGAL point source
catalogue \citep{omont03,schuller03}.  In this Chapter, I emphasise
the importance of having several recent infrared surveys, such as
DENIS, 2MASS, ISOGAL and MSX, in a common effort to unveil the
overall structure of the Milky Way and in particular of its central
and most obscured regions.  These surveys require a huge amount of
technical work which is of primary importance to obtain a reliable
point source catalogue that can be used to perform such studies.

It is thanks to these new catalogues that the SiO maser project,
i.e. the present thesis, could be performed.

\setlength{\bibsep}{0.5mm}

\addcontentsline{toc}{section}{Bibliography}
 

   \chapter{
   86 GHz SiO maser survey of late-type stars in the Inner Galaxy
   I. Observational data}
\chaptermark{SiO maser survey I. Observational data}

\begin{authorline}  
	M.\ Messineo, H.\ J.\ Habing, L.\ O.\ Sjouwerman,
        A.\ Omont, and  K.\ M.\ Menten
	\journal{Astronomy and Astrophysics (2002), 393, 115; (astro-ph/0207284)}
\end{authorline}

  \begin{abstract}
  We present 86 GHz ($v = 1, J = 2 \rightarrow 1$) SiO maser line
  observations with the IRAM 30-m telescope
  of a  sample of 441 late-type stars in the Inner Galaxy ($ -4 \degr <
  l < +30 \degr$).
  These stars were selected on basis of their infrared magnitudes 
  and colours from the ISOGAL and MSX catalogues. 
  SiO maser emission was detected in 271 sources, and their
  line-of-sight velocities indicate that the stars are 
  located in the Inner Galaxy. 
  These new detections double the number 
  of line-of-sight velocities  available from previous SiO and OH
   maser observations
  in the area covered by our survey and are, together with other samples
  of e.g.\ OH/IR stars, useful for kinematic studies of the
  central parts of the Galaxy.
  \end{abstract}
  
  This chapter is available at:\\
  ${\tt http://lanl.arxiv.org/abs/astro-ph/0207284}$ or \\
  ${\tt http://www.edpsciences.org/articles/aa/pdf/2002/37/aa2744.pdf}$.
  

%
   \chapter{
   86 GHz SiO maser survey of late-type stars in the Inner Galaxy
   II. Infrared photometry}

\chaptermark{SiO maser survey II. Infrared photometry}

\begin{authorline}  
	M.\ Messineo, H.\ J.\ Habing, K.\ M.\ Menten, 
        L.\ O.\ Sjouwerman and A.\ Omont  
	\journal{Astronomy and Astrophysics (2004), 418, 103; (astro-ph/0401262)}
\end{authorline}

   \begin{abstract}We present a compilation and study of DENIS, 2MASS,
ISOGAL, MSX and IRAS 1--25$\mu$m photometry for a sample of 441
late-type stars in the inner Galaxy, which we previously searched for
86 GHz SiO maser emission  \defcitealias{messineo02}{Chapter\,II}
 \citepalias{messineo02}.  The comparison of the
DENIS and 2MASS $J$ and \ks\ magnitudes shows that most of the SiO
targets are indeed variable stars.  The MSX colours and the IRAS
$[12]-[25]$ colour of our SiO targets are consistent with those of
Mira type stars with dust silicate feature at 9.7 $\mu$m feature in
emission, indicating only a moderate mass-loss rate.  
  \end{abstract}
  This chapter is available at:\\
  ${\tt http://lanl.arxiv.org/abs/astro-ph/0401262}$ or \\
  ${\tt http://www.edpsciences.org/articles/aa/full/2004/16/aa0488/aa0488.html}$.

   \chapter{
   86 GHz SiO maser survey of late-type stars in the Inner Galaxy
   III. Interstellar extinction and colours }
\chaptermark{Interstellar extinction and colours }
 
\begin{authorline}
        M.\ Messineo, H.\ J.\ Habing, K.\ M.\ Menten,
        A.\ Omont,\\ L.\ O.\ Sjouwerman and  F.\ Bertoldi
        \journal{Astronomy and Astrophysics (2004), submitted}
\end{authorline}
 
\begin{abstract} 
 We have computed extinction corrections for a sample of 441
late-type stars in the inner Galaxy using the 2MASS near-infrared
photometry of the surrounding stars and assuming the intrinsic source
colours.  From this, the near-infrared power law is found to be
A$_\lambda \propto \lambda^{\mathrm -1.9\pm0.1}$.  Near- and
mid-infrared colour-colour properties of known Mira stars are also
reviewed.  From the distribution of the dereddened infrared colours of
the SiO target stars we infer mass-loss rates between $10^{-7}$ and
$10^{-5}$ M$_\odot$ yr$^{-1}$.
\end{abstract}

In this article we study the interstellar extinction toward a sample
of evolved late-type stars in the inner Galaxy (-4\degr\ $< l <
+30$\degr, $|b|<1$\degr) which were searched for SiO maser emission
\citep[``SiO targets'' hereafter; ][ Chapter\,II] {messineo02}.
\defcitealias{messineo02}{Chapter\,II} The maser emission reveals the
stellar line of sight velocities with an accuracy of a few \kms,
making the maser stars ideal for Galactic kinematics studies.

The combination of the kinematic information with the physical
properties of the SiO targets, e.g.\ their intrinsic colours and
bolometric magnitudes, will enable a revised kinematic study of the
inner Galaxy, revealing which Galactic component and which epoch of
Galactic star formation the SiO targets are tracing.

A proper correction for interstellar extinction is of primary
importance for our photometric study of the stellar population of the
inner Galaxy, where extinction can be significant even at infrared
wavelengths. The extinction hampers an accurate determination of the
stellar intrinsic colours and bolometric magnitudes.

This is especially critical in the central Bulge region where
interstellar extinction is larger than 30 visual magnitudes and the
uncertainty in the stellar bolometric luminosities of evolved
late-type stars is at least 1 magnitude due to the current uncertainty
in the near-infrared extinction law (30\%).

The available near- and mid-infrared photometry of the SiO
targets from the DENIS\footnote{DEep Near-Infrared Survey of the
southern sky; see\\ ${\tt http://www{\rm-}denis.iap.fr/}$. }
\citep{epchtein94}, 2MASS\footnote{Two Micron All Sky Survey; see\\
${\tt http://www.ipac.caltech.edu/2mass/}$.} \citep{2massES},
ISOGAL\footnote{A deep survey of the obscured inner Milky Way with ISO
at 7\um\ and at 15\um; see ${\tt http://www{\rm-}isogal.iap.fr/}$.}
\citep{omont03,schuller03} and MSX\footnote{The Midcourse Space
Experiment; see\\ ${\tt
http://www.ipac.caltech.edu/ipac/msx/msx.html}$.}
\citep{egan99,price01} surveys were already presented by
\citet[][ Chapter\,III] {messineo03_2}. \defcitealias{messineo03_2}{Chapter\,III}
  
The corrections for interstellar extinction of the photometric
measurements of each SiO target will enable us to derive the spectral
energy distributions and bolometric magnitudes of the SiO targets.
The bolometric magnitudes will be presented in a subsequent paper
\citep[][ Chapter\,V] {messineo03_4}.
\defcitealias{messineo03_4}{Chapter\,V}

Our sample consists mainly of large-amplitude variable AGB stars
\citepalias{messineo02,messineo03_2}.  The estimates of interstellar
extinction toward this class of objects are complicated by the
presence of a circumstellar envelope which may have various thickness.
Therefore, in order to disentangle circumstellar and interstellar
extinction one needs to study the dust distribution along the line of
sight toward each AGB star of interest.  For each SiO target we adopt
the median extinction derived from near-infrared field stars (mainly
giants) close to the line of sight of the target. Then the dereddened
colour-colour distribution of our targets is compared to those of
local Mira stars in order to iteratively improve the extinction
correction and to statistically estimate the mass-loss rates of our
targets.

In Sect.\ \ref{law} we discuss the uncertainty of the extinction law
at near- and mid-infrared wavelengths, and the consequent uncertainty
of the stellar luminosities. In Sect.\ \ref{fieldextinction} we
describe the near-infrared colour-magnitude diagrams of field stars
toward the inner Galaxy and we use the latter to derive the median
extinction toward each target. In Sect.\ \ref{miras} we review the
location of Mira stars on the colour-magnitude (CMD) and colour-colour
diagrams. In Sects.\ \ref{siotargets} and \ref{massloss} we use the
median extinction from surrounding field stars to deredden our SiO
targets and we discuss their colours and mass-loss rates. The main
conclusions are given in Sect.\ \ref{conclusion}.

\section{Interstellar extinction law}\label{law}
The composition and abundance of interstellar dust and its detailed
extinction properties remain unclear, limiting the accuracy of stellar
population studies in the inner Galaxy.  In the following we discuss
the near- and mid-infrared extinction law, in order to assess the
uncertainty in the extinction correction.

\subsection{Near-infrared interstellar extinction}\label{nearextinction}
Interstellar extinction at near-infrared wavelengths (1-5 $\mu$m)
is dominated by graphite grains. Although for historical reasons the
near-infrared extinction law is normalised in the visual, practically
it is possible to derive near-infrared extinction by measuring the
near-infrared reddening of stars of known colour.

Near-infrared photometric studies have shown that the
wavelength-dependence of the extinction may be expressed by a power
law, {\it \Al} $\propto \lambda^{-\alpha}$, where $\alpha$ was found
to range between 1.6 \citep{rieke85} and 1.9 \citep{glass99,
landini84,vandehulst46}.

When deriving the extinction from broad-band photometric measurements,
one needs to properly account for the bandpass, stellar spectral
shape, and the wavelength-dependence of the extinction.  We have
therefore computed an ``effective extinction'' for the
DENIS $I$ and 2MASS  $J,~H$ and $K_{\rm s}$
passbands, as a function of the \ks\ band extinction. This effective
extinction was computed by reddening an M0 III stellar spectrum
\citep{fluks94} with a power law extinction curve and integrating it
over the respective filter transmission curves. When we convolve the
filter response with a stellar sub-type spectrum different from the M0
III, the effective I-band extinction slightly differs, e.g.\
decreasing by 3\% for a M7 III spectrum \citep[see
also][]{vanloon03}.

The \ks-band extinction \Aks\ can then be found from
$$A_{\rm K_{\rm s}} = C_{JK} \times E(J-K_{\rm s}),$$
$$A_{\rm K_{\rm s}} = C_{HK} \times E(H-K_{\rm s}),$$ where
$E(J-K_{\rm s})$ and $E(H-K_{\rm s})$ are the reddening in the
$J-K_{\rm s}$ and $H-K_{\rm s}$ colour, respectively, and the $C$ are
constants.  These relations are independent of visual extinction and
of the coefficient of selective extinction, $R_{\rm V} = A_{\rm
V}/E(B-V)$, but they depend on the slope of the near-infrared power
law (see Table \ref{table:nearextinction}).  However, to provide the
reader with the traditionally used ratios between near-infrared
effective extinction and visual extinction, we also used the commonly
adopted extinction law of \citet{cardelli89}.  Such ratios may be
useful in low-extinction Bulge windows, where visual data are also
available. \citet{cardelli89} proposed an analytic expression, which
depends only on the parameter $R_{\rm V}$, based on multi-wavelength
stellar colour excess measurements from the violet to 0.9$\mu$m, and
extrapolating to the near-infrared using the power law of
\citet{rieke85}. We extrapolated Cardelli's extinction law to
near-infrared wavelengths using a set of different power laws. The
results are listed in Table \ref{table:nearextinction}.

The uncertainty in the slope of the extinction law produces an
uncertainty in the estimates of the near-infrared extinction of
typically 30\% in magnitude (see Table \ref{table:nearextinction}).
For a \ks\ band extinction of \Aks$=3$ mag the uncertainty may be up
to 0.9 mag, which translates into an uncertainty in the stellar
bolometric magnitudes of the same magnitude.

In Sect.\ \ref{fieldextinction} we show that a power law index
$\alpha=1.6$ is inconsistent with the observed colours of field
giant stars toward the inner Galaxy, and that the most likely value of
$\alpha$ is $1.9\pm 0.1$.

\begin{sidewaystable}
\caption{\label{table:nearextinction} Near-infrared effective
extinction, $\langle$A$\rangle$/A$_{V} \propto \lambda^{-\alpha},$ for
various filters.  A different value of $R_{\rm V}$ does not affect 
{\rm A}$_{\rm K_{\rm s}}/{\rm E(H-K}_{\rm s})$ and {\rm A}$_{\rm
K_{\rm s}}/{\rm E(H-K}_{\rm s})$, but the slope of the
extinction curve does. Our findings favour a model with $\alpha=1.9$
(see Sect.\ \ref{fieldextinction}).}
\begin{tabular}{cccccclll}
\hline
\hline
{\rm A$_{\rm I}/{\rm A}_{\rm V}$}&{\rm A$_{\rm J}/{\rm A}_{\rm V}$}&{\rm A$_{\rm H}/{\rm A}_{\rm V}$}&{\rm A$_{\rm K_S}/{\rm A}_{\rm V}$}& {\rm A}$_{\rm K_S}/{\rm E(J-K}_{\rm s})$ &
{\rm A}$_{\rm K_{\rm s}}/{\rm E(H-K}_{\rm s})$&$\alpha$&{\rm R$_{\rm V}$} & Ref.\ \\
\hline
0.592&0.256&0.150&0.089& 0.533&1.459&1.85&    &\citet{glass99}\\
0.584&0.270&0.165&0.103& 0.617&1.661&1.73&3.08&\citet{he95}\\
0.482&0.282&0.175&0.112& 0.659&1.778&1.61&3.09&\citet{rieke85}\\
0.606&0.287&0.182&0.118& 0.696&1.842&1.61&3.10&\citet{cardelli89} \\
0.563&0.259&0.164&0.106& 0.696&1.842&1.61&2.50$^{\mathrm{*}}$&''\\
0.606&0.277&0.169&0.106& 0.623&1.684&1.73&3.10&\citet{cardelli89}$^{\mathrm{+}}$\\
0.563&0.249&0.152&0.096& 0.623&1.684&1.73&2.50&''\\
0.606&0.267&0.158&0.096& 0.561&1.548&1.85&3.10&''\\
0.563&0.240&0.142&0.086& 0.561&1.548&1.85&2.50&''\\
0.606&0.263&0.153&0.092& 0.537&1.496&1.90&3.10&''\\
0.563&0.237&0.138&0.083& 0.537&1.496&1.90&2.50&''\\
0.606&0.255&0.144&0.084& 0.493&1.401&2.00&3.10&''\\
0.563&0.229&0.130&0.076& 0.494&1.400&2.00&2.50&''\\
0.606&0.238&0.127&0.070& 0.420&1.236&2.20&3.10&''\\
0.563&0.213&0.114&0.063& 0.420&1.236&2.20&2.50&''\\
\hline
\end{tabular}
\begin{list}{}{}
\item[$^{\mathrm{*}}$] recent determination toward the Bulge 
\citep[e.g.][]{udalski03}.
\item[$^{\mathrm{+}}$] parametric expression modified to 
extrapolate to $\lambda >0.9~ \mu$m with $\lambda^{-\alpha}$.
\end{list}
\end{sidewaystable}

\subsection{Mid-infrared interstellar extinction}
Mid-infrared extinction (5-25 $\mu$m) is characterised by the 9.7
and 18 $\mu$m silicate features. The strength and profile of these
features are uncertain and appear to vary from one line of sight to
another.  In the inner Galaxy silicate grains may be more abundant due
to the outflows from oxygen-rich AGB stars.  Another uncertainty
is the minimum of $A_{\lambda}/A_{2.12}$ at 7$\mu$m, which is
predicted for standard graphite-silicate mixes, though not observed to
be very pronounced toward the Galactic Centre \citep{lutz96,lutz99}.

\begin{sidewaystable}
\caption{\label{table:extinction} Effective extinction,
$\langle$A$\rangle$/A$_{K_S}$, using M-giant spectra \citep{fluks94},
for different bands defined by the ISOCAM and MSX filters (see
Fig.\ \ref{fig:filter.ps}). $A_{K_S}/A_{2.12}=0.97$. }
\begin{tabular}{lrrcccc}
\hline
\hline
{\rm Filter}& $\lambda_{\rm ref}$&$\Delta \lambda$ & {\rm Curve 1 (Mathis)} &{\rm Curve 2 }&{\rm Curve 3 (Lutz)}\\
            &                    &                 & {\rm  ($A_{9.7}/A_{2.12}=0.54$)}   &{\rm  ($A_{9.7}/A_{2.12}=1.00$)}&{\rm ($A_{9.7}/A_{2.12}=1.00$ \& no minimum)}\\
{\rm }&{\rm $\mu$m }& {\rm  $\mu$m}&{\rm $\langle$A$\rangle$/A$_{K_S}$ }& {\rm $\langle$A$\rangle$/A$_{K_S}$ }&{\rm $\langle$A$\rangle$/A$_{K_S}$}\\
\hline
{\rm LW}2& 6.7&3.5& 0.21&  0.21&0.41\\
{\rm LW}5& 6.8&0.5& 0.18&  0.15&0.41\\
{\rm LW}6& 7.7&1.5& 0.21&  0.26&0.43\\
{\rm LW}3&14.3&6.0& 0.18&  0.34&0.34\\
{\rm LW}9&14.9&2.0& 0.14&  0.29&0.29\\
{\rm A}  &8.28&4.0& 0.26&  0.38&0.55\\
{\rm C}  &12.1&2.1& 0.25&  0.49&0.49\\
{\rm D}  &14.6&2.4& 0.14&  0.29&0.29\\
{\rm E}  &21.3&6.9& 0.17&  0.41&0.41\\
\hline
\end{tabular}
\end{sidewaystable}

The commonly adopted mid-infrared extinction curve is that of
\citet{mathis90}, which is a combination of a power law and the
astronomical silicate profile from \citet{draine84}, with
$A_{9.7}/A_{2.2} \simeq 0.54$ -- a value found in the diffuse
interstellar medium toward Wolf-Rayet stars \citep[e.g.][ and
references therein]{mathis98}. However, using hydrogen recombination
lines, \citet{lutz99} found $A_{9.7}/A_{2.2} \simeq 1.0$ in the
direction of the Galactic centre, and analysing the observed H$_2$
level populations toward Orion OMC-1, \citet{rosenthal00} derived
$A_{9.7}/A_{2.12} = 1.35$. It seems that the mid-infrared extinction
law is not universal.

\begin{figure}[th]
\begin{center}
\resizebox{0.7\hsize}{!}{\includegraphics{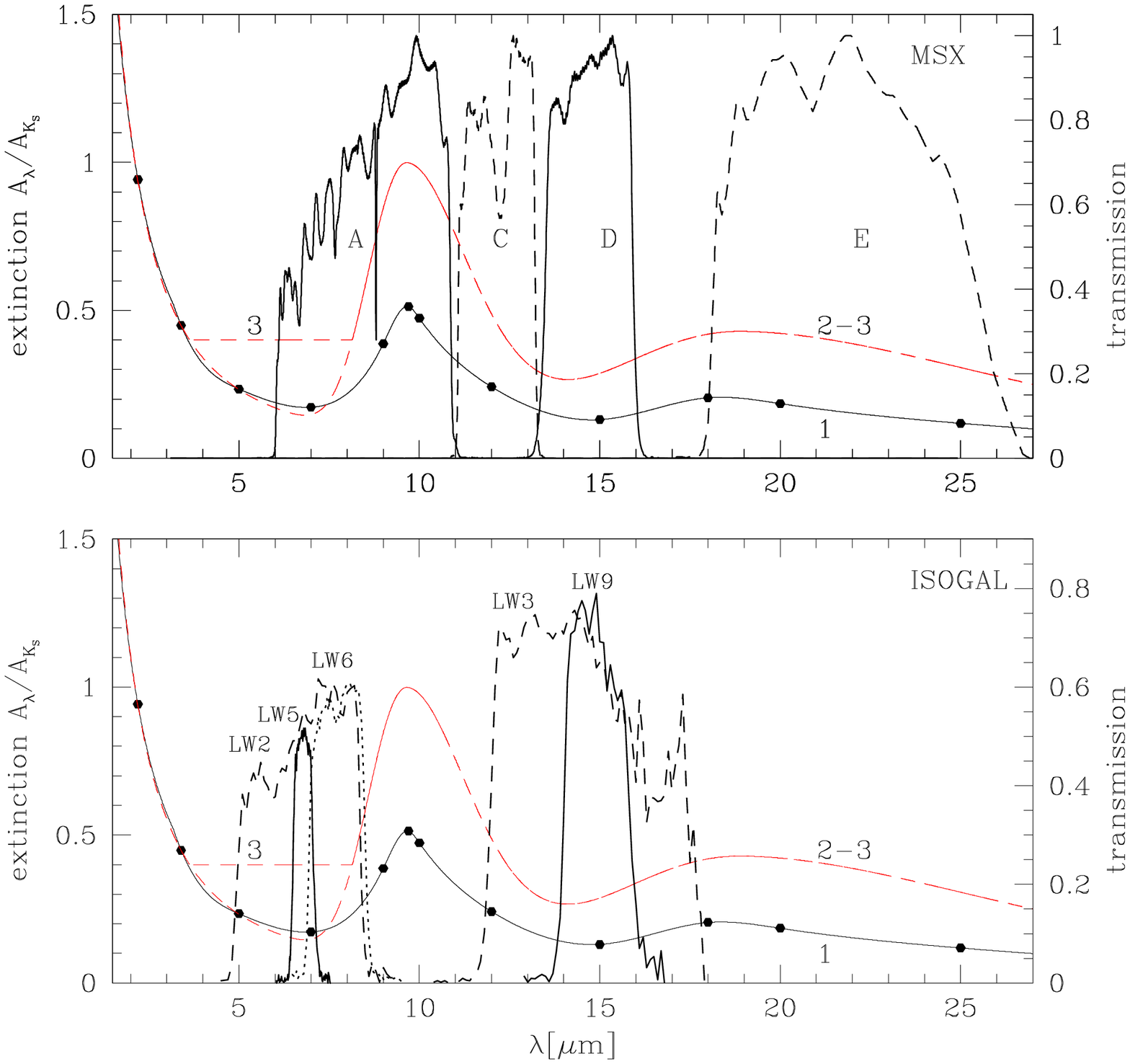}}
\end{center}
\caption{\label{fig:filter.ps} Filter transmission curves and
extinction laws as function of wavelength. The continuous line shows
the curve (Curve 1) obtained by fitting the values (dots) given by
\citet{mathis90}; the dashed curve shows the parametric expression
given by \citet{rosenthal00} plotted using a value of the silicate
peak $A_{9.7}/A_{2.12}=1.0$ (Curve 2). The latter is also shown
without the minimum around 4-8 $\mu$m (Curve 3), following
\citet{lutz99}.  {\bf In the top panel} the transmission curves of the
MSX $A, C, D$ and $E$ filters are also overplotted, while {\bf in the
bottom panel} the transmission curves of the ISOCAM $LW2, LW3, LW5,
LW6$ and $LW9$ filters used in the ISOGAL survey are shown.}
\end{figure}

In order to derive the extinction ratios for all  ISOGAL and MSX
filters \citep[for definitions see ] []{isocam03,price01}, and to
analyse the effect of an increase of the depth of the 9.7 $\mu$m
silicate feature on the extinction ratios, we built a set of
extinction curves with different silicate peak strengths at 9.7
$\mu$m. We use a parametric mid-infrared extinction curve given by
\citet{rosenthal00}, where the widths of the 9.7 and 18 $\mu$m
silicate features are those calculated by \citet{draine84} and the
depth of the 18 $\mu$m feature is assumed to be 0.44 times that of the
9.7 $\mu$m feature.  Using this parametric fit we constructed two
different extinction curves with $A_{9.7}/A_{2.2}$ equal to 1.0, one
in combination with the minimum predicted by the models at 4-8 $\mu$m
(Curve 2) and one without it as suggested by \citet{lutz99}
(Curve 3).  The two curves are shown together with the Mathis curve
(Curve 1, $A_{9.7}/A_{2.2}=0.54$) in Fig.\ \ref{fig:filter.ps}.

Using the various extinction curves detailed in Table
\ref{table:extinction}, we reddened the M-type synthetic spectra from
\citet{fluks94} (beyond 12.5$\mu$m a blackbody extrapolation is used),
and convolved the resulting spectra with the ISOCAM and MSX filter
transmission curves.  The effective extinctions $\langle A
\rangle$/\Aks\ in the various ISOCAM and MSX filters are listed in
Table \ref{table:extinction}.  They are not sensitive to the stellar
sub-type used.  An increase of the ratio $A_{9.7}/A_{2.2}$ from 0.54
to 1.0 results in an increase between 0.15 and 0.20$\times$\Aks\ of
the average attenuation in the $LW3,~LW9,~C,~D$ and $E$ spectral
bands.  The spectral bands of the $LW2$ and $LW5$ filters are not very
sensitive to the intensity of the silicate feature, but to the minimum
of the extinction curve in the 4-8 $\mu$m region. Although $\langle A
\rangle$/\Aks\ varies with \Aks, these variations are small compared
to those arising from different choices of the mid-infrared extinction
law.

\citet{hennebelle01} obtained observational constraints on
mid-infrared extinction ratios from observations of infrared dark
clouds within the ISOGAL survey. Using observations in the $LW2$ and
$LW3$ bands in the inner Galactic disk they obtained $A_{\rm
LW2}/A_{\rm LW3} = 0.7$, and using observations in the two bands $LW6$
and $LW9$ in the region ($|l|<1^\circ,~ 0.2^\circ < |b| <0.4^\circ$)
they found $A_{\rm LW6}/A_{\rm LW9}=0.8$.  Both these values are in
good agreement with the extinction curve calculated by
\citet{draine84} with a silicate peak at 9.7 $\mu$m of 1.0 (Curve 2 in
Table \ref{table:extinction}).  However, for the clouds located at
($|l|<1^\circ, |b|<0.2^\circ$) observed in the $LW5$ and $LW9$ bands
\citet{hennebelle01} found  $A_{\rm LW5}/A_{\rm LW9} = 1.07$, which
is twice the value predicted by Curve 2 in Table
\ref{table:extinction}, but would be consistent with Curve 1 or 3.

For stars in the inner disk \citet{jiang03} derived (\Aks$-A_{\rm
LW2}) / (A_{\rm J}-$\Aks) $ = 0.35$ and (\Aks$-A_{\rm LW3})$ $/ (A_{\rm
J}-$\Aks) $ = 0.39$. These ratios when combined with the near-infrared
extinction law imply that $A_{\rm LW2}/$\Aks\ must range from 0.35 to
0.47 and $A_{\rm LW3}/$\Aks\ from 0.28 to 0.41, which are higher
values than those produced by Curve 1 and suggest an attenuation of
the minimum at 4-8 $\mu$m, consistent with Curve 3.

Concluding, there is some uncertainty in the mid-infrared
extinction law that is in part due to uncertainties in the photometric
measurements and possibly due to spatial variations in the strength of
the silicate features.  In the most obscured regions (\Aks$=3$) 
 uncertainties for the ISOGAL and MSX filters
range from  0.45 mag ($LW3,LW9,D$) to 0.85 mag ($A$).  However,
this has a negligible effect (0.1 mag in average) on the calculated
$M_{\mathrm {bol}}$ of the SiO targets because their energy is emitted
mostly at near-infrared wavelengths, and has therefore  also a
negligible effect on the mass-loss rate estimates (see Sect.\
\ref{massloss}).

In the following we will use the Lutz law (Curve 3) to deredden the
colours of our SiO targets, since this law ensures a consistency
between mid-infrared and near-infrared stellar colours as found by
\citet{jiang03}.

\section{Interstellar extinction of field stars from 
near-infrared colour-magnitude diagrams}
\label{fieldextinction}

\begin{figure*}[th!]
\resizebox{0.5\hsize}{!}{\includegraphics{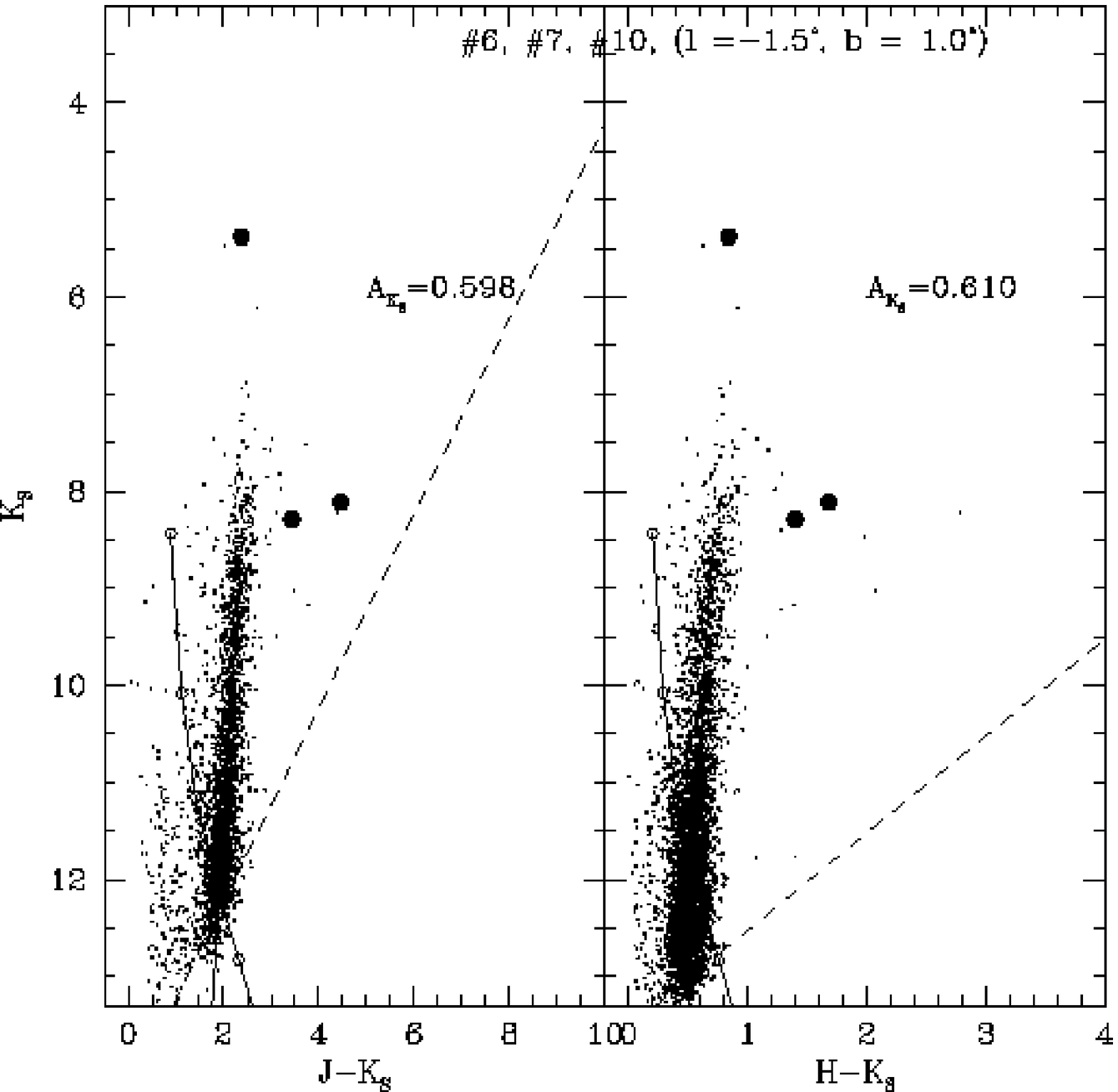}}
\resizebox{0.5\hsize}{!}{\includegraphics{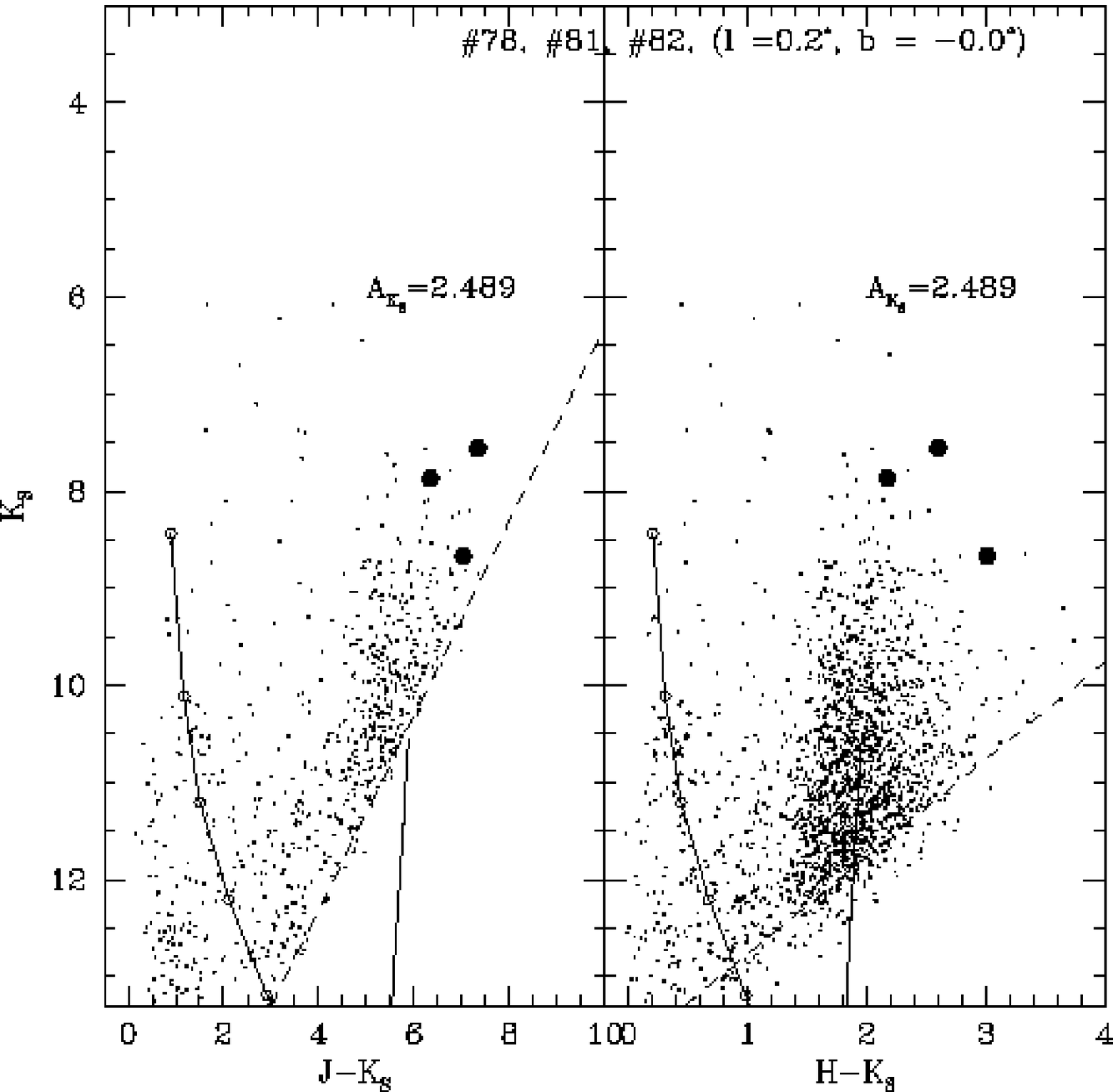}}
\resizebox{0.5\hsize}{!}{\includegraphics{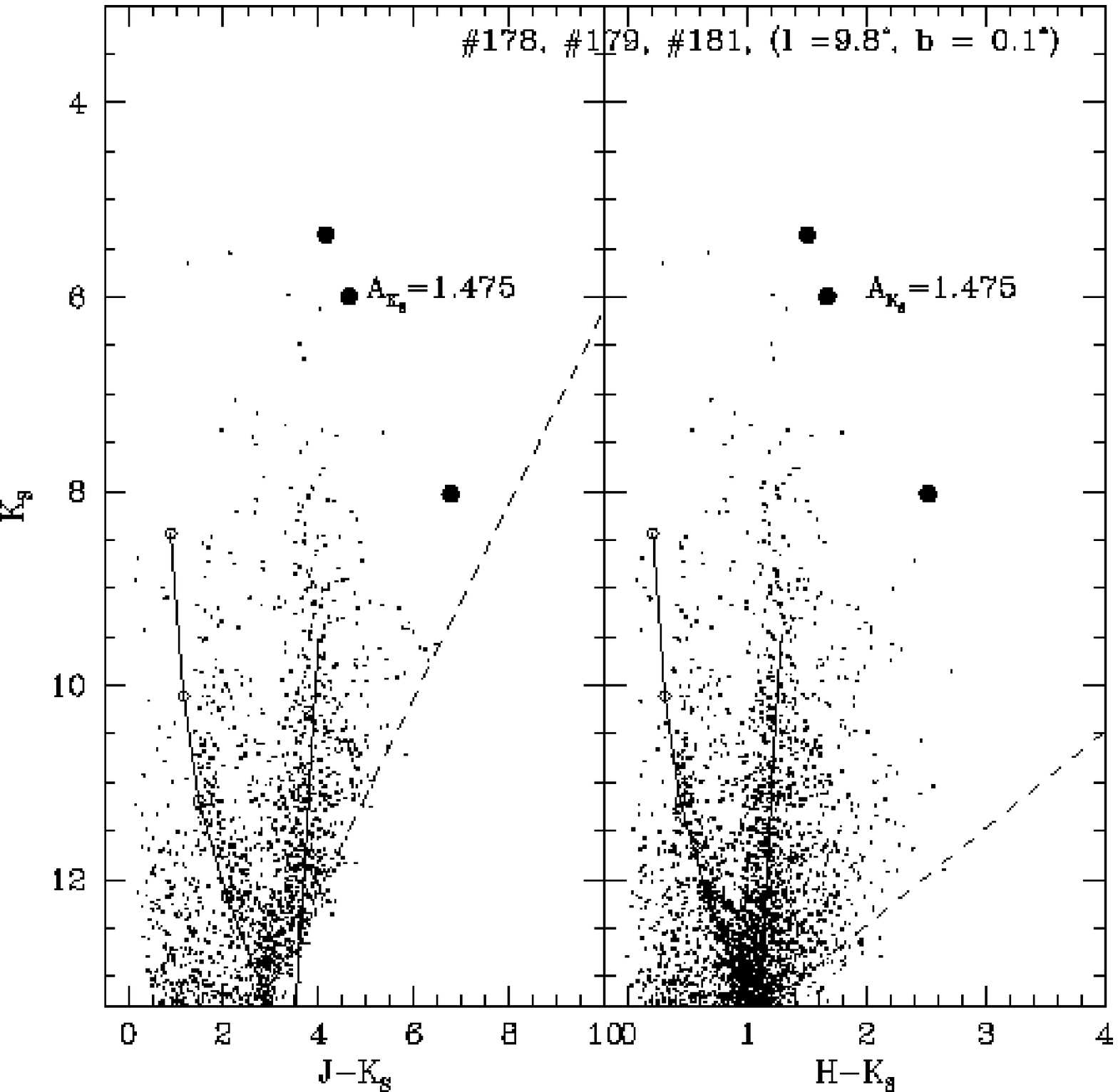}}
\resizebox{0.5\hsize}{!}{\includegraphics{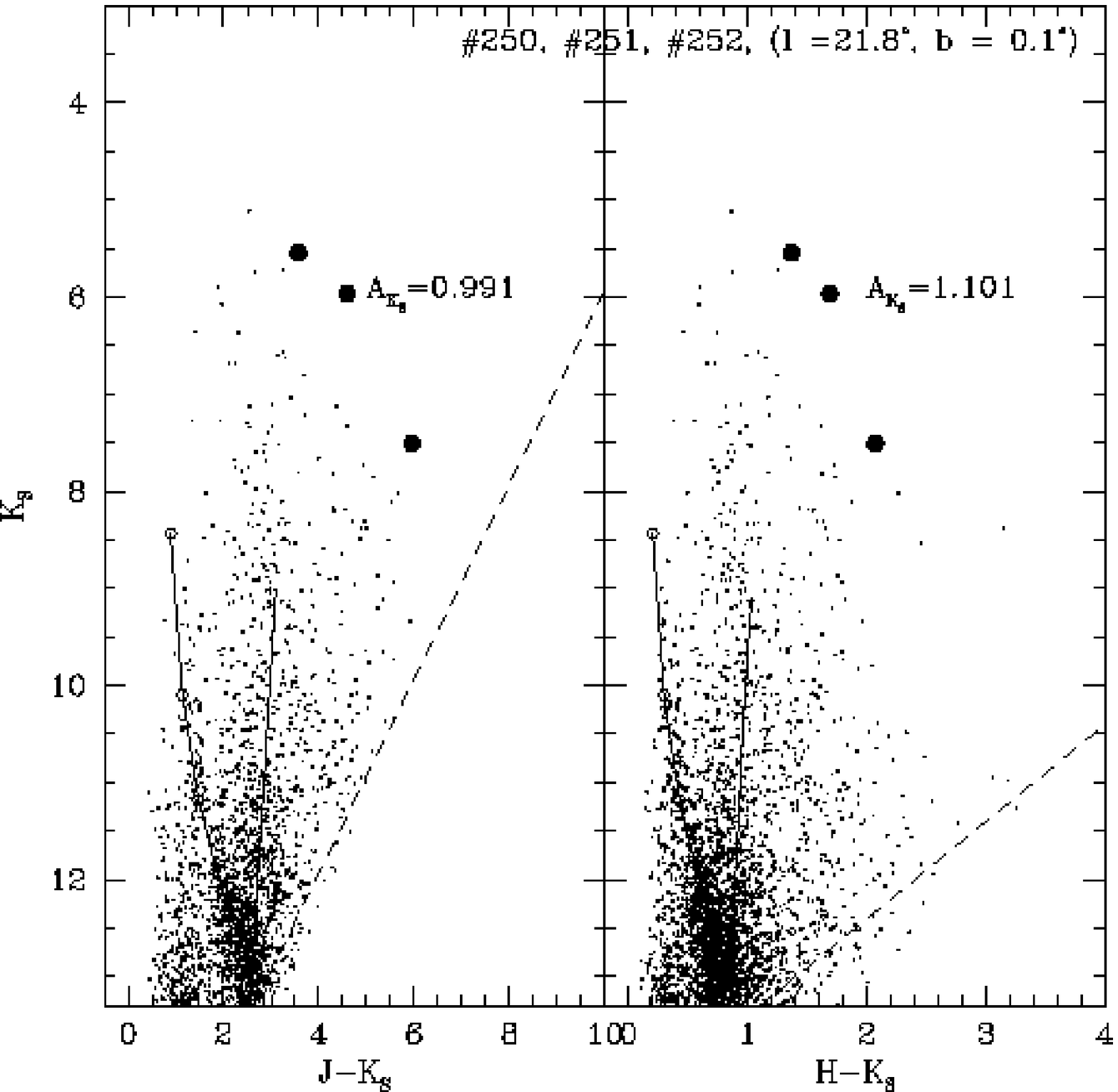}}
\caption{\label{fig:panels} Colour-magnitude diagrams of 2MASS
datapoints (small dots) of good quality located within 4\arcmin\ from
the position of the SiO target (big dot). Three fields at equal median
extinction are combined in each panel. The right-hand continuous line
indicates the locus of the reference RGB curve (see Sect.\ \ref{rgb}),
adopting a distance of 8 kpc and reddening it with the median
extinction of field stars (\Aks). The left-hand continuous curve shows
the trace of clump stars for increasing distance and extinction along
a given line of sight (see Sect.\ \ref{outside}), obtained using the
extinction model by \citet{drimmel03} and the absolute magnitudes from
\citet{wainscoat92}. Dashed lines indicate the diagonal cut-offs
due to the detection limits in $J$ and $H$. Circles on the clump
trace mark a distance from 1 to 5 kpc with a step of 1 kpc downward.}
\end{figure*}

Most of the sources detected by DENIS and 2MASS toward the inner
Galaxy are red giants and asymptotic giant branch stars. Because the
intrinsic $(J-K_{S})_0$ colours of giants are well known and steadily
increase from 0.6 to 1.5 mag with increasing luminosity, one can study
the colour-magnitude diagrams (CMDs) of $(J-K_{\rm s})$ versus $K_{\rm
s}$ and of $(H-K_{\rm s})$ versus $K_{\rm s}$ to estimate the average
extinction toward a given line of sight for a population of such
stars.

Under the assumption that our SiO targets are spatially well mixed
with the red giant stars, and that the interstellar extinction is
uniform over a 4\arcmin$ \times 4$\arcmin\ field (corresponding to
9$\times$9 pc$^2$ at the distance of the Galactic Centre), we estimate
the extinction, \Aks, towards our 441 SiO targets by examining the
CMDs of 2MASS sources in field of 2-4\arcmin\ radius around each SiO
target. Figure 2 shows a representative sample of these CMDs.  We
assume that the red giant branch (RGB) has the same intrinsic shape
for all red giants in the inner Galaxy: the absolute magnitude of the
tip of the RGB, $M_{\rm K_{\rm s}}(tip)$, and the RGB colour-magnitude
relation does not vary.  This means that at a given distance, $d$,
along the line of sight the observed RGB extends towards fainter
magnitudes from the tip at magnitude $K_{\rm s}(tip) = M_{\rm K_{\rm
s}}(tip)+DM+A_{\rm K_{\rm s}}$; here $DM$ is the distance modulus
corresponding to $d$ and \Aks\ the corresponding extinction in the
\ks\ band.  With increasing distance along a given line of sight the
RGB becomes redder, due to the increase of interstellar
extinction. The reddening is proportional to the extinction, \Aks, and
the shift to fainter magnitudes equals $DM+$\Aks.  In principle we
could thus derive both $DM$ and \Aks\ as a function of distance by
locating discrete features of the RGB.  Due to small number
fluctuations, it is difficult to estimate $K_{\rm s}(tip)$ and thus
$DM+$\Aks. Some of the CMDs contain also the so-called ``red clump''
stars which all have the same absolute magnitude ($M_{\rm K_{\rm s}}=
-1.65$), so that they can be used to trace the stellar distribution
and that of the dust along the line of sight. However, close to the
Galactic centre the distance modulus and the extinction shift 
clump stars below the detection limits of DENIS and 2MASS.

The average field extinction can be estimated by assuming a reference
isochrone (colour-magnitude relation) for the RGB (Sect.\
\ref{rgb}), and by fitting the isochrone to the observed giants. This
approach was used by \citet{schultheis99} and \citet{dutra03} to map
the extinction in the central region of the Galaxy ($|l|<10$\degr).
\citet{schultheis99} obtained an extinction map within 8$^\circ$ of
the Galactic Centre by comparing DENIS ($J,K_{\rm s}$) photometry with an
isochrone from \citet{bertelli94} (metallicity $Z=0.02$, age 10 Gyr,
distance 8 kpc), adopting the extinction law of \citet{glass99}. A
similar map was also produced by \citet{dutra03} using 2MASS
($J,K_{\rm s}$) data together with an empirical reference RGB
isochrone, which is a linear fit to the giants in Baade's windows, and
adopting the extinction law of \citet{mathis90}.

The SiO targets are located at longitudes $l$ between 0\degr\ and
30\degr\ and mostly at latitude $b<0.5^\circ$.
In this region of high extinction even at near-infrared wavelengths,
fits to the apparent (\ks, $J-$\ks) RGB may underestimate the
extinction, due to observational bias as explained in Sect.\
\ref{planes} \citep[see also][]{dutra03,cotera00,figer04}. Therefore
it is useful to also consider the (\ks, $H-$\ks) plane, which is deep
and not sensitive to extinction and therefore less affected by bias.

The extinction toward each of the SiO targets was calculated from 
individual field stars in both the (\ks, $J-$\ks) and (\ks, $H-$\ks)
CMDs by shifting the datapoints on the reference RGB (see Sect.\
\ref{rgb}) along the reddening vector.  Then, the median extinction of
the field was determined in both the (\ks, $J-$\ks) and (\ks, $H-$\ks)
planes, applying an iterative 2$\sigma$ clipping to the extinction
distribution in order to exclude foreground stars \citep{dutra03}.  A
comparison of the extinction estimates derived from both diagrams, and
possible selection effects are described in Sect.\ \ref{planes}.

SiO targets usually appear redder than  neighbouring stars
(Fig.\ \ref{fig:panels}), which implies that they are
intrinsically obscured if we assume that the spatial distribution of
SiO targets is the same as that of red giant branch stars.  About
fifty of our 441 SiO target stars are brighter in \ks\ and bluer than
sources in the field, so those must be nearer than the median.

The CMDs contain much information on the distribution of stars and
dust in the inner Galaxy. Here we have used them only to estimate a
median extinction, \Aks.  In a future study we hope to make a more
complete analysis of these diagrams with a more self-consistent
model. In the following some general remarks from the analysis of the
CMDs are summarised:

\begin{itemize}
\item We can determine \Aks\ for individual stars in each CMD and the
statistical properties of the extinction values within a given CMD.
In most CMDs the interstellar extinction, \Aks, shows a strong
concentration, which reflects the Bulge and  the Galactic centre.
In a minority of CMDs the histogram is broad without clear peaks.

\item  Broad, diffuse extinction distributions are found at
longitudes 20\degr$< l <$ 30\degr.  This suggests that 
 stars and  dust are spread along these
line of sight.

\item Lines of sight which pass through complex star forming regions
such as M17 are easily identified as regions of anomalously high
extinction compared to their surrounding regions.

\item  Toward some lines of sight, especially at latitudes above
$|b|\approx$ 0.6 \degr, a sharp edge at the high end is found in the
extinction distribution.  These lines of sight apparently extend to
above the dust layer.
\end{itemize}

\begin{table*}[h!] 
\caption{\label{table:final} Extinction values.  The identification
number (ID) of the SiO target, as in Table 2 and 3 of
\citetalias{messineo02}, is followed by the field extinction \Aks, by
the corresponding dispersion of individual extinctions of field stars,
and by the total extinction in \ks-band toward the target star
(tot). Finally, a flag (Fg) is listed, which is equal to unity when
the SiO target is classified as a ``foreground object''. Stars for
which not all $J,H,$\ks\ measurements were available have no total
extinction.}  {\scriptsize
\begin{center}
\begin{tabular}{@{\extracolsep{-.07in}}rrrrr|rrrrr|rrrrr|rrrrr}
\hline 
\hline 
{\rm ID} & {\Aks} &{\sig} &{\rm tot}& {\rm Fg}&{\rm ID} & {\Aks} &{\sig}&{\rm tot}&{\rm Fg}&{\rm ID} & {\Aks} &{\sig} &{\rm tot}& {\rm Fg}&{\rm ID} & {\Aks} &{\sig} & {\rm tot}&{\rm Fg}\\
{\rm   } & {\rm mag} &{\rm mag}&{\rm mag} &   &{\rm   } & {\rm mag} &{\rm mag} &{\rm mag}&   &{\rm   } & {\rm mag} &{\rm mag} &{\rm mag}&   &{\rm   } & {\rm mag} &{\rm mag} &{\rm mag}&   \\
\hline 

  1& 0.96& 0.18& 1.63&  & 61& 1.72& 0.29& 3.46&  &121& 1.65& 0.28& 1.91&  &181& 1.38& 0.59& 2.77&  \\ 
  2& 1.23& 0.31& 2.81&  & 62& 1.70& 0.28& 2.47&  &122& 1.55& 0.25& 2.10&  &182& 1.29& 0.19& 1.46&  \\ 
  3& 1.90& 0.33& 1.79&  & 63& 2.44& 0.42& 3.18&  &123& 1.26& 0.36& 1.51&  &183& 1.32& 0.28& 1.45&  \\ 
  4& 2.16& 0.52& 3.00&  & 64& 2.28& 0.42& 3.67&  &124& 1.43& 0.30& 2.18&  &184& 1.44& 0.29& 1.62&  \\ 
  5& 1.51& 0.25& 1.75&  & 65& 2.28& 0.42& 3.67&  &125& 1.65& 0.35& 2.26&  &185& 1.37& 0.24& 0.44&1 \\ 
  6& 0.59& 0.09& 1.16&  & 66& 1.71& 0.21& 1.21&1 &126& 0.75& 0.11& 1.59&  &186& 1.57& 0.18& 1.51&  \\ 
  7& 0.57& 0.07& 0.51&  & 67& 1.62& 0.26& 2.09&  &127& 0.91& 0.19& 1.78&  &187& 0.97& 0.33& 1.65&  \\ 
  8& 1.98& 0.41& 1.53&1 & 68& 1.68& 0.23& 1.09&1 &128& 1.51& 0.24& 1.60&  &188& 0.85& 0.22& 1.06&  \\ 
  9& 2.14& 0.32& 2.59&  & 69& 2.17& 0.43& 2.53&  &129& 0.93& 0.24& 0.36&1 &189& 0.21& 0.04& 1.14&  \\ 
 10& 0.61& 0.11& 1.48&  & 70& 1.65& 0.36& 2.02&  &130& 0.79& 0.09& 0.52&1 &190& 1.04& 0.59& 1.10&  \\ 
 11& 0.94& 0.17& 1.04&  & 71& 1.84& 0.35& 2.12&  &131& 1.69& 0.44& 1.94&  &191& 2.22& 0.75& 3.39&  \\ 
 12& 1.21& 0.17& 1.61&  & 72& 1.86& 0.23& 2.24&  &132& 1.61& 0.42& 2.53&  &192& 0.93& 0.33& 1.69&  \\ 
 13& 1.11& 0.23& 2.15&  & 73& 2.63& 0.30& 3.94&  &133& 0.99& 0.19& 0.86&  &193& 0.97& 0.34& 1.60&  \\ 
 14& 1.38& 0.14& 1.66&  & 74& 1.42& 0.18& 1.92&  &134& 1.52& 0.26& 2.60&  &194& 1.37& 0.20& 1.94&  \\ 
 15& 1.34& 0.27& 0.89&1 & 75& 1.36& 0.23& 1.19&  &135& 1.20& 0.27& 1.46&  &195& 1.42& 0.14& 1.71&  \\ 
 16& 1.46& 0.15& 1.99&  & 76& 2.71& 0.38& 3.58&  &136& 1.27& 0.13& 1.48&  &196& 1.85& 0.79& 2.84&  \\ 
 17& 1.45& 0.15& 1.86&  & 77& 2.49& 0.43& 2.22&  &137& 1.21& 0.24& 1.73&  &197& 1.91& 0.53& 2.02&  \\ 
 18& 1.53& 0.27& 1.89&  & 78& 2.49& 0.43& 3.74&  &138& 1.11& 0.19& 2.17&  &198& 1.76& 0.47& 2.46&  \\ 
 19& 0.60& 0.06& 0.71&  & 79& 1.88& 0.34& 1.93&  &139& 0.21& 0.04& 0.95&  &199& 1.22& 0.31& 0.57&1 \\ 
 20& 1.47& 0.21& 1.81&  & 80& 2.12& 0.28& 2.81&  &140& 1.03& 0.13& 0.38&1 &200& 2.10& 1.07& 1.97&  \\ 
 21& 1.40& 0.29& 1.35&  & 81& 2.53& 0.38& 2.56&  &141& 1.22& 0.17& 2.49&  &201& 0.92& 0.24& 1.49&  \\ 
 22& 1.40& 0.29& 1.82&  & 82& 2.45& 0.40& 3.33&  &142& 1.22& 0.16& 1.26&  &202& 0.92& 0.24& 0.14&1 \\ 
 23& 1.73& 0.63& 1.06&1 & 83& 1.58& 0.23& 1.50&  &143& 1.30& 0.19& 1.96&  &203& 0.91& 0.26& 3.15&  \\ 
 24& 2.17& 0.36& 3.07&  & 84& 2.34& 0.39& 2.90&  &144& 1.49& 0.32& 1.36&  &204& 0.97& 0.25& 3.04&  \\ 
 25& 1.83& 0.26& 1.10&1 & 85& 2.24& 0.44& 2.25&  &145& 1.02& 0.18& 0.98&  &205& 1.19& 0.32& 1.46&  \\ 
 26& 2.26& 0.40& 2.73&  & 86& 2.21& 0.68& 2.88&  &146& 0.83& 0.15& 1.06&  &206& 2.36& 0.48& 0.45&1 \\ 
 27& 1.93& 0.24& 2.19&  & 87& 1.69& 0.14& 1.24&1 &147& 1.34& 0.27& 0.89&1 &207& 1.15& 0.29& 1.17&  \\ 
 28& 2.09& 0.36& 2.77&  & 88& 2.45& 0.51& 3.07&  &148& 0.80& 0.21& 0.57&1 &208& 0.91& 0.33& 1.41&  \\ 
 29& 2.56& 0.32& 4.31&  & 89& 2.60& 0.41& 2.97&  &149& 1.17& 0.25& 1.43&  &209& 1.91& 0.63& 2.70&  \\ 
 30& 2.82& 0.36& 3.92&  & 90& 1.90& 0.34& 2.12&  &150& 1.33& 0.31& 2.23&  &210& 0.88& 0.24& 1.64&  \\ 
 31& 1.80& 0.19& 1.72&  & 91& 2.82& 0.52& 3.09&  &151& 1.17& 0.32& 1.41&  &211& 1.42& 0.53& 2.50&  \\ 
 32& 1.61& 0.24& 1.78&  & 92& 1.66& 0.28& 1.40&  &152& 1.08& 0.39& 1.88&  &212& 0.82& 0.31& 2.02&  \\ 
 33& 1.62& 0.22& 2.69&  & 93& 1.87& 0.35& 1.85&  &153& 0.14& 0.15& 0.34&  &213& 2.56& 1.05& 4.03&  \\ 
 34& 1.78& 0.25& 2.67&  & 94& 2.07& 0.47& 3.62&  &154& 1.73& 0.64& 5.16&  &214& 1.00& 0.19& 1.13&  \\ 
 35& 1.89& 0.23& 2.50&  & 95& 2.11& 0.76& 4.65&  &155& 1.20& 0.30& 1.55&  &215& 1.30& 0.36& 3.56&  \\ 
 36& 1.47& 0.22& 0.80&1 & 96& 1.92& 0.32& 1.88&  &156& 0.97& 0.25& 1.75&  &216& 0.89& 0.35& 1.43&  \\ 
 37& 1.87& 0.25& 3.37&  & 97& 2.31& 0.61& 3.38&  &157& 1.31& 0.33& 1.42&  &217& 1.02& 0.34& 1.89&  \\ 
 38& 2.98& 0.44& 3.52&  & 98& 2.37& 0.61& 2.94&  &158& 1.42& 0.32& 1.60&  &218& 1.30& 0.65& 1.40&  \\ 
 39& 2.91& 0.44& 3.64&  & 99& 2.41& 0.56& 2.34&  &159& 1.46& 0.19& 2.15&  &219& 1.48& 0.31& 2.15&  \\ 
 40& 2.29& 0.35& 2.79&  &100& 2.30& 0.56& 3.55&  &160& 1.35& 0.23& 0.81&1 &220& 2.21& 0.57& 3.39&  \\ 
 41& 2.68& 0.40& 3.53&  &101& 2.11& 0.39& 3.89&  &161& 1.84& 0.35& 2.12&  &221& 1.10& 0.35& 2.24&  \\ 
 42& 2.19& 0.41& 2.37&  &102& 2.24& 0.48& 2.24&  &162& 1.66& 0.41& 3.35&  &222& 1.93& 0.37& 2.29&  \\ 
 43& 1.66& 0.26& 3.29&  &103& 0.99& 0.16&&  &163& 1.75& 0.47& 2.25&  &223& 1.02& 0.29& 1.87&  \\ 
 44& 1.75& 0.21& 1.86&  &104& 0.92& 0.10& 1.21&  &164& 1.32& 0.30& 1.49&  &224& 2.04& 0.57&&  \\ 
 45& 2.67& 0.34& 3.76&  &105& 1.62& 0.31& 2.32&  &165& 1.08& 0.21& 2.67&  &225& 1.53& 0.48& 2.24&  \\ 
 46& 2.89& 0.47& 4.44&  &106& 1.23& 0.54& 2.10&  &166& 1.21& 0.26& 2.44&  &226& 0.84& 0.28& 1.10&  \\ 
 47& 1.34& 0.11& 1.15&1 &107& 1.70& 0.42& 2.56&  &167& 1.26& 0.18& 1.58&  &227& 1.83& 0.26& 0.61&1 \\ 
 48& 2.25& 0.39& 2.67&  &108& 2.01& 0.56& 0.77&1 &168& 1.35& 0.16& 1.52&  &228& 0.92& 0.32& 1.94&  \\ 
 49& 2.33& 0.41& 2.69&  &109& 1.88& 0.39& 2.98&  &169& 1.11& 0.48& 1.98&  &229& 1.32& 0.43& 2.29&  \\ 
 50& 1.10& 0.16& 1.75&  &110& 1.78& 0.53& 2.64&  &170& 1.19& 0.19& 1.96&  &230& 0.87& 0.26& 1.61&  \\ 
 51& 2.31& 0.39& 3.70&  &111& 1.81& 0.38& 2.10&  &171& 1.57& 0.31& 1.67&  &231& 0.70& 0.22& 1.54&  \\ 
 52& 2.33& 0.38& 3.86&  &112& 1.35& 0.30& 1.32&  &172& 1.09& 0.33& 2.79&  &232& 1.10& 0.35& 1.45&  \\ 
 53& 2.68& 0.34& 2.65&  &113& 1.70& 0.28& 1.14&1 &173& 1.61& 0.63& 1.20&  &233& 1.64& 0.54& 1.56&  \\ 
 54& 2.31& 0.38& 2.33&  &114& 1.25& 0.28& 1.47&  &174& 1.02& 0.53& 1.49&  &234& 1.18& 0.38& 2.17&  \\ 
 55& 1.71& 0.27& 2.94&  &115& 1.92& 0.43& 2.00&  &175& 1.20& 0.39& 2.09&  &235& 1.00& 0.36& 1.77&  \\ 
 56& 2.34& 0.36& 2.49&  &116& 1.75& 0.46& 1.41&  &176& 1.42& 0.50& 1.46&  &236& 0.81& 0.30& 1.90&  \\ 
 57& 2.55& 0.45& 3.49&  &117& 1.85& 0.41& 3.25&  &177& 1.27& 0.30& 1.51&  &237& 0.75& 0.23& 2.00&  \\ 
 58& 2.20& 0.33& 3.59&  &118& 2.78& 0.59& 3.97&  &178& 1.50& 0.37& 1.48&  &238& 0.95& 0.37& 1.34&  \\ 
 59& 2.25& 0.34& 2.38&  &119& 1.63& 0.60& 1.54&  &179& 1.43& 0.32& 1.79&  &239& 0.83& 0.30& 1.49&  \\ 
 60& 2.11& 0.29& 2.75&  &120& 1.62& 0.29& 2.27&  &180& 1.20& 0.38& 1.05&  &240& 0.76& 0.30& 2.00&  \\ 
\hline 
\end{tabular}
\end{center}
}
\end{table*} 
\afterpage{\clearpage}
\begin{table*} 
\addtocounter{table}{-1}
\caption{(continued) }
{\scriptsize
\begin{center}
\begin{tabular}{@{\extracolsep{-.07in}}rrrrr|rrrrr|rrrrr|rrrrr}
\hline 
\hline 
{\rm ID} & {\Aks} &{\sig} &{\rm tot}& {\rm Fg}&{\rm ID} & {\Aks} &{\sig}&{\rm tot}&{\rm Fg}&{\rm ID} & {\Aks} &{\sig} &{\rm tot}& {\rm Fg}&{\rm ID} & {\Aks} &{\sig} & {\rm tot}&{\rm Fg}\\
{\rm   } & {\rm mag} &{\rm mag}&{\rm mag} &   &{\rm   } & {\rm mag} &{\rm mag} &{\rm mag}&   &{\rm   } & {\rm mag} &{\rm mag} &{\rm mag}&   &{\rm   } & {\rm mag} &{\rm mag} &{\rm mag}&   \\
\hline 

241& 0.92& 0.29& 2.05&  &301& 2.13& 0.33& 1.88&  &361& 1.60& 0.49& 3.67&  &421& 1.06& 0.45& 3.64&  \\ 
242& 0.88& 0.20& 0.93&  &302& 2.56& 0.33& 2.50&  &362& 1.35& 0.18&&  &422& 1.25& 0.44& 1.86&  \\ 
243& 0.92& 0.32& 2.42&  &303& 1.36& 0.11& 0.63&1 &363& 1.87& 0.41& 3.00&  &423& 0.72& 0.68& 0.32&  \\ 
244& 0.96& 0.39& 0.99&  &304& 2.07& 0.35& 3.41&  &364& 0.61& 0.09& 0.96&  &424& 1.34& 0.40& 2.28&  \\ 
245& 0.67& 0.22& 0.89&  &305& 1.79& 0.22& 2.53&  &365& 1.72& 0.26& 3.62&  &425& 1.72& 0.71& 1.37&  \\ 
246& 1.15& 0.44& 1.65&  &306& 1.88& 0.24& 2.06&  &366& 0.60& 0.15& 1.21&  &426& 1.07& 0.47& 2.53&  \\ 
247& 1.26& 0.36& 0.79&1 &307& 1.24& 0.13& 1.48&  &367& 1.16& 0.32& 2.95&  &427& 0.90& 0.46&&  \\ 
248& 1.26& 0.39& 1.61&  &308& 2.20& 0.36& 2.92&  &368& 1.57& 0.31& 0.63&1 &428& 0.94& 0.41& 1.53&  \\ 
249& 0.98& 0.42& 2.31&  &309& 2.26& 0.35& 2.52&  &369& 1.35& 0.22& 2.58&  &429& 0.74& 0.39& 1.70&  \\ 
250& 0.93& 0.42& 2.36&  &310& 2.79& 0.41& 2.42&  &370& 1.11& 0.23& 0.53&1 &430& 0.88& 0.32& 1.63&  \\ 
251& 1.39& 0.42& 1.74&  &311& 2.93& 0.51& 4.32&  &371& 0.83& 0.14& 1.49&  &431& 0.83& 0.31& 1.17&  \\ 
252& 0.92& 0.31& 1.23&  &312& 1.70& 0.28& 2.14&  &372& 1.24& 0.16& 2.13&  &432& 0.97& 0.35& 1.17&  \\ 
253& 0.96& 0.32& 2.34&  &313& 1.75& 0.28& 2.25&  &373& 1.24& 0.11& 1.12&1 &433& 0.72& 0.29& 1.84&  \\ 
254& 1.27& 0.55& 1.82&  &314& 2.18& 0.46& 3.18&  &374& 0.95& 0.20& 1.36&  &434& 1.71& 0.60& 1.84&  \\ 
255& 1.13& 0.30& 1.22&  &315& 1.05& 0.13& 1.47&  &375& 1.03& 0.21& 1.70&  &435& 1.02& 0.45& 0.40&1 \\ 
256& 0.84& 0.61& 2.80&  &316& 1.54& 0.22& 1.77&  &376& 1.33& 0.22& 2.97&  &436& 0.97& 0.36& 0.00&1 \\ 
257& 0.90& 0.38& 4.34&  &317& 1.32& 0.15& 1.53&  &377& 1.11& 0.22& 1.80&  &437& 1.01& 0.34& 3.31&  \\ 
258& 1.64& 0.70& 2.88&  &318& 1.63& 0.25& 2.15&  &378& 0.80& 0.24& 1.00&  &438& 0.99& 0.30& 0.38&1 \\ 
259& 0.85& 0.48& 1.07&  &319& 0.61& 0.08& 1.16&  &379& 0.84& 0.19& 0.68&  &439& 0.99& 0.34& 2.42&  \\ 
260& 1.08& 0.43& 1.67&  &320& 2.20& 0.33& 2.60&  &380& 1.03& 0.31& 2.52&  &440& 1.03& 0.32& 0.45&1 \\ 
261& 1.62& 0.37& 2.44&  &321& 2.28& 0.43&&  &381& 1.79& 0.55& 2.60&  &441& 1.02& 0.44& 1.34&  \\ 
262& 1.45& 0.34& 1.71&  &322& 2.23& 0.41& 3.79&  &382& 0.86& 0.16& 0.95&  &442& 1.17& 0.53& 2.69&  \\ 
263& 0.91& 0.49& 1.80&  &323& 1.87& 0.25& 2.92&  &383& 1.13& 0.31& 1.52&  &443& 0.96& 0.57&&  \\ 
264& 1.11& 0.36& 1.78&  &324& 2.41& 0.49& 3.49&  &384& 1.33& 0.35& 0.86&1 &444& 1.16& 0.70& 0.88&  \\ 
265& 0.84& 0.33& 1.48&  &325& 2.13& 0.43& 2.08&  &385& 1.17& 0.39& 1.67&  \\                             
266& 0.75& 0.30& 1.52&  &326& 1.74& 0.21& 2.62&  &386& 1.61& 0.20& 1.23&1 \\                             
267& 0.72& 0.34& 1.04&  &327& 2.17& 0.25& 3.09&  &387& 2.00& 0.55& 5.02&  \\                             
268& 0.74& 0.30& 2.59&  &328& 1.39& 0.21& 1.95&  &388& 1.77& 0.56& 3.17&  \\                             
269& 0.82& 0.30& 1.32&  &329& 1.74& 0.24& 2.06&  &389& 1.28& 0.23& 2.15&  \\                             
270& 0.88& 0.31& 2.22&  &330& 1.90& 0.28& 2.59&  &390& 1.56& 0.32& 1.36&  \\                             
271& 1.00& 0.54& 2.46&  &331& 2.15& 0.27& 3.22&  &391& 1.58& 0.31& 2.36&  \\                             
272& 1.48& 0.30& 1.51&  &332& 2.58& 0.37& 2.89&  &392& 1.39& 0.29& 2.19&   \\                            
273& 1.80& 0.45&&  &333& 2.41& 0.44& 3.12&  &393& 0.75& 0.19& 1.39&   \\                            
274& 0.49& 0.05& 0.96&  &334& 2.59& 0.45& 2.27&  &394& 1.16& 0.46& 2.31&   \\                            
275& 0.47& 0.05& 0.40&1 &335& 2.33& 0.52& 2.13&  &395& 0.92& 0.22& 0.63&1 \\                             
276& 2.27& 0.60& 3.76&  &336& 2.14& 0.23& 2.82&  &396& 1.24& 0.26& 0.37&1  \\                            
277& 0.57& 0.07& 0.49&1 &337& 1.37& 0.15& 1.52&  &397& 1.03& 0.29& 1.04&  \\                             
278& 2.12& 0.40& 2.92&  &338& 2.23& 0.34& 2.23&  &398& 1.55& 0.29& 0.38&1 \\                             
279& 0.98& 0.17& 1.23&  &339& 2.28& 0.33& 2.28&  &399& 1.08& 0.29& 1.45&  \\                             
280& 1.18& 0.21& 1.60&  &340& 1.26& 0.19& 1.50&  &400& 1.39& 0.34& 4.32&  \\                             
281& 0.93& 0.14& 1.34&  &341& 2.83& 0.58& 4.58&  &401& 1.24& 0.21& 1.34&    \\                           
282& 1.38& 0.19& 2.09&  &342& 2.62& 0.39& 3.31&  &402& 1.79& 0.27& 0.00&1  \\                            
283& 1.21& 0.11& 1.51&  &343& 2.60& 0.52& 2.56&  &403& 1.89& 0.51&&  \\                             
284& 0.83& 0.12& 1.54&  &344& 2.69& 0.50& 2.47&  &404& 3.43& 0.48& 3.90&  \\                             
285& 1.15& 0.14& 1.50&  &345& 2.13& 0.61& 2.61&  &405& 2.49& 1.05& 0.80&1   \\                           
286& 0.61& 0.06& 0.32&1 &346& 2.51& 0.56& 2.90&  &406& 1.56& 0.21& 2.01&   \\                            
287& 0.55& 0.06& 0.41&1 &347& 2.89& 0.62& 0.63&1 &407& 0.54& 0.19& 2.59&   \\                            
288& 1.43& 0.21& 2.15&  &348& 2.04& 0.35& 3.73&  &408& 1.41& 0.16& 3.58&  \\                             
289& 1.72& 0.31& 2.06&  &349& 1.63& 0.27& 3.79&  &409& 1.00& 0.35& 2.43&  \\                             
290& 1.51& 0.22& 1.70&  &350& 1.85& 0.41& 1.94&  &410& 1.27& 0.38& 1.85&  \\                             
291& 1.41& 0.14& 0.96&1 &351& 1.38& 0.28& 2.73&  &411& 1.02& 0.34& 3.10&  \\                             
292& 1.63& 0.21& 2.07&  &352& 1.25& 0.24& 1.69&  &412& 1.77& 0.46& 1.76& \\                              
293& 1.62& 0.30& 2.21&  &353& 1.71& 0.26& 2.28&  &413& 2.13& 0.44& 3.65&   \\                            
294& 2.23& 0.50& 2.37&  &354& 1.36& 0.25& 2.63&  &414& 0.78& 0.25& 1.57&  \\                             
295& 1.67& 0.22& 2.18&  &355& 1.55& 0.23& 2.17&  &415& 0.95& 0.25&&  \\                             
296& 1.54& 0.18& 1.54&  &356& 0.82& 0.11& 1.08&  &416& 0.73& 0.27& 0.64& \\                              
297& 2.35& 0.28& 1.07&1 &357& 1.84& 0.46& 3.40&  &417& 0.93& 0.27&& \\                              
298& 1.50& 0.17&&  &358& 1.64& 0.31& 0.96&1 &418& 0.94& 0.34& 3.61&\\                               
299& 1.54& 0.15& 1.99&  &359& 1.38& 0.27& 2.26&  &419& 0.79& 0.37& 3.41&\\                               
300& 2.28& 0.32& 3.73&  &360& 1.46& 0.29& 1.36&  &420& 1.16& 0.39& 3.27& \\                             
\hline  			     		
\end{tabular}
\end{center}			     		
}				     
\end{table*} 

\afterpage{\clearpage}

\subsection{Reference red giant branch}\label{rgb}

\begin{figure}[t!]
\begin{center}
\resizebox{0.7\hsize}{!}{\includegraphics{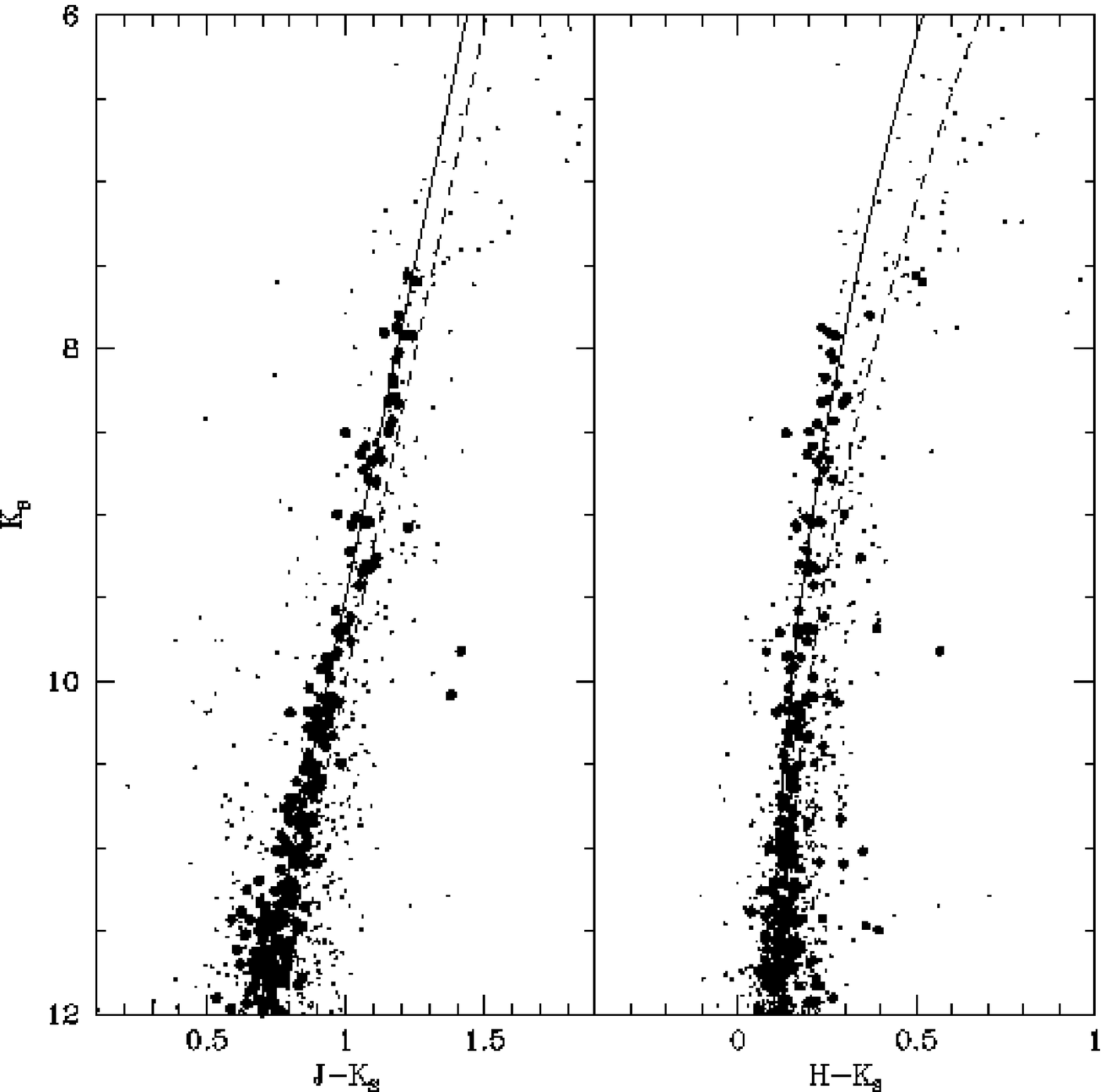}}
\caption{\label{fig:47tuc.ps} Magnitudes versus colours of 2MASS
stars.  Large dots are 2MASS sources within 4\arcmin\ from the centre
of the globular cluster 47 Tuc, brought at the distance of 8 kpc
adopting a distance modulus of 13.32 \citep{ferraro99}. Small dots are
dereddened 2MASS Sgr--I field sources, selected within 30\arcsec\ from
Mira stars \citep{glass95} in order to overpopulate the Mira
region above the RGB tip.  Continuous lines indicate the mean
ridge lines for the cluster red giant branch, while  dashed lines
those of the Sgr--I giants.}
\end{center}
\end{figure}
In Figure \ref{fig:47tuc.ps} we plot the extinction-corrected 2MASS
point sources within 30\arcsec\ from the positions of  Mira stars
in the Sgr--I field \citep{glass95}.  Objects above the RGB tip
(\ks$\sim8.2$ mag) are AGB stars: 63 Mira variables from
\citep{glass95} and 24 other stars, most probably semiregular (SR)
variables \citep{alard01}. In the same diagram we also plot 2MASS
sources within 4\arcmin\ from the centre of the Galactic globular
cluster 47 Tuc, moved to a distance of 8 kpc \citep[cluster distance
modulus, DM=13.32 mag, from][]{ferraro99}.

The upper part (\ks$<12$ mag) of the 47 Tuc RGB in the $K_{\rm
S0},(J-K_{\rm s})_0$ CMD is well represented by a linear fit:
\begin{equation}
({\rm J-K_{\rm s}})_0 = 2.19(\pm0.02)-0.125(\pm0.002){\rm K_{S0}}.
\end{equation}

The 47 Tuc giants appear bluer by $0.05$ mag in $(J-K_{\rm s})_0$ than the
colours of Sgr--I giants, which are more metal rich. However, the
cluster $(J-K_{\rm s})_0$ vs. $K_{S0}$ RGB has a slope identical to that
found in Sgr--I, confirming that the slope does not vary
significantly with metallicity \citep[see][]{frogel99,dutra03}.

To assess the uncertainty of our extinction estimates, we examined the
model RGB colours of \citet{bertelli00} with two extreme values of
metallicity, Z=0.04 and 0.30. These models do not show a significant
difference in the CMD slope, and they differ in $(J-K_{\rm s})_0$ by only
0.1 mag. A similarly small difference is found for  models of 2 Gyr
and 16 Gyr old populations.  Nominal 2MASS photometric errors are
smaller than 0.04 mag for $J<15$ and \ks$<13$ mag, but toward the
Galactic centre  uncertainties are larger because of crowding.
However, all the uncertainties have a small impact on the extinction
estimates: a change of 0.1 in $(J-K_{\rm s})_0$ implies a change of 0.05 in
\Aks\ (0.6 in \Av).

The right-hand panel of Fig.\ \ref{fig:47tuc.ps} shows the $(H-K_{\rm s})_0$
vs. $K_{S0}$ diagram of  Sgr--I giants and of the 47 Tuc giants.
There is again a well defined RGB sequence. A second order polynomial
fit well fits  47 Tuc giants with \ks$_0 < 12$ mag:
\begin{equation}
({\rm H-K_{\rm s}})_0 = 1.73(\pm0.22)-0.268(\pm0.035)\times{\rm K_{S0}} 
\end{equation} 
$$+0.011(\pm0.002)\times{\rm K_{S0}^2}.$$

At \ks$_0 < 8$ mag this fit
traces the blue boundary toward Mira stars.  A fit of only  Sgr--I 
giants is slightly more shallow:
\begin{equation}
({\rm H-K_{\rm s}})_0 = 2.25(\pm0.13)-0.347(\pm0.029)\times{\rm K_{S0}}
\end{equation}
$$+0.014(\pm0.002)\times{\rm K_{S0}^2}.$$

The $K_{\rm s},(H-K_{\rm s})$ plane has a lower sensitivity to
extinction than the $K_{\rm s},(J-K_{\rm s})$ plane: here a colour
change of 0.1 mag implies a change of $\sim 0.15$ in \Aks.  The models
of \citet{bertelli00} predict variations within $(H-K_{\rm s}) = 0.04$ mag
with metallicity and age variations.  The nominal 2MASS photometric
errors are smaller than 0.04 for $H<14$ and \ks$<13$.

The uncertainty in distance yields only a minor uncertainty in the
extinction. A shift in distance modulus  of the reference RGB  of $\pm
2$ mag results in a change in the extinction of \Aks$\mp 0.15$ mag.

\subsection{Determination of extinction value and extinction law 
in the $J$, $H$, $K_{\rm s}$ CMD}
\label{planes}
Assuming a colour-magnitude relation for red giants, the apparent
near-infrared colours of field stars yield information on, both,
the magnitude of the extinction and on the shape of the extinction
law.

In principle, one can try to fit the observed field star colours to
the red giant branch colours in the near-infrared colour-magnitude
diagrams, and optimise the fit for both the absolute average
extinction in the $K_{\rm s}$ band along the line of sight and the
spectral index of the extinction power law.  To do that,
one needs to consider only the region of the colour-magnitude diagram
where the upper RGB is well defined, i.e.  not affected by large
photometric errors or by the diagonal cut-off from the 2MASS detection
limits (see Fig.\ \ref{fig:panels}), which would bias the calculation
of the median extinction toward a lower value. In the inner Galaxy,
the photometric error is typically below 0.04 for stars with $K_{\rm
s}<12$ mag.  To quantify the incompleteness due to the diagonal
cut-off, with zero extinction, our average 2MASS detection limits of
$J=16.0$ and $H=14.0$ correspond to a RGB \ks\ magnitude of 15.2 and
13.0 mag, respectively, at a distance of 8 kpc. Accounting for a
scatter in the observed colours of $\pm 0.5$ mag, the RGB would be
sampled well to $J=15.5$ and $H=13.5$ mag, corresponding to \ks$<
14.6$ and 12.6, respectively.  With a \ks\ extinction of 3 mag ($\sim
5.0$ mag in $H$, $\sim 8.6$ mag in $J$), a typical value in the
direction of the Galactic centre, these RGB completeness limits would
rise to $K_{\rm s}=7.1$ and 10.8 mag in the $(K_{\rm s},J-K_{\rm s})$
and $(K_{\rm s},H-K_{\rm S})$ planes, respectively. In the $J$ band we
would therefore be left with variable AGB stars well above the RGB tip
and foreground stars, and only the $H$ band would provide a sufficient
number of red giant stars to match the reference RGB. We conclude from
this that with the 2MASS data $(K_{\rm s},J-K_{\rm s})$
colour-magnitude diagrams are useful for extinction determinations
only to a \ks\ extinction of about 1.6 mag, and one must always make
sure that only stars above the completeness limit are matched to the
RGB.  Because of the larger reddening, the $(K_{\rm s},J-K_{\rm s})$
plane would in principle give more accurate extinction estimates,
would it not be affected by the selection effect due to the
relatively bright detection limit.

\begin{figure}[h!]
\begin{center}
\resizebox{0.7\hsize}{!}{\includegraphics{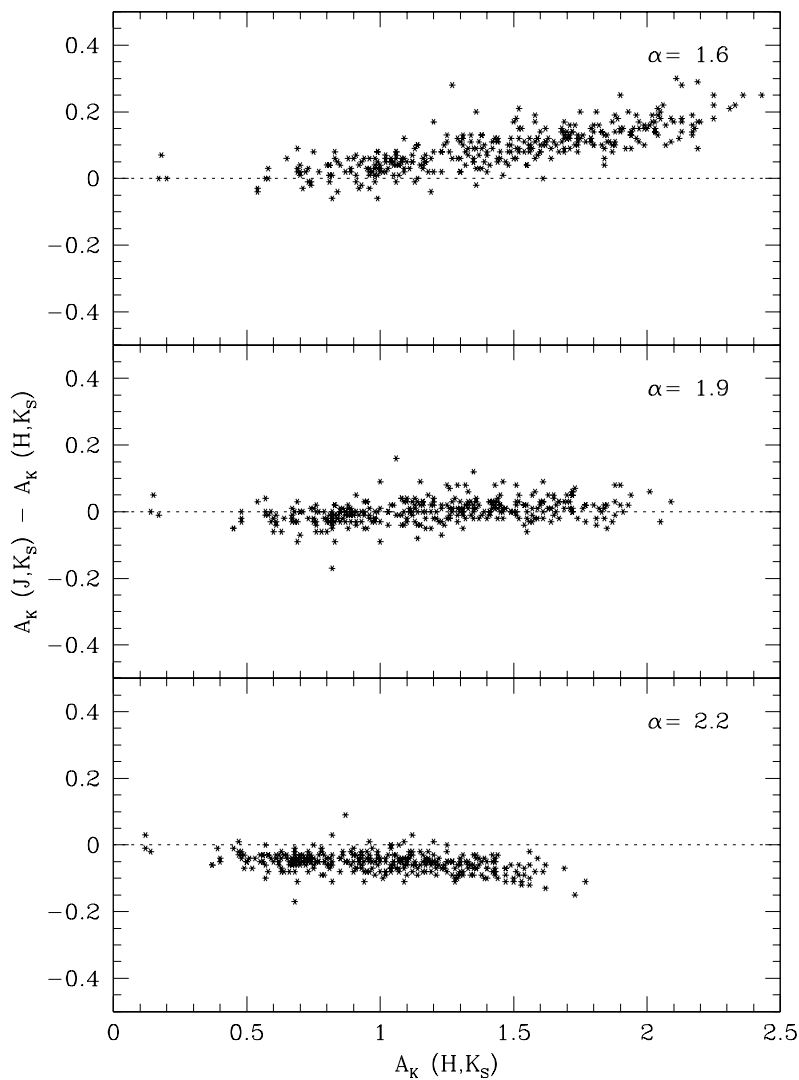}}
\caption{\label{extall.ps} Comparison between the median extinction
values toward our fields obtained from the (\ks,$J-$\ks) plane and
those from the (\ks,$H-$\ks) plane.  Only sources detected in
$J,H,$\ks\ above the \ks\ completeness limits are used.  In each panel
a different value of the spectral index of the extinction power law,
$\alpha$, is adopted. }
\end{center}
\end{figure}

To determine the slope, $\alpha$, of the near-infrared extinction law
we examined the CMDs for field stars which were detected in all three
bands, $J$, $H$, and \ks, brighter than the \ks\ completeness limits
for the RGB at the extinction of each field. We determined the
extinction toward each of our fields separately in the $(K_{\rm
S},J-K_{\rm s})$ and $(K_{\rm s},H-K_{\rm s})$ CMDs, as the median of
the extinctions from individual field stars.  The difference of the
extinction determinations from the two planes must agree independent
from \Aks within the dispersion. 

Therefore, we vary $\alpha$ until we get an overall agreement between
the extinction estimates from the two planes.  Figure \ref{extall.ps}
shows the differences of the median extinction values
\Aks($J$,\ks)$-$\Aks($H$,\ks) plotted against the \Aks\ -calculated-
for different values of $\alpha$.  The discrepancy between the
extinction values increases with \Aks\ if the assumed value of $\alpha$ is
too small. We find that for $\alpha=1.9$ the
two extinction estimates do yield consistent values, within the
photometric uncertainties, over the entire range of \Aks. 

The main uncertainty in the determination of the 
extinction power law  arises from the uncertainty in the
slope of the RGB. Using a fit to the colour-magnitude distribution of
the giants in the Sgr--I field ($l=1.4$\degr,b$=-2.6$) instead of the 47
Tuc globular cluster giants, that is somewhat steeper in the
(\ks,$H-$\ks) plane, we find that the best value for alpha increases
to 2.2.

The slope of the RGB decreases with increasing metallicity, leading to
higher values of $\alpha$.  However, since 47 Tuc has a lower
metallicity ($-0.7$ dex) than the average Bulge stars
\citep{frogel99,rich94},  $\alpha=1.9$  may be
taken as a lower limit to the actual value.

We furthermore find that the (\ks,$H-$\ks) distribution of the giants
in the low extinction region (\Aks=0.2) at $l=0.2$\degr and b$=-2.1$
\citep{dutra02,stanek98} match the distribution of the 47 Tuc giants
better than that of Sgr--I.

\begin{figure}[th]
\begin{center}
\resizebox{0.7\hsize}{!}{\includegraphics{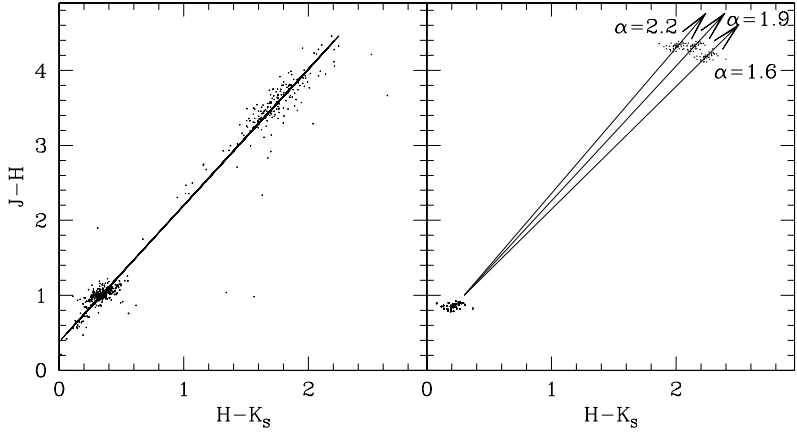}}
\caption{\label{fig:lala.ps} $J-H$ vs. $H-$\ks\ colours. The arrows
are the reddening vectors for a visual extinction of 35 mag and a
near-infrared power law slopes $\alpha$ of 1.6, 1.9 and 2.0.  {\bf
Left-hand panel:} Bulge giant stars with $10<K_{\rm s}<8$ mag taken
from several Bulge fields at different median extinction. {\bf
Right-hand panel:} Giant stars from 47 Tuc with $10<K_{\rm s}<8$ mag
when brought at the distance of the Galactic centre. The same stars
are plotted again reddened using the near-infrared power laws with
slope $\alpha=$ 1.6, 1.9 and 2.2 and \Ak=2.9.}
\end{center}
\end{figure}

Although complicated by the intrinsic colour-magnitude relation that
giant stars follow, the value of the $\alpha$ parameter can be tested
using a $J-H$ vs. $H-$\ks\ diagram. Using the values from Table
\ref{table:nearextinction} for power laws with
$\alpha = 1.6$, 1.85, 1.9, 2.0, 2.2, the slopes of the reddening vector in
the $J-H$ vs. $H-$\ks\ diagram are 1.64, 1.75, 1.80, 1.83, 1.94,
respectively.  Identical slopes are found when reddening artificially
the 47 Tuc giants (or the Sgr--I giants) and linearly fitting the
non-reddened giants plus the reddened ones.  A $J-H$ vs. $H-$\ks\
diagram of giant stars from fields with median extinction between
\Aks$ = 2.0$ and 2.3 mag and from Sgr--I field is shown in Fig.\
\ref{fig:lala.ps}. The best fit to the datapoints gives a slope of
$1.81\pm0.03$, also suggesting that $\alpha = 1.9\pm0.1$.

The value $\alpha = 1.9$ is in agreement with the work of
\citet{glass99} and \citet{landini84} and the historical Curve 15 of
\citet{vandehulst46}.

In the rest of the paper we use the extinction calculated assuming
$\alpha = 1.9$. For fields with \Aks $<1.6$ mag we adopt the extinction
values determined from the (\ks,$J-$\ks) plane, otherwise we will use
the values from the (\ks,$H-$\ks) plane.

\subsection{Outside the Bulge}\label{outside}
K2 giant stars are the dominant population of late-type stars seen
along the disk \citep[e.g.][]{drimmel03,lopez03}. They correspond to
 red clump stars in metal-rich globular clusters such as 47 Tuc.
The location of clump stars on the CMD depends on extinction and
distance.  This trace  was modelled taking the absolute
magnitudes of clump stars from \citet{wainscoat92} and the
distribution of dust and stars in the Galaxy found by
\citet{drimmel03}.  The trace of the clump stars is shown by the left
curve in Fig. \ref{fig:panels}, and it appears more populated by stars
and distinct from the Bulge RGB in the CMD of the field at
$l = \sim10$\degr.

Toward the Bulge the Bulge RGB population is dominant and therefore
the median interstellar extinction is practically not affected by
possible foreground clump stars.  This is not the case in the disk,
where one must eliminate the foreground clump stars before fitting the
RGB in order to properly calculate the median extinction of field
giants.

We therefore identified as likely clump stars those located within 0.3
mag from the $J-$\ks\ colour of the clump 
trace, and we identify as
giants those stars redder than the clump stars
\citep[e.g.][]{lopez03}.

\subsection{Dispersion  of the extinctions along a line of sight}
Toward a given target star together with the median extinction of
field stars \Aks\ we determined the standard deviation of the
distribution of the individual extinctions, \sig.

The patchy nature of the extinction is visible even within the
$2-4\arcmin $ radius sampling area. This patchiness integrated over a
longer path generates larger \sig\ with increasing extinction for
Bulge line of sights.  The 1\sig\ uncertainty in the field extinction
varies from $\sim0.2$ mag when \Aks$ = 0.6$ mag up to $\sim0.7$ mag in
the regions with the largest extinction (\Aks$>2.0$).  In fields at
longitudes longer that 10\degr\ a larger \sig\ is found than in Bulge
fields of similar median extinction. This is probably due to the
presence of several Galactic components, e.g.\ the disk, arms, bar and
molecular ring, along these line of sights.

\section{Near-infrared properties of known Mira stars}
\label{miras}
As indicated by their variability, their strong 15 $\mu$m emission
\citepalias{messineo03_2}, and their SiO maser emission
\citepalias{messineo02}, our SiO targets are AGB stars in the thermal
pulsing phase. At the present time their pulsation periods and
amplitudes are not known. However, most of our SiO targets must be
large amplitude variables \citepalias{messineo03_2}.

Though they are 20 times less numerous than semiregular AGB stars
(SRs) \citep{alard01},  Mira stars are among the best studied pulsating
variable stars. They are regular long period variables (LPV) with
visual light amplitude over 2.5 mag, or $K$ band amplitude over $\sim
0.3$ mag. Since large amplitudes tend to be associated with the most
regular light curves \citep{cioni03}, the amplitude remains the main
parameter for the classification of a Mira star.

To analyse the colours of our SiO targets, in particular to check the
quality of the extinction corrections, it is useful to have a
comparison sample of large amplitude LPV AGB stars, well studied and
covering a wide range of colours.

Therefore, we examined various comparison samples of known Mira stars
free of extinction: two samples in the solar vicinity, taken from
\citet{olivier01} and \citet{whitelock-hyp}, plus one sample toward
the Galactic Cap taken from \citet{whitelock94}.  To account for
possible changes in the colour properties of the Mira stars with
Galactic position, we also looked at two samples of Bulge Mira stars
from regions of low extinction: 18 Mira stars detected by IRAS
\afterpage{\clearpage}
\noindent
\citep[][]{glass95} in the Sgr--I field, and 104 IRAS Mira stars at
latitude $6^\circ<b<7^\circ$ and $|l|<15^\circ$ \citep{whitelock91};
for comparison a sample of LPV in the Large 
Magellanic Cloud is also
considered \citep{whitelock03}.  All these stars have IRAS 12 $\mu$m
magnitude, [12], and mean $J,H,K$ magnitudes in the SAAO system
\citep{carter90}.

The stellar fluxes are given already corrected for reddening only in
the work of \citet{olivier01}.  For the LPV stars analysed by
\citet{whitelock03, whitelock-hyp,whitelock94} the effects of
interstellar extinction are negligible because these stars are nearby
or outside of the Galactic plane and we therefore did not correct
these for extinction. We dereddened the Baade Sgr--I window data
\citep{glass95} adopting our favourite extinction curve ($\alpha = 1.9$)
and \Aks\ $= 0.15$ mag \citep[consistently with the extinction value
adopted in][]{glass95}. We corrected for reddening the magnitudes of
the outer Bulge Miras \citep{whitelock91} adopting values of \Aks\
derived from their surrounding stars (see next section).

Next we analyse the location of these well-known Mira stars in the
near-infrared CMDs and colour-colour diagrams.

\afterpage{\clearpage}

\subsection{Colour-magnitude diagram of outer Bulge Mira
stars and  surrounding field stars}

Long period variable stars in the outer Bulge as studied by
\citet{whitelock91} are interesting in several aspects: they were
selected on the basis of their IRAS fluxes and colours according to
criteria similar to those with which we selected our MSX targets
\citepalias{messineo02,messineo03_2}; since their main period ranges
from 170 to 722 days and their $K$ amplitudes from 0.4 to 2.7 mag,
they are classical Mira stars; their distances were estimated from the
period-luminosity relation \citep{whitelock91}, resulting in a
distribution of the distance moduli peaking at 14.7 mag with a
$\sigma = \sim0.5$ mag; since they are at latitudes between 6 and
7\degr, they are in regions of low interstellar extinction. All this
makes them ideal objects for a comparison with our SiO targets, the
study of which is complicated by the large interstellar extinction at
their low latitudes.

\begin{figure}[h]
\begin{center}
\resizebox{0.65\hsize}{!}{\includegraphics{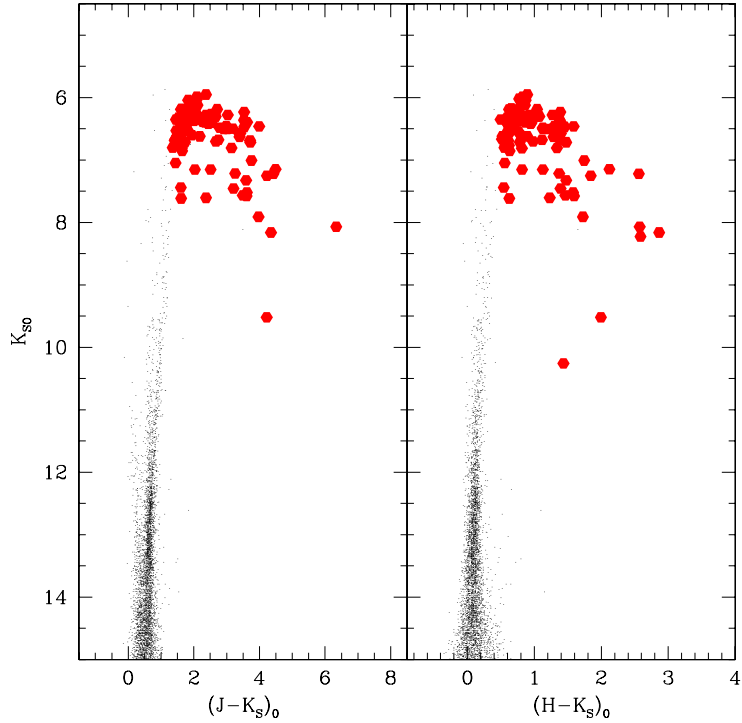}}
\caption{\label{fig:whitelock91.ps} Dereddened colour-magnitude
diagrams. Big dots represent the outer Bulge Mira stars found by
\citet{whitelock91}; the magnitudes plotted are mean magnitudes at the
equal distance of 8 kpc, adopting the distances of
\citet{whitelock91}. Small dots represent the point sources detected
by 2MASS within 1\arcmin\ from each Mira star.}
\end{center}
\end{figure}

Fig.\ \ref{fig:whitelock91.ps} shows the 2MASS point sources within
1\arcmin\ of each Mira star.  A giant branch is clearly apparent.
From an isochrone fitting (see Sect.\ \ref{fieldextinction}) we
derived the median extinction toward each field, resulting in values
of \Aks\ ranging from 0.01 to 0.30 mag, with a typical dispersion of
0.01-0.08 mag.  On the colour-magnitude diagrams the Mira stars appear
mostly brighter than the RGB tip of the field stars \citep[$K = 8.2$
mag at a distance of 8 kpc, see][]{frogel87}.  Due to the presence of
a circumstellar envelope, Mira stars have red colours (up to
$(H-K_{\rm s})_0 = 3$ mag ) and lie on the red-side of the giant
sequence.


It is therefore not possible to derive the interstellar extinction
toward these Mira stars from their colours relative to the RGB.  On
the other hand, we do not have reasons to assume that the Mira stars
are spatially distributed differently than the other giants stars.
Therefore, the extinction of its surrounding field stars may serve as
an approximation for that of the respective Mira star.

When going to lower latitude fields, however, the extinction increases
and the RGB becomes broader. A worry in the assumption that a Mira
star is at the median extinction of the field is the lack of knowledge
of the actual distribution of extinction along the line of sight which
does not warrant that the extinction of a given Mira star is the
median extinction of field stars.

\begin{figure}[h!]
\begin{center}
\resizebox{0.5\hsize}{!}{\includegraphics{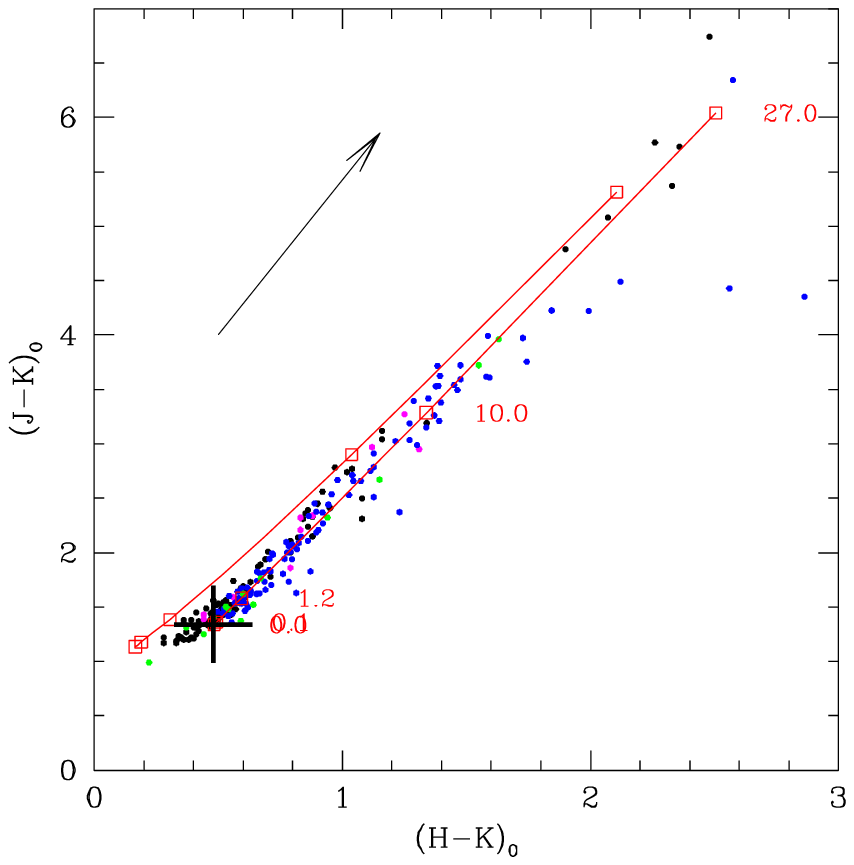}}
\caption{\label{fig:jhkLPV.ps} Dereddened colours of known dusty Mira
stars \citep{olivier01,whitelock-hyp,whitelock94,glass95,whitelock91}.
Near-infrared mean magnitudes are used.  The two curves represent M3
(upper) and an M10 (lower) type stars with increasing mass-loss rates
(indicated by squares and labels $\times 10^{-6}$ M$_\odot$/year), as
modelled by \citet{groenewegen93}.  The cross indicates the position
of an M10 star without mass-loss ($(H-K)_0 = 0.48$ and
$(J-K)_0 = 1.34$). The arrow shows the reddening vector for \Ak$ = 1$
mag.}
\end{center}
\end{figure}

\subsection{Colour-colour diagram of Mira stars}

\begin{figure}[h!]
\begin{center}
\resizebox{0.5\hsize}{!}{\includegraphics{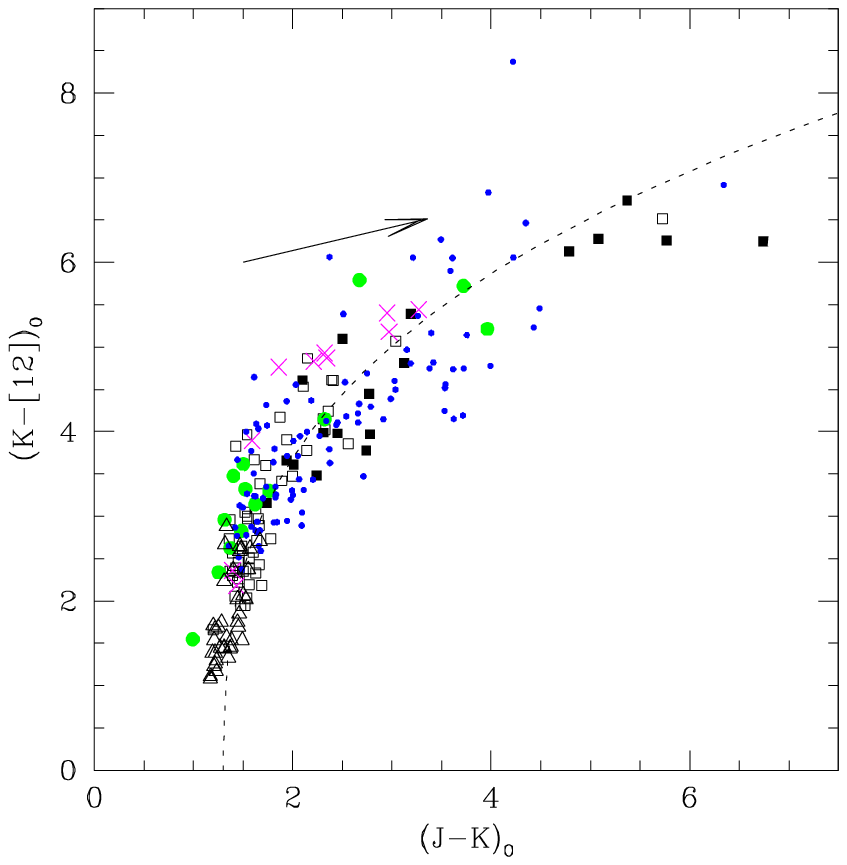}}
\caption{\label{fig:jk12LPV.ps} Dereddened colours of
         infrared-monitored Mira stars (based on the IRAS 12 $\mu$m
         and near-infrared mean magnitudes): in the solar vicinity
         (filled squares) \citep{olivier01}; detected by Hipparcos
         (open triangles) \citep{whitelock-hyp}; toward the South
         Galactic Cap (open squares) \citep{whitelock94}; in the Baade
         Sgr--I window (big dots) \citep{glass95}; in the outer Bulge
         (small dots) \citep{whitelock91}; in the Large Magellanic
         Cloud (crosses) \citep{whitelock94}.  The dotted line is the
         best fit to an IRAS sample of oxygen-rich AGB stars
         \citet{vanloon97}.  The arrow shows the reddening vector for
         \Ak$ = 1$ mag.  }
\end{center}
\end{figure}

\begin{figure}[h!]
\begin{center}
\resizebox{0.5\hsize}{!}{\includegraphics{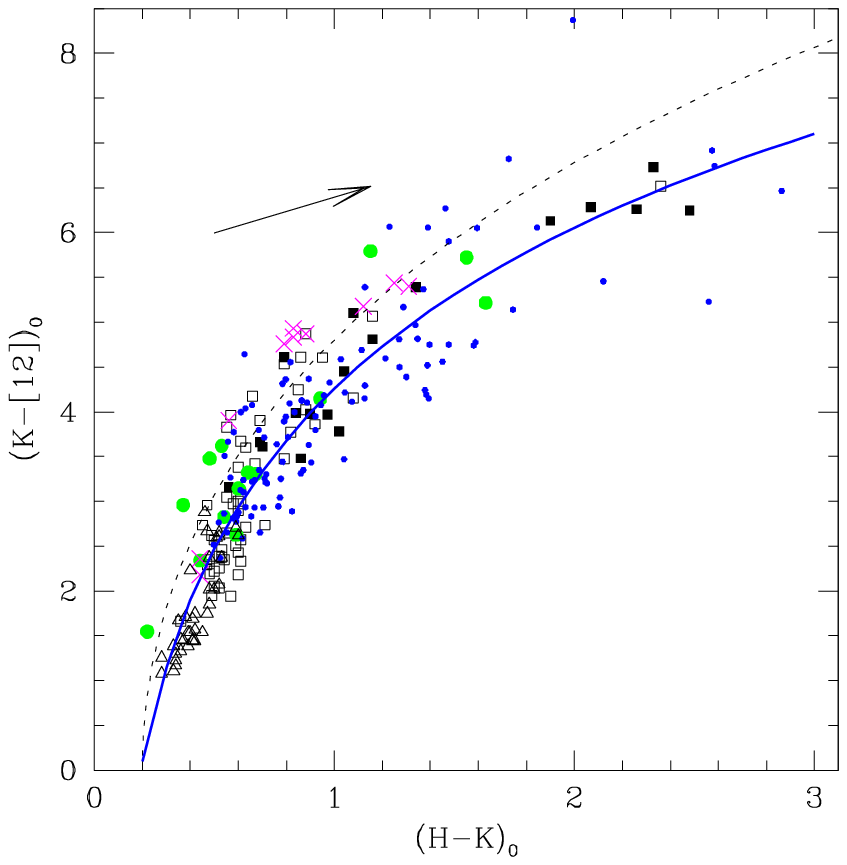}}
\caption{\label{fig:hk12LPV.ps} Dereddened colours of
         infrared-monitored Mira stars, based on the IRAS 12 $\mu$m
         and near-infrared mean magnitudes. Symbols are the same as in
         Fig.\ \ref{fig:jk12LPV.ps}.  The dotted line is the best fit
         to an IRAS sample of oxygen-rich AGB stars \citep{vanloon98}. 
         The continuous curve is our best fit for
         Galactic Mira stars. The arrow shows the reddening vector for
         \Ak$ = 1$ mag.}
\end{center}
\end{figure}

The $(J-K)_0$ vs. $(H-K)_0$ colours of  Mira stars are shown in
Fig.\ \ref{fig:jhkLPV.ps}.  For stars with low mass-loss rate
($<10^{-7}$ M$_\odot$ yr$^{-1}$) $(J-K)_{0}$ ranges between 1.2 and
1.6 \citep{whitelock-hyp}.  Dust-enshrouded IRAS AGB stars with
mass-loss rates of $10^{-6} - 10^{-4}$ M$_\odot$ yr$^{-1}$
\citep{olivier01} are much redder, $(J-K)_{0}$ ranging from 2 to 6.5
mag. The overall distribution appears to form a sequence of ever
redder colours with increasing mass-loss rate, a trend that is well
reproduced, e.g., by a model for an M10 type AGB star with increasing
shell opacity \citep{groenewegen93}. Thus, a higher mass loss has the
same effect on $(J-K)_{0}$ and $(H-K)_0$ colours as more interstellar
absorption/reddening, making a distinction between intrinsic and
interstellar reddening in these colours impossible.

In, both, the $(J-K)_0$ vs. $(K-[12])_0$ and the $(H-K)_0$
vs. $(K-[12])_0$ planes Mira stars are distributed along a broad
sequence compared to that seen in the $(J-K)_0$ vs. $(H-K)_0$ plane,
as shown in Figs.\ \ref{fig:jk12LPV.ps} and \ref{fig:hk12LPV.ps}.

A comparison with the best fit to the colour-colour sequence of
IRAS-selected oxygen-rich AGB stars \citep{vanloon98,vanloon97} shows
that in the $(H-K)_0$ vs. $(K-[12])_0$ plane  Mira stars lie below
that curve, while in the $(J-K)_0$ vs. $(K-[12])_0$ plane there
appears no such offset.  The offset could be due to water absorption
bands in the $H$ band \citep{frogel87,glass95}, that are found strong
in large amplitude variable AGB stars, although we could not find any
correlation between colour and variability index for the IRAS stars
used by van Loon et al.  \citep{fouque92,guglielmo93}.  The offset
might then indicate that for such cold stars the \citet{carter90}
transformations between the ESO photometry
\citep{fouque92,guglielmo93} and the SAAO system
\citep{vanloon98,vanloon97} are not adequate.

Mira stars with $0.2< (H-K)_0 < 3$ mag fit
$$(K-[12])_0 = 4.26(\pm 0.04) + 5.95(\pm0.17) {\rm ~log}(H-K)_0,$$
with an rms deviation of 0.5 mag.  Although we use the SAAO system
\citep{carter90}, in Appendix A we show that this relation is also
valid in the 2MASS photometric system.

\section{Interstellar extinction and intrinsic colours of the SiO targets}
\label{siotargets}
\begin{figure*}[th!]
\begin{center}
\resizebox{0.9\hsize}{!}{\includegraphics{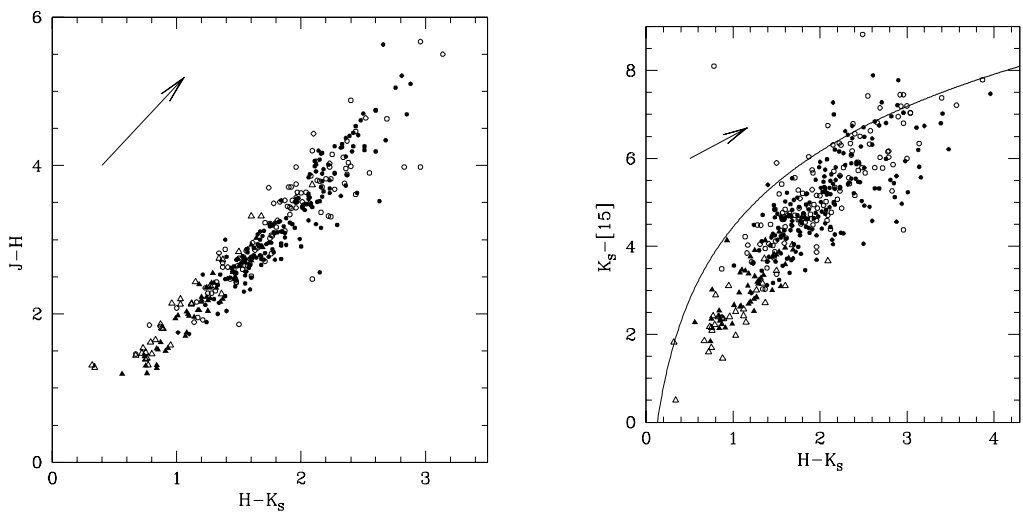}}
\caption{\label{fig:undcol.ps} {\bf Left panel:} 2MASS $J-H$ versus
$H-$\ks\ colours.  Stars with upper magnitude limits are not shown.
Dots and triangles represent objects with \ks\ smaller and larger than
6.0 mag, respectively. Filled and open symbols represent SiO
detections and non-detections, respectively.  The arrow shows the
reddening vector for \Aks\ $= 1$ mag.  {\bf Right panel:} 2MASS
\ks$-[15]$ versus $H-$\ks\ colours. Symbols are as in the left panel.
The curve represents the best fit to the colours of known Mira stars
(see Sect.\ \ref{miras}).}
\end{center}
\end{figure*}

\begin{figure*}[th!]
\begin{center}
\resizebox{0.9\hsize}{!}{\includegraphics{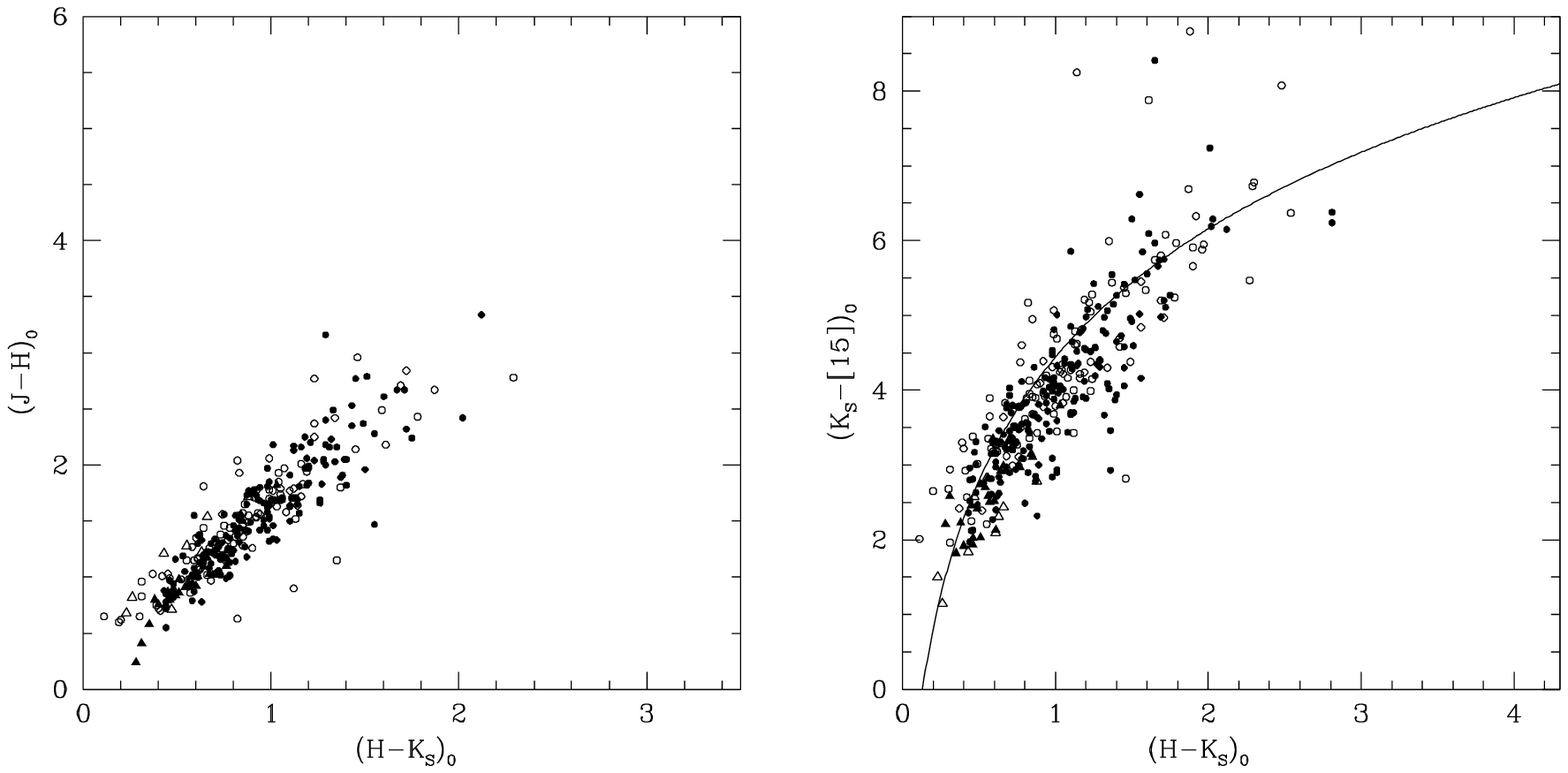}}
\caption{\label{fig:dercolfin.ps} {\bf Left panel:} Dereddened 2MASS
$(J-H)_0$ versus ($H-$\ks$)_0$ colours.  {\bf Right panel:} Dereddened
2MASS (\ks$-[15])_0$ versus ($H-$\ks$)_0$ colours. Symbols are as in
Fig.\ \ref{fig:undcol.ps}. ``Foreground objects'' have been removed.}
\end{center}
\end{figure*}

The observed colours of the SiO target stars are shown in Fig.\
\ref{fig:undcol.ps}.  For a given \ks$-[15]$, the $H-$\ks\ colours of
the SiO target stars are redder than those expected from the
colour-colour relation of known Mira stars, and some stars show excess
larger than $E(H-$\ks) $= 1$ mag. This is due to  interstellar
extinction along the line of sight.

Since observations of Mira stars may suffer from several magnitudes of
circumstellar reddening, it not possible to estimate the interstellar
extinction toward the SiO target stars by simply assuming an
intrinsic colour for a given star.

As seen from Figs.\ \ref{fig:jhkLPV.ps}, \ref{fig:jk12LPV.ps} and
\ref{fig:hk12LPV.ps}, Mira stars follow near- and mid-infrared
colour-colour relations.  In the first figure, Fig.\
\ref{fig:jhkLPV.ps}, the reddening vector is parallel to the
distribution of observed stellar colours and to those of models of AGB
stars of increasing shell opacity.  In contrast, the reddening vectors
in Figs.\ \ref{fig:jk12LPV.ps} and \ref{fig:hk12LPV.ps} have a slope
smaller than that of the distribution of observed stellar points and
in principle can permit a separation between interstellar and
circumstellar extinction. However, the dispersion of the Mira stars
around the colour-colour fiducial sequence is large (0.5 mag), due to
the non-contemporaneity of the near- and mid-infrared data and because
of the dependence of such relations on metallicity and stellar
spectral type. The uncertainty of the interstellar extinction
estimates by shifting a star along the reddening vector onto the
$H-K_{\rm s},K_{\rm s}-[15]$ curve is larger than \Aks $= 1$ mag for
$K-[15]>3.5$ mag.

Therefore, to deredden our targets we prefer to use the ``field''
extinction values (reported in Table \ref{table:final}), i.e.\ we
assume that a given SiO target star is located at the distance of the
median extinction along the line of sight.  Though the dispersion of
individual field star extinctions along a given line of sight is
considerable (from 0.1 to 0.8 \sig), the distribution is strongly
peaked, especially in the Bulge region. This means also to assume that
the SiO targets are located in the region with the highest stellar
density along the line of sight. This assumption is justified by the
fact that the lifetime of a star on the AGB evolutionary phase is very
short \citep[$\sim 5$\% of the time spent on the helium core burning
phase and from 0.1 to 2\% of the time spent on the main sequence phase
of such a star;][]{vassiliadis93} and we therefore expect most of the
AGB stars to be located in those regions with higher stellar density.

The distribution of the SiO targets in the dereddened $(H-K_{\rm
S})_0,(K_{\rm}-[15])_0$ diagram (Fig.\ \ref{fig:dercolfin.ps})
approaches that of known Mira stars. This confirms that the ``field''
median extinction is a good approximation of the interstellar
extinction for most of the SiO targets and that the SiO targets do
mostly belong to the highest stellar density region, i.e.\ the inner
Galaxy.  There is still an asymmetry in the distribution of the SiO
targets around the fiducial colour line of known Miras, which suggests
that we could have underestimated the interstellar extinction for part
of the sample. In regions of high extinction, due to their high
near-infrared luminosity, Mira-like stars are detectable to larger
distances than ordinary field stars.  Deeper infrared observations are
needed to obtain more accurate extinction estimates \citep{figer04}.

\subsection{``Foreground objects''}
Dereddening all observed points in the ($H-K_{\rm s})$ vs. $(J-H)$
diagram  (cf. Fig.\ \ref{fig:jhkLPV.ps}) to the \Aks $= 0.0$ position,
we can measure the total (circumstellar plus interstellar) extinction
of each SiO target (Table \ref{table:final}).  We assumed a stellar
photospheric $(J-K_{\rm s})_0$ colour of 1.4 mag and a $H_{0}-K_{\rm
S0}$ colour of 0.5 mag.

The difference of such total extinction estimates from 2MASS and DENIS
$J,K_{\rm s}$ data gives an rms of $\Delta$\Aks\ $ = 0.2$ mag, while the
estimates from 2MASS $H-K_{\rm s}$ and $J-K_{\rm s}$ colours gives an
rms difference of $\Delta$\Ak\ $ = 0.13$ mag. For only 12 SiO targets we
do not have any observed $H-K_{\rm s}$ or $J-K_{\rm s}$ values, and
therefore total extinction values are not determined (Table
\ref{table:final}).

The median interstellar extinction of the field stars surrounding a
target star was compared with the individual total extinction of the
target star.  As expected, on average the target stars show larger
total extinctions than their field stars. This is not the case,
however, for a group of $\sim50$ mostly very bright target stars (\ks\
$< 6.0$), at various longitudes, marked with flag 1 in Table
\ref{table:final}, which have total extinctions lower (a least 1\sig)
than the ``field'' extinction and are therefore likely to be
significantly less distant.  Thereby the extinction was used to
identify foreground objects. We dereddened the ``foreground objects''
by directly shifting them on the $H,K_{\rm s},[15]$ colour-colour
sequence.

\section{Intrinsic colours and mass-loss rates}\label{massloss}
The stellar mass-loss rate is best estimated from measurements of CO
rotational lines.  The CO emission arises in the circumstellar shell.
However, because of confusion with interstellar CO emission it is
difficult to obtain such measurements toward stars in the inner
Galaxy \citep{winnberg91}.  Although infrared emission also arises
from the stellar photosphere, stellar outflows may be studied from the
infrared emission of dust grains which form in the cool circumstellar
envelopes.  Relations between the infrared colours (e.g.  $J-K$,
$K-L$, $K-[12]$ or $K-[15]$) of O-rich AGB stars and their mass-loss
rate have been established empirically
\citep[e.g.][]{whitelock94,olivier01,alard01} and supported by theoretical
models \citep[e.g.][]{groenewegen93,ivezic99,jeong03,ojha03}.

The empirical relation between the $(K-[15])_0$ colour and the mass
loss rate, $\dot M$, is very useful to study stars detected in the
2MASS, DENIS, ISOGAL or MSX surveys toward the most obscured regions
of the Galaxy.

The uncertainties arising from the variability of the stars and the
temporal difference between the \ks\ and 15$\mu$m measurements, is
somewhat alleviated by using an average of the 2MASS and DENIS \ks\
fluxes and of the ISOGAL and MSX 15$\mu$m measurements \citep{ojha03}.
The remaining r.m.s. uncertainty of the mass-loss rate is thus a
factor $\sim$2 for $\dot M > 10^{-6}$ M$_\odot$ yr$^{-1}$.

\begin{figure}[t]
\begin{center}
\resizebox{0.6\hsize}{!}{\includegraphics{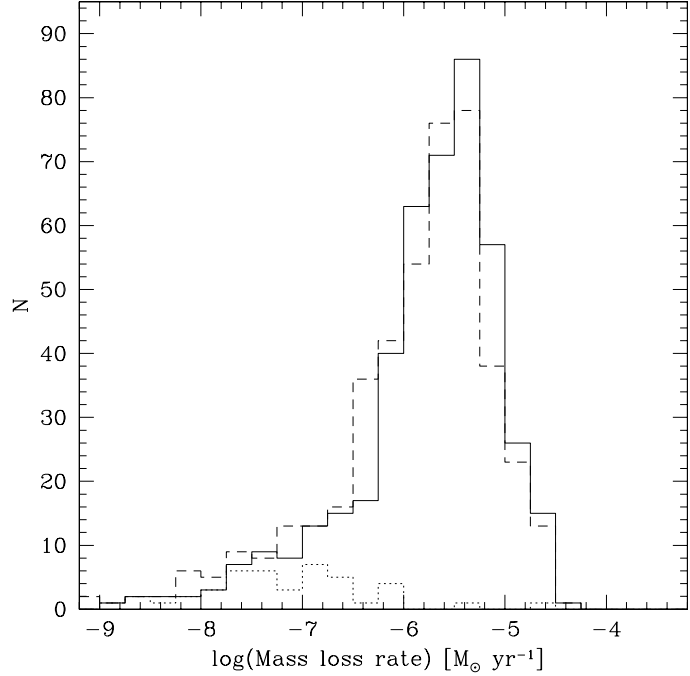}}
\caption{\label{fig:jeong.ps} Distribution of mass-loss rates derived
from the $K-[15]$ vs.\ $\dot M$ relation \citet{jeong03}. The
continuous line shows the distribution for all SiO targets dereddened
using Lutz's extinction law (Curve 3), and the dotted line that of
 foreground stars.  The dashed line is the
distribution of all SiO targets using the mid-infrared extinction law
of Mathis (Curve 1).}
\end{center}
\end{figure}

Following the prescription of \citet{jeong03} and \citet{ojha03} we
obtained mass-loss rates for the SiO targets, the distribution of
which is shown in Fig. \ref{fig:jeong.ps}. 90\% of the sources have
implied mass loss rates between $10^{-7}$ and $2 \times 10^{-5}$
M$_\odot$ yr$^{-1}$, with a peak in the range $10^{-6}$ -- $10^{-5}$
M$_\odot$ yr$^{-1}$, although the apparent distribution is widened by
the uncertainty of the mass-loss--colour relation, by the photometric
uncertainty and by the effects of variability.  Note that the
selection criterion on (\ks$-[15])_0$ for the ISOGAL sample completely
eliminated sources with $\dot M >10^{-5}$, and that the elimination of
OH/IR sources and the criteria on $A-D$ and $C-E$ colours for the MSX
sample also considerably reduced the proportion of sources with large
mass-loss rates \citepalias[see][]{messineo03_2,messineo02}.  The same
results are obtained considering the subsample of targets with the
best extinction corrections, i.e.\ with $\sigma _{\rm A_{\rm
K_{\rm s}}}<0.2$ mag.  The distribution of the mass-loss rates of the SiO
targets with detected SiO maser emission appears similar to that of
the SiO targets non-detected \citepalias{messineo02}.

Adopting Mathis' mid-infrared extinction law rather than Luts's one,
the distribution of the mass-loss rates only slightly shift toward
lower values (see Fig.\ \ref{fig:jeong.ps}).

\section{Conclusion}\label{conclusion}
 
 We estimated the interstellar extinction toward each of our 441 SiO
 target stars. For all 2MASS stars within 2-4\arcmin\ radius field of
 each target we shifted the $(J-K_{\rm s})$ and $(H-K_{\rm s})$ colour
 versus $K_{\rm s}$ magnitude along the reddening vector onto the
 reference red giant branch. The use of both 
 colour-magnitude planes enabled us to obtain  a mean extinction
 for each field and new constraints on the index of the near-infrared
 extinction power law, $\alpha$.  We found that a value of $\alpha =
 1.6$ is inconsistent with the colours of inner Galactic stars, and,
 taking 47 Tuc as a reference for the RGB, we determine 
 $\alpha = 1.9 \pm 0.1$.
  
 For \ks-band extinctions larger than 1.6 mag the 2MASS $(K_{\rm
 S},J-K_{\rm s})$ CMD yields too low extinction estimates, due to a
 selection effect from the $J$-band dropout of more distant sources.
 The 2MASS ($K_{\rm s},H-K_{\rm s}$) CMD suffers less from this bias.

 We reviewed  near- and mid-infrared dereddened colour-colour
 relations of Mira stars and use them to test the quality of the
 extinction corrections for each  SiO target.  

 Under the assumption that SiO targets are spatially distributed
 similarly to surrounding field stars, we corrected the photometric
 measurements of the SiO targets adopting the median extinction of
 their surrounding field stars.  Derredened colours of the SiO targets
 are not perfectly symmetrically distributed around the fiducial
 colour-colour line derived from known Mira stars, suggesting that for
 part of the SiO targets we may still be underestimating the
 interstellar extinction of up to about 15\%. About 50 SiO targets lie
 significantly in the ``foreground'' of the mean stellar distribution.
 
 Using the relation between mass-loss rate and (\ks$-15)_0$ colour
 given by \citet{jeong03}, we estimated that most of the SiO targets
 have mass-loss rates in the range 10$^{-7}$ to 10$^{-5}$ M$_{\odot}$
 yr$^{-1}$.

\begin{acknowledgements}
The MSX transmission curves were kindly provided by M. Egan.  MM
thanks P. Popowski, J. van Loon, S. Ganesh, and M. Schultheis for
useful discussions on interstellar extinction, and M. Sevenster for
her constructive criticism.  This paper uses and partly depends on the
studies of Mira stars conducted at the SAAO observatory by Patricia
Whitelock and her collaborators.  The DENIS project was carried out in
the context of EARA, the European Association for Research in
Astronomy.  This publication makes use of data products from the IRAS
data base server,
from the Two Micron All Sky Survey,
%
from the Midcourse Space Experiment, 
and from the SIMBAD data base.
The work of MM is funded by the Netherlands Research School for
Astronomy (NOVA) through a {\it netwerk 2, Ph.D. stipend}.
\end{acknowledgements}
    

\bibliographystyle{aa}
\bibliography{phy}

\begin{thebibliography}{30}
\expandafter\ifx\csname natexlab\endcsname\relax\def\natexlab#1{#1}\fi

\bibitem[{{Beaulieu} {et~al.}(2000){Beaulieu}, {Freeman}, {Kalnajs}, {Saha}, \&
  {Zhao}}]{beaulieu00}
{Beaulieu}, S.~F., {Freeman}, K.~C., {Kalnajs}, A.~J., {Saha}, P., \& {Zhao},
  H. 2000, \aj, 120, 855

\bibitem[{{Binney} {et~al.}(1991){Binney}, {Gerhard}, {Stark}, {Bally}, \&
  {Uchida}}]{binney91}
{Binney}, J., {Gerhard}, O.~E., {Stark}, A.~A., {Bally}, J., \& {Uchida}, K.~I.
  1991, \mnras, 252, 210

\bibitem[{{Dame} {et~al.}(2001){Dame}, {Hartmann}, \& {Thaddeus}}]{dame01}
{Dame}, T.~M., {Hartmann}, D., \& {Thaddeus}, P. 2001, \apj, 547, 792

\bibitem[{{Debattista} {et~al.}(2002){Debattista}, {Gerhard}, \&
  {Sevenster}}]{debattista02}
{Debattista}, V.~P., {Gerhard}, O., \& {Sevenster}, M.~N. 2002, \mnras, 334,
  355

\bibitem[{{Deguchi} {et~al.}(2000{\natexlab{a}}){Deguchi}, {Fujii}, {Izumiura},
  {Kameya}, {Nakada}, \& {Nakashima}}]{deguchi00b}
{Deguchi}, S., {Fujii}, T., {Izumiura}, H., {et~al.} 2000{\natexlab{a}}, \apjs,
  130, 351

\bibitem[{{Deguchi} {et~al.}(2000{\natexlab{b}}){Deguchi}, {Fujii}, {Izumiura},
  {Kameya}, {Nakada}, {Nakashima}, {Ootsubo}, \& {Ukita}}]{deguchi00a}
{Deguchi}, S., {Fujii}, T., {Izumiura}, H., {et~al.} 2000{\natexlab{b}}, \apjs,
  128, 571

\bibitem[{{Elitzur}(1992)}]{elitzur92}
{Elitzur}, M. 1992, \araa, 30, 75

\bibitem[{{Feast} {et~al.}(1989){Feast}, {Glass}, {Whitelock}, \&
  {Catchpole}}]{feast89}
{Feast}, M.~W., {Glass}, I.~S., {Whitelock}, P.~A., \& {Catchpole}, R.~M. 1989,
  \mnras, 241, 375

\bibitem[{{Fux}(1999)}]{fux99}
{Fux}, R. 1999, \aap, 345, 787

\bibitem[{{Ghez} {et~al.}(2000){Ghez}, {Morris}, {Becklin}, {Tanner}, \&
  {Kremenek}}]{ghez00}
{Ghez}, A.~M., {Morris}, M., {Becklin}, E.~E., {Tanner}, A., \& {Kremenek}, T.
  2000, \nat, 407, 349

\bibitem[{{Gilmore} \& {Reid}(1983)}]{gilmore83}
{Gilmore}, G. \& {Reid}, N. 1983, \mnras, 202, 1025

\bibitem[{{Gilmore} {et~al.}(2002){Gilmore}, {Wyse}, \& {Norris}}]{gilmore02}
{Gilmore}, G., {Wyse}, R.~F.~G., \& {Norris}, J.~E. 2002, \apjl, 574, L39

\bibitem[{{Habing}(1996)}]{habing96}
{Habing}, H.~J. 1996, \aapr, 7, 97

\bibitem[{{Helmi} {et~al.}(1999){Helmi}, {White}, {de Zeeuw}, \&
  {Zhao}}]{helmi99}
{Helmi}, A., {White}, S.~D.~M., {de Zeeuw}, P.~T., \& {Zhao}, H. 1999, \nat,
  402, 53

\bibitem[{{Ibata} {et~al.}(1994){Ibata}, {Gilmore}, \& {Irwin}}]{ibata94}
{Ibata}, R.~A., {Gilmore}, G., \& {Irwin}, M.~J. 1994, \nat, 370, 194

\bibitem[{{Iben}(1985)}]{iben85}
{Iben}, I. 1985, \qjras, 26, 1

\bibitem[{{Izumiura} {et~al.}(1999){Izumiura}, {Deguchi}, {Fujii}, {Kameya},
  {Matsumoto}, {Nakada}, {Ootsubo}, \& {Ukita}}]{izumiura99}
{Izumiura}, H., {Deguchi}, S., {Fujii}, T., {et~al.} 1999, \apjs, 125, 257

\bibitem[{{Launhardt} {et~al.}(2002){Launhardt}, {Zylka}, \&
  {Mezger}}]{mezger02}
{Launhardt}, R., {Zylka}, R., \& {Mezger}, P.~G. 2002, \aap, 384, 112

\bibitem[{{Lewis}(1989)}]{lewis89}
{Lewis}, B.~M. 1989, \apj, 338, 234

\bibitem[{{McWilliam} \& {Rich}(1994)}]{mcwilliam94}
{McWilliam}, A. \& {Rich}, R.~M. 1994, \apjs, 91, 749

\bibitem[{{Omont} {et~al.}(2003){Omont}, {Gilmore}, {Alard}, {Aracil},
  {August}, {Baliyan}, {Beaulieu}, {B{\' e}gon}, {Bertou}, {Blommaert},
  {Borsenberger}, {Burgdorf}, {Caillaud}, {Cesarsky}, {Chitre}, {Copet}, {de
  Batz}, {Egan}, {Egret}, {Epchtein}, {Felli}, {Fouqu{\' e}}, {Ganesh},
  {Genzel}, {Glass}, {Gredel}, {Groenewegen}, {Guglielmo}, {Habing},
  {Hennebelle}, {Jiang}, {Joshi}, {Kimeswenger}, {Messineo}, {Miville-Desch{\^
  e}nes}, {Moneti}, {Morris}, {Ojha}, {Ortiz}, {Ott}, {Parthasarathy}, {P{\'
  e}rault}, {Price}, {Robin}, {Schultheis}, {Schuller}, {Simon}, {Soive},
  {Testi}, {Teyssier}, {Tiph{\` e}ne}, {Unavane}, {van Loon}, \&
  {Wyse}}]{omont03}
{Omont}, A., {Gilmore}, G.~F., {Alard}, C., {et~al.} 2003, \aap, 403, 975

\bibitem[{{Reid} \& {Menten}(1997)}]{reid97}
{Reid}, M.~J. \& {Menten}, K.~M. 1997, \apj, 476, 327

\bibitem[{{Reid} \& {Moran}(1988)}]{reid88}
{Reid}, M.~J. \& {Moran}, J.~M. 1988, in Galactic and Extragalactic Radio
  Astronomy, 255--294

\bibitem[{{Sch{\" o}del} {et~al.}(2003){Sch{\" o}del}, {Ott}, {Genzel},
  {Eckart}, {Mouawad}, \& {Alexander}}]{schodel03}
{Sch{\" o}del}, R., {Ott}, T., {Genzel}, R., {et~al.} 2003, \apj, 596, 1015

\bibitem[{{Schuller} {et~al.}(2003){Schuller}, {Ganesh}, {Messineo}, {Moneti},
  {Blommaert}, {Alard}, {Aracil}, {Miville-Desch{\^ e}nes}, {Omont},
  {Schultheis}, {Simon}, {Soive}, \& {Testi}}]{schuller03}
{Schuller}, F., {Ganesh}, S., {Messineo}, M., {et~al.} 2003, \aap, 403, 955

\bibitem[{{Sevenster}(1999)}]{sevenster99}
{Sevenster}, M.~N. 1999, \mnras, 310, 629

\bibitem[{{Sevenster} {et~al.}(1997{\natexlab{a}}){Sevenster}, {Chapman},
  {Habing}, {Killeen}, \& {Lindqvist}}]{sevenster97a}
{Sevenster}, M.~N., {Chapman}, J.~M., {Habing}, H.~J., {Killeen}, N.~E.~B., \&
  {Lindqvist}, M. 1997{\natexlab{a}}, \aaps, 122, 79

\bibitem[{{Sevenster} {et~al.}(1997{\natexlab{b}}){Sevenster}, {Chapman},
  {Habing}, {Killeen}, \& {Lindqvist}}]{sevenster97b}
{Sevenster}, M.~N., {Chapman}, J.~M., {Habing}, H.~J., {Killeen}, N.~E.~B., \&
  {Lindqvist}, M. 1997{\natexlab{b}}, \aaps, 124, 509

\bibitem[{{Sevenster} {et~al.}(2001){Sevenster}, {van Langevelde}, {Moody},
  {Chapman}, {Habing}, \& {Killeen}}]{sevenster01}
{Sevenster}, M.~N., {van Langevelde}, H.~J., {Moody}, R.~A., {et~al.} 2001,
  \aap, 366, 481

\bibitem[{{Wyse} {et~al.}(1997){Wyse}, {Gilmore}, \& {Franx}}]{wyse97}
{Wyse}, R.~F.~G., {Gilmore}, G., \& {Franx}, M. 1997, \araa, 35, 637

\end{thebibliography}

\begin{thebibliography}{62}
\expandafter\ifx\csname natexlab\endcsname\relax\def\natexlab#1{#1}\fi

\bibitem[{{Alard} {et~al.}(2001){Alard}, {Blommaert}, {Cesarsky}, {Epchtein},
  {Felli}, {Fouque}, {Ganesh}, {Genzel}, {Gilmore}, {Glass}, {Habing}, {Omont},
  {Perault}, {Price}, {Robin}, {Schultheis}, {Simon}, {van Loon}, {Alcock},
  {Allsman}, {Alves}, {Axelrod}, {Becker}, {Bennett}, {Cook}, {Drake},
  {Freeman}, {Geha}, {Griest}, {Lehner}, {Marshall}, {Minniti}, {Nelson},
  {Peterson}, {Popowski}, {Pratt}, {Quinn}, {Sutherland}, {Tomaney},
  {Vandehei}, \& {Welch}}]{alard01}
{Alard}, C., {Blommaert}, J.~A.~D.~L., {Cesarsky}, C., {et~al.} 2001, \apj,
  552, 289

\bibitem[{{Bertelli} {et~al.}(1994){Bertelli}, {Bressan}, {Chiosi}, {Fagotto},
  \& {Nasi}}]{bertelli94}
{Bertelli}, G., {Bressan}, A., {Chiosi}, C., {Fagotto}, F., \& {Nasi}, E. 1994,
  \aaps, 106, 275

\bibitem[{{Blommaert} {et~al.}(2003){Blommaert}, {Siebenmorgen}, {Coulais},
  {Metcalfe}, {Miville-Deschenes}, {Okumura}, {Ott}, {Pollock}, {Sauvage}, \&
  {Starck}}]{isocam03}
{Blommaert}, J.~A.~D.~L., {Siebenmorgen}, R., {Coulais}, A., {et~al.}, eds.
  2003, {The ISO Handbook, Volume II - CAM - The ISO Camera}

\bibitem[{{Cardelli} {et~al.}(1989){Cardelli}, {Clayton}, \&
  {Mathis}}]{cardelli89}
{Cardelli}, J.~A., {Clayton}, G.~C., \& {Mathis}, J.~S. 1989, \apj, 345, 245

\bibitem[{{Carpenter}(2001)}]{carpenter01}
{Carpenter}, J.~M. 2001, \aj, 121, 2851

\bibitem[{{Carter}(1990)}]{carter90}
{Carter}, B.~S. 1990, \mnras, 242, 1

\bibitem[{{Cioni} {et~al.}(2003){Cioni}, {Blommaert}, {Groenewegen}, {Habing},
  {Hron}, {Kerschbaum}, {Loup}, {Omont}, {van Loon}, {Whitelock}, \&
  {Zijlstra}}]{cioni03}
{Cioni}, M.-R.~L., {Blommaert}, J.~A.~D.~L., {Groenewegen}, M.~A.~T., {et~al.}
  2003, \aap, 406, 51

\bibitem[{{Cotera} {et~al.}(2000){Cotera}, {Simpson}, {Erickson}, {Colgan},
  {Burton}, \& {Allen}}]{cotera00}
{Cotera}, A.~S., {Simpson}, J.~P., {Erickson}, E.~F., {et~al.} 2000, \apjs,
  129, 123

\bibitem[{{Cutri} {et~al.}(2003){Cutri}, {Skrutskie}, \& {Van Dyk}}]{2massES}
{Cutri}, C.~M., {Skrutskie}, M.~F., \& {Van Dyk}, S. 2003, avalaible on line at
  \\ {\tt http://www.ipac.caltech.edu/2mass/}

\bibitem[{{Draine} \& {Lee}(1984)}]{draine84}
{Draine}, B.~T. \& {Lee}, H.~M. 1984, \apj, 285, 89

\bibitem[{{Drimmel} {et~al.}(2003){Drimmel}, {Cabrera-Lavers}, \& {L{\'
  o}pez-Corredoira}}]{drimmel03}
{Drimmel}, R., {Cabrera-Lavers}, A., \& {L{\' o}pez-Corredoira}, M. 2003, \aap,
  409, 205

\bibitem[{{Dutra} {et~al.}(2002){Dutra}, {Santiago}, \& {Bica}}]{dutra02}
{Dutra}, C.~M., {Santiago}, B.~X., \& {Bica}, E. 2002, \aap, 381, 219

\bibitem[{{Dutra} {et~al.}(2003){Dutra}, {Santiago}, {Bica}, \&
  {Barbuy}}]{dutra03}
{Dutra}, C.~M., {Santiago}, B.~X., {Bica}, E.~L.~D., \& {Barbuy}, B. 2003,
  \mnras, 338, 253

\bibitem[{{Egan} {et~al.}(1999){Egan}, {Price}, {Moshir}, {Cohen}, {Tedesco},
  {Murdock}, {Zweil}, {Burdick}, {Bonito}, {Gugliotti}, \& J.}]{egan99}
{Egan}, M.~P., {Price}, S.~D., {Moshir}, M.~M., {et~al.} 1999, AFRL-VS-TR-1999,
  1522

\bibitem[{{Epchtein} {et~al.}(1994){Epchtein}, {de Batz}, {Copet}, {Fouque},
  {Lacombe}, {Le Bertre}, {Mamon}, {Rouan}, {Tiphene}, {Burton}, {Deul},
  {Habing}, {Boersenberger}, {Dennefeld}, {Omont}, {Renault},
  {Rocca-Volmerange}, {Kimeswenger}, {Appenzeller}, {Bender}, {Forveille},
  {Garzon}, {Hron}, {Persi}, {Ferrari-Toniolo}, \& {Vauglin}}]{epchtein94}
{Epchtein}, N., {de Batz}, B., {Copet}, E., {et~al.} 1994, \apss, 217, 3

\bibitem[{{Ferraro} {et~al.}(1999){Ferraro}, {Messineo}, {Fusi Pecci}, {de
  Palo}, {Straniero}, {Chieffi}, \& {Limongi}}]{ferraro99}
{Ferraro}, F.~R., {Messineo}, M., {Fusi Pecci}, F., {et~al.} 1999, \aj, 118,
  1738

\bibitem[{{Figer} {et~al.}(2004){Figer}, {Rich}, {Kim}, {Morris}, \&
  {Serabyn}}]{figer04}
{Figer}, D.~F., {Rich}, R.~M., {Kim}, S.~S., {Morris}, M., \& {Serabyn}, E.
  2004, \apj, 601, 319

\bibitem[{{Fluks} {et~al.}(1994){Fluks}, {Plez}, {The}, {de Winter},
  {Westerlund}, \& {Steenman}}]{fluks94}
{Fluks}, M.~A., {Plez}, B., {The}, P.~S., {et~al.} 1994, \aaps, 105, 311

\bibitem[{{Fouque} {et~al.}(1992){Fouque}, {Le Bertre}, {Epchtein},
  {Guglielmo}, \& {Kerschbaum}}]{fouque92}
{Fouque}, P., {Le Bertre}, T., {Epchtein}, N., {Guglielmo}, F., \&
  {Kerschbaum}, F. 1992, \aaps, 93, 151

\bibitem[{{Frogel} {et~al.}(1999){Frogel}, {Tiede}, \& {Kuchinski}}]{frogel99}
{Frogel}, J.~A., {Tiede}, G.~P., \& {Kuchinski}, L.~E. 1999, \aj, 117, 2296

\bibitem[{{Frogel} \& {Whitford}(1987)}]{frogel87}
{Frogel}, J.~A. \& {Whitford}, A.~E. 1987, \apj, 320, 199

\bibitem[{{Girardi} {et~al.}(2000){Girardi}, {Bressan}, {Bertelli}, \&
  {Chiosi}}]{bertelli00}
{Girardi}, L., {Bressan}, A., {Bertelli}, G., \& {Chiosi}, C. 2000, \aaps, 141,
  371

\bibitem[{{Glass}(1999)}]{glass99}
{Glass}, I.~S. 1999, {Book Review: Handbook of infrared astronomy} (Cambridge
  U. Press)

\bibitem[{{Glass} {et~al.}(1995){Glass}, {Whitelock}, {Catchpole}, \&
  {Feast}}]{glass95}
{Glass}, I.~S., {Whitelock}, P.~A., {Catchpole}, R.~M., \& {Feast}, M.~W. 1995,
  \mnras, 273, 383

\bibitem[{{Groenewegen} \& {de Jong}(1993)}]{groenewegen93}
{Groenewegen}, M.~A.~T. \& {de Jong}, T. 1993, \aap, 267, 410

\bibitem[{{Guglielmo} {et~al.}(1993){Guglielmo}, {Epchtein}, {Le Bertre},
  {Fouque}, {Hron}, {Kerschbaum}, \& {Lepine}}]{guglielmo93}
{Guglielmo}, F., {Epchtein}, N., {Le Bertre}, T., {et~al.} 1993, \aaps, 99, 31

\bibitem[{{He} {et~al.}(1995){He}, {Whittet}, {Kilkenny}, \& {Spencer
  Jones}}]{he95}
{He}, L., {Whittet}, D.~C.~B., {Kilkenny}, D., \& {Spencer Jones}, J.~H. 1995,
  \apjs, 101, 335

\bibitem[{{Hennebelle} {et~al.}(2001){Hennebelle}, {P{\' e}rault}, {Teyssier},
  \& {Ganesh}}]{hennebelle01}
{Hennebelle}, P., {P{\' e}rault}, M., {Teyssier}, D., \& {Ganesh}, S. 2001,
  \aap, 365, 598

\bibitem[{{Ivez\'\i c} {et~al.}(1999){Ivez\'\i c}, {Nenkova}, \&
  {Elitzur}}]{ivezic99}
{Ivez\'\i c}, Z., {Nenkova}, M., \& {Elitzur}, M. 1999, User Manual for DUSTY
  (Lexington: Univ. Kentucky)

\bibitem[{{Jeong} {et~al.}(2003){Jeong}, {Winters}, {Le Bertre}, \&
  {Sedmayr}}]{jeong03}
{Jeong}, K.~S., {Winters}, J.~M., {Le Bertre}, T., \& {Sedmayr}, E. 2003,
  Proceeding of WS on Mass-Losing Pulsating Stars and their Circumstellar
  Matter, Sendai, Japan, Y.Nakada \& M.Honma (eds), Kluwer ASSL series

\bibitem[{{Jiang} {et~al.}(2003){Jiang}, {Omont}, {Ganesh}, {Simon}, \&
  {Schuller}}]{jiang03}
{Jiang}, B.~W., {Omont}, A., {Ganesh}, S., {Simon}, G., \& {Schuller}, F. 2003,
  \aap, 400, 903

\bibitem[{{L{\' o}pez-Corredoira} {et~al.}(2002){L{\' o}pez-Corredoira},
  {Cabrera-Lavers}, {Garz{\' o}n}, \& {Hammersley}}]{lopez03}
{L{\' o}pez-Corredoira}, M., {Cabrera-Lavers}, A., {Garz{\' o}n}, F., \&
  {Hammersley}, P.~L. 2002, \aap, 394, 883

\bibitem[{{Landini} {et~al.}(1984){Landini}, {Natta}, {Salinari}, {Oliva}, \&
  {Moorwood}}]{landini84}
{Landini}, M., {Natta}, A., {Salinari}, P., {Oliva}, E., \& {Moorwood},
  A.~F.~M. 1984, \aap, 134, 284

\bibitem[{{Lutz}(1999)}]{lutz99}
{Lutz}, D. 1999, in ESA SP-427: The Universe as Seen by ISO, Vol. 427, 623

\bibitem[{{Lutz} {et~al.}(1996){Lutz}, {Feuchtgruber}, {Genzel}, {Kunze},
  {Rigopoulou}, {Spoon}, {Wright}, {Egami}, {Katterloher}, {Sturm},
  {Wieprecht}, {Sternberg}, {Moorwood}, \& {de Graauw}}]{lutz96}
{Lutz}, D., {Feuchtgruber}, H., {Genzel}, R., {et~al.} 1996, \aap, 315, L269

\bibitem[{{Mathis}(1990)}]{mathis90}
{Mathis}, J.~S. 1990, \araa, 28, 37

\bibitem[{{Mathis}(1998)}]{mathis98}
---. 1998, \apj, 497, 824

\bibitem[{{McWilliam} \& {Rich}(1994)}]{rich94}
{McWilliam}, A. \& {Rich}, R.~M. 1994, \apjs, 91, 749

\bibitem[{{Messineo} {et~al.}(2004{\natexlab{a}}){Messineo}, {Habing},
  {Menten}, {Omont}, \& {Sjouwerman}}]{messineo03_4}
{Messineo}, M., {Habing}, H.~J., {Menten}, K.~M., {Omont}, A., \& {Sjouwerman},
  L.~O. 2004{\natexlab{a}}, \aap\ in preparation (Chapter\,V)

\bibitem[{{Messineo} {et~al.}(2004{\natexlab{b}}){Messineo}, {Habing},
  {Menten}, {Omont}, \& {Sjouwerman}}]{messineo03_2}
---. 2004{\natexlab{b}}, \aap, 418, 103 (Chapter\,III)

\bibitem[{{Messineo} {et~al.}(2002){Messineo}, {Habing}, {Sjouwerman}, {Omont},
  \& {Menten}}]{messineo02}
{Messineo}, M., {Habing}, H.~J., {Sjouwerman}, L.~O., {Omont}, A., \& {Menten},
  K.~M. 2002, \aap, 393, 115 (Chapter\,II)

\bibitem[{{Ojha} {et~al.}(2003){Ojha}, {Omont}, {Schuller}, {Simon}, {Ganesh},
  \& {Schultheis}}]{ojha03}
{Ojha}, D.~K., {Omont}, A., {Schuller}, F., {et~al.} 2003, \aap, 403, 141

\bibitem[{{Olivier} {et~al.}(2001){Olivier}, {Whitelock}, \&
  {Marang}}]{olivier01}
{Olivier}, E.~A., {Whitelock}, P., \& {Marang}, F. 2001, \mnras, 326, 490

\bibitem[{{Omont} {et~al.}(2003){Omont}, {Gilmore}, {Alard}, {Aracil},
  {August}, {Baliyan}, {Beaulieu}, {B{\' e}gon}, {Bertou}, {Blommaert},
  {Borsenberger}, {Burgdorf}, {Caillaud}, {Cesarsky}, {Chitre}, {Copet}, {de
  Batz}, {Egan}, {Egret}, {Epchtein}, {Felli}, {Fouqu{\' e}}, {Ganesh},
  {Genzel}, {Glass}, {Gredel}, {Groenewegen}, {Guglielmo}, {Habing},
  {Hennebelle}, {Jiang}, {Joshi}, {Kimeswenger}, {Messineo}, {Miville-Desch{\^
  e}nes}, {Moneti}, {Morris}, {Ojha}, {Ortiz}, {Ott}, {Parthasarathy}, {P{\'
  e}rault}, {Price}, {Robin}, {Schultheis}, {Schuller}, {Simon}, {Soive},
  {Testi}, {Teyssier}, {Tiph{\` e}ne}, {Unavane}, {van Loon}, \&
  {Wyse}}]{omont03}
{Omont}, A., {Gilmore}, G.~F., {Alard}, C., {et~al.} 2003, \aap, 403, 975

\bibitem[{{Price} {et~al.}(2001){Price}, {Egan}, {Carey}, {Mizuno}, \&
  {Kuchar}}]{price01}
{Price}, S.~D., {Egan}, M.~P., {Carey}, S.~J., {Mizuno}, D.~R., \& {Kuchar},
  T.~A. 2001, \aj, 121, 2819

\bibitem[{{Rieke} \& {Lebofsky}(1985)}]{rieke85}
{Rieke}, G.~H. \& {Lebofsky}, M.~J. 1985, \apj, 288, 618

\bibitem[{{Rosenthal} {et~al.}(2000){Rosenthal}, {Bertoldi}, \&
  {Drapatz}}]{rosenthal00}
{Rosenthal}, D., {Bertoldi}, F., \& {Drapatz}, S. 2000, \aap, 356, 705

\bibitem[{{Schuller} {et~al.}(2003){Schuller}, {Ganesh}, {Messineo}, {Moneti},
  {Blommaert}, {Alard}, {Aracil}, {Miville-Desch{\^ e}nes}, {Omont},
  {Schultheis}, {Simon}, {Soive}, \& {Testi}}]{schuller03}
{Schuller}, F., {Ganesh}, S., {Messineo}, M., {et~al.} 2003, \aap, 403, 955

\bibitem[{{Schultheis} {et~al.}(1999){Schultheis}, {Ganesh}, {Simon}, {Omont},
  {Alard}, {Borsenberger}, {Copet}, {Epchtein}, {Fouqu{\' e}}, \&
  {Habing}}]{schultheis99}
{Schultheis}, M., {Ganesh}, S., {Simon}, G., {et~al.} 1999, \aap, 349, L69

\bibitem[{{Stanek}(1998)}]{stanek98}
{Stanek}, K.~Z. 1998, Using the DIRBE/IRAS All-Sky Reddening Map to Select
  Low-Reddening Windows Near the Galactic Plane, preprint [astro-ph/9802307]

\bibitem[{{Udalski}(2003)}]{udalski03}
{Udalski}, A. 2003, \apj, 590, 284

\bibitem[{{van de Hulst}(1946)}]{vandehulst46}
{van de Hulst}, H.~C. 1946, {Optics of spherical particles.} (Amsterdam,
  Drukkerij J.~F.~Duwaer, 1946.), 1

\bibitem[{{van Loon} {et~al.}(2003){van Loon}, {Gilmore}, {Omont}, {Blommaert},
  {Glass}, {Messineo}, {Schuller}, {Schultheis}, {Yamamura}, \&
  {Zhao}}]{vanloon03}
{van Loon}, J.~T., {Gilmore}, G.~F., {Omont}, A., {et~al.} 2003, \mnras, 338,
  857

\bibitem[{{van Loon} {et~al.}(1998){van Loon}, {Zijlstra}, {Whitelock}, {Te
  Lintel Hekkert}, {Chapman}, {Loup}, {Groenewegen}, {Waters}, \&
  {Trams}}]{vanloon98}
{van Loon}, J.~T., {Zijlstra}, A.~A., {Whitelock}, P.~A., {et~al.} 1998, \aap,
  329, 169

\bibitem[{{van Loon} {et~al.}(1997){van Loon}, {Zijlstra}, {Whitelock},
  {Waters}, {Loup}, \& {Trams}}]{vanloon97}
---. 1997, \aap, 325, 585

\bibitem[{{Vassiliadis} \& {Wood}(1993)}]{vassiliadis93}
{Vassiliadis}, E. \& {Wood}, P.~R. 1993, \apj, 413, 641

\bibitem[{{Wainscoat} {et~al.}(1992){Wainscoat}, {Cohen}, {Volk}, {Walker}, \&
  {Schwartz}}]{wainscoat92}
{Wainscoat}, R.~J., {Cohen}, M., {Volk}, K., {Walker}, H.~J., \& {Schwartz},
  D.~E. 1992, \apjs, 83, 111

\bibitem[{{Whitelock} {et~al.}(1991){Whitelock}, {Feast}, \&
  {Catchpole}}]{whitelock91}
{Whitelock}, P., {Feast}, M., \& {Catchpole}, R. 1991, \mnras, 248, 276

\bibitem[{{Whitelock} {et~al.}(2000){Whitelock}, {Marang}, \&
  {Feast}}]{whitelock-hyp}
{Whitelock}, P., {Marang}, F., \& {Feast}, M. 2000, \mnras, 319, 728

\bibitem[{{Whitelock} {et~al.}(1994){Whitelock}, {Menzies}, {Feast}, {Marang},
  {Carter}, {Roberts}, {Catchpole}, \& {Chapman}}]{whitelock94}
{Whitelock}, P., {Menzies}, J., {Feast}, M., {et~al.} 1994, \mnras, 267, 711

\bibitem[{{Whitelock} {et~al.}(2003){Whitelock}, {Feast}, {van Loon}, \&
  {Zijlstra}}]{whitelock03}
{Whitelock}, P.~A., {Feast}, M.~W., {van Loon}, J.~T., \& {Zijlstra}, A.~A.
  2003, \mnras, 342, 86

\bibitem[{{Winnberg} {et~al.}(1991){Winnberg}, {Lindqvist}, {Olofsson}, \&
  {Henkel}}]{winnberg91}
{Winnberg}, A., {Lindqvist}, M., {Olofsson}, H., \& {Henkel}, C. 1991, \aap,
  245, 195

\end{thebibliography}

\begin{thebibliography}{74}
\expandafter\ifx\csname natexlab\endcsname\relax\def\natexlab#1{#1}\fi

\bibitem[{{Alcolea} {et~al.}(1990){Alcolea}, {Bujarrabal}, \&
  {Gomez-Gonzalez}}]{alcolea90}
{Alcolea}, J., {Bujarrabal}, V., \& {Gomez-Gonzalez}, J. 1990, \aap, 231, 431

\bibitem[{{Alcolea} {et~al.}(1999){Alcolea}, {Pardo}, {Bujarrabal},
  {Bachiller}, {Barcia}, {Colomer}, {Gallego}, {G{\' o}mez-Gonz{\' a}lez}, {del
  Pino Cisneros}, {Planesas}, {del R{\'{\i}}o}, {Rodr{\'{\i}}guez-Franco}, {del
  Romero}, {Tafalla}, \& {de Vicente}}]{alcolea99}
{Alcolea}, J., {Pardo}, J.~R., {Bujarrabal}, V., {et~al.} 1999, \aaps, 139, 461

\bibitem[{{Bally} {et~al.}(1988){Bally}, {Stark}, {Wilson}, \&
  {Henkel}}]{bally88}
{Bally}, J., {Stark}, A.~A., {Wilson}, R.~W., \& {Henkel}, C. 1988, \apj, 324,
  223

\bibitem[{{Baud} {et~al.}(1979){Baud}, {Habing}, {Matthews}, \&
  {Winnberg}}]{baud79}
{Baud}, B., {Habing}, H.~J., {Matthews}, H.~E., \& {Winnberg}, A. 1979, \aaps,
  35, 179

\bibitem[{{Bedijn}(1987)}]{bedijn87}
{Bedijn}, P.~J. 1987, \aap, 186, 136

\bibitem[{{Beichman} {et~al.}(1988){Beichman}, {Neugebauer}, {Habing}, {Clegg},
  \& {Chester}}]{iras}
{Beichman}, C.~A., {Neugebauer}, G., {Habing}, H.~J., {Clegg}, P.~E., \&
  {Chester}, T.~J. 1988, in NASA RP-1190, Vol. 1 (1988)

\bibitem[{{Binney} {et~al.}(1991){Binney}, {Gerhard}, {Stark}, {Bally}, \&
  {Uchida}}]{binney91}
{Binney}, J., {Gerhard}, O.~E., {Stark}, A.~A., {Bally}, J., \& {Uchida}, K.~I.
  1991, \mnras, 252, 210

\bibitem[{{Blitz} \& {Spergel}(1991)}]{blitz91}
{Blitz}, L. \& {Spergel}, D.~N. 1991, \apj, 379, 631

\bibitem[{{Blommaert} {et~al.}(1994){Blommaert}, {van Langevelde}, \&
  {Michiels}}]{blommaert94}
{Blommaert}, J.~A.~D.~L., {van Langevelde}, H.~J., \& {Michiels}, W.~F.~P.
  1994, \aap, 287, 479

\bibitem[{{Burton}(1988)}]{burton88}
{Burton}, W.~B. 1988, in Galactic and Extragalactic Radio Astronomy, 295--358

\bibitem[{{Burton} \& {Liszt}(1978)}]{burton78}
{Burton}, W.~B. \& {Liszt}, H.~S. 1978, \apj, 225, 815

\bibitem[{{Carter}(1990)}]{carter90}
{Carter}, B.~S. 1990, \mnras, 242, 1

\bibitem[{{Cho} {et~al.}(1998){Cho}, {Chung}, {Kim}, {Oh}, {Lee}, \&
  {Han}}]{cho98}
{Cho}, S., {Chung}, H., {Kim}, H., {et~al.} 1998, \apjs, 115, 277

\bibitem[{{Cole} \& {Weinberg}(2002)}]{cole02}
{Cole}, A.~A. \& {Weinberg}, M.~D. 2002, \apjl, 574, L43

\bibitem[{{Cutri} {et~al.}(2003){Cutri}, {Skrutskie}, \& {Van Dyk}}]{2massES}
{Cutri}, C.~M., {Skrutskie}, M.~F., \& {Van Dyk}, S. 2003, avalaible on line at
  \\ {\tt http://www.ipac.caltech.edu/2mass/}

\bibitem[{{de Vaucouleurs}(1964)}]{devaucouleurs64}
{de Vaucouleurs}, G. 1964, in IAU Symp. 20: The Galaxy and the Magellanic
  Clouds, 195

\bibitem[{{Deguchi} {et~al.}(2000{\natexlab{a}}){Deguchi}, {Fujii}, {Izumiura},
  {Kameya}, {Nakada}, \& {Nakashima}}]{deguchi00b}
{Deguchi}, S., {Fujii}, T., {Izumiura}, H., {et~al.} 2000{\natexlab{a}}, \apjs,
  130, 351

\bibitem[{{Deguchi} {et~al.}(2000{\natexlab{b}}){Deguchi}, {Fujii}, {Izumiura},
  {Kameya}, {Nakada}, {Nakashima}, {Ootsubo}, \& {Ukita}}]{deguchi00a}
---. 2000{\natexlab{b}}, \apjs, 128, 571

\bibitem[{{Drimmel} {et~al.}(2003){Drimmel}, {Cabrera-Lavers}, \& {L{\'
  o}pez-Corredoira}}]{drimmel03}
{Drimmel}, R., {Cabrera-Lavers}, A., \& {L{\' o}pez-Corredoira}, M. 2003, \aap,
  409, 205

\bibitem[{{Egan} {et~al.}(1999){Egan}, {Price}, {Moshir}, {Cohen}, {Tedesco},
  {Murdock}, {Zweil}, {Burdick}, {Bonito}, {Gugliotti}, \& J.}]{egan99}
{Egan}, M.~P., {Price}, S.~D., {Moshir}, M.~M., {et~al.} 1999, AFRL-VS-TR-1999,
  1522

\bibitem[{{Englmaier} \& {Gerhard}(1999)}]{englmaier99}
{Englmaier}, P. \& {Gerhard}, O. 1999, \mnras, 304, 512

\bibitem[{{Epchtein} {et~al.}(1994){Epchtein}, {de Batz}, {Copet}, {Fouque},
  {Lacombe}, {Le Bertre}, {Mamon}, {Rouan}, {Tiphene}, {Burton}, {Deul},
  {Habing}, {Boersenberger}, {Dennefeld}, {Omont}, {Renault},
  {Rocca-Volmerange}, {Kimeswenger}, {Appenzeller}, {Bender}, {Forveille},
  {Garzon}, {Hron}, {Persi}, {Ferrari-Toniolo}, \& {Vauglin}}]{epchtein94}
{Epchtein}, N., {de Batz}, B., {Copet}, E., {et~al.} 1994, \apss, 217, 3

\bibitem[{{Feast} {et~al.}(1989){Feast}, {Glass}, {Whitelock}, \&
  {Catchpole}}]{feast89}
{Feast}, M.~W., {Glass}, I.~S., {Whitelock}, P.~A., \& {Catchpole}, R.~M. 1989,
  \mnras, 241, 375

\bibitem[{{Feltzing} \& {Gilmore}(2000)}]{feltzing00}
{Feltzing}, S. \& {Gilmore}, G. 2000, \aap, 355, 949

\bibitem[{{Ferraro} {et~al.}(2000){Ferraro}, {Montegriffo}, {Origlia}, \& {Fusi
  Pecci}}]{ferraro00}
{Ferraro}, F.~R., {Montegriffo}, P., {Origlia}, L., \& {Fusi Pecci}, F. 2000,
  \aj, 119, 1282

\bibitem[{{Frogel} \& {Whitford}(1987)}]{frogel87}
{Frogel}, J.~A. \& {Whitford}, A.~E. 1987, \apj, 320, 199

\bibitem[{{Girardi} {et~al.}(2000){Girardi}, {Bressan}, {Bertelli}, \&
  {Chiosi}}]{bertelli00}
{Girardi}, L., {Bressan}, A., {Bertelli}, G., \& {Chiosi}, C. 2000, \aaps, 141,
  371

\bibitem[{{Glass} {et~al.}(2001){Glass}, {Matsumoto}, {Carter}, \&
  {Sekiguchi}}]{glass01}
{Glass}, I.~S., {Matsumoto}, S., {Carter}, B.~S., \& {Sekiguchi}, K. 2001,
  \mnras, 321, 77

\bibitem[{{Glass} {et~al.}(1995){Glass}, {Whitelock}, {Catchpole}, \&
  {Feast}}]{glass95}
{Glass}, I.~S., {Whitelock}, P.~A., {Catchpole}, R.~M., \& {Feast}, M.~W. 1995,
  \mnras, 273, 383

\bibitem[{{Groenewegen} \& {de Jong}(1993)}]{groenewegen93}
{Groenewegen}, M.~A.~T. \& {de Jong}, T. 1993, \aap, 267, 410

\bibitem[{{Guarnieri} {et~al.}(1997){Guarnieri}, {Renzini}, \&
  {Ortolani}}]{guarnieri97}
{Guarnieri}, M.~D., {Renzini}, A., \& {Ortolani}, S. 1997, \apjl, 477, L21

\bibitem[{{Habing}(1996)}]{habing96}
{Habing}, H.~J. 1996, \aapr, 7, 97

\bibitem[{{Haikala}(1990)}]{haikala90}
{Haikala}, L.~K. 1990, \aaps, 85, 875

\bibitem[{{Hughes} \& {Wood}(1990)}]{hughes90}
{Hughes}, S.~M.~G. \& {Wood}, P.~R. 1990, \aj, 99, 784

\bibitem[{{Iben} \& {Renzini}(1983)}]{iben83}
{Iben}, I. \& {Renzini}, A. 1983, \araa, 21, 271

\bibitem[{{Imai} {et~al.}(2002){Imai}, {Deguchi}, {Fujii}, {Glass}, {Ita},
  {Izumiura}, {Kameya}, {Miyazaki}, {Nakada}, \& {Nakashima}}]{imai02}
{Imai}, H., {Deguchi}, S., {Fujii}, T., {et~al.} 2002, \pasj, 54, L19

\bibitem[{{Izumiura} {et~al.}(1999){Izumiura}, {Deguchi}, {Fujii}, {Kameya},
  {Matsumoto}, {Nakada}, {Ootsubo}, \& {Ukita}}]{izumiura99}
{Izumiura}, H., {Deguchi}, S., {Fujii}, T., {et~al.} 1999, \apjs, 125, 257

\bibitem[{{Josselin} {et~al.}(1996){Josselin}, {Loup}, {Omont}, {Barnbaum}, \&
  {Nyman}}]{josselin96}
{Josselin}, E., {Loup}, C., {Omont}, A., {Barnbaum}, C., \& {Nyman}, L.-A.
  1996, \aap, 315, L23

\bibitem[{{Le Bertre} \& {Nyman}(1990)}]{lebertre90}
{Le Bertre}, T. \& {Nyman}, L.-A. 1990, \aap, 233, 477

\bibitem[{{Lewis}(1989)}]{lewis89}
{Lewis}, B.~M. 1989, \apj, 338, 234

\bibitem[{{Lindqvist} {et~al.}(1992){Lindqvist}, {Winnberg}, {Habing}, \&
  {Matthews}}]{lindqvist92}
{Lindqvist}, M., {Winnberg}, A., {Habing}, H.~J., \& {Matthews}, H.~E. 1992,
  \aaps, 92, 43

\bibitem[{{Loup} {et~al.}(2004){Loup}, {Groenewegen}, {Cioni}, {Blommaert},
  {Cohen}, {Fouque}, \& {Wood}}]{loup03}
{Loup}, C., {Groenewegen}, M.~A.~T., {Cioni}, M.~R., {et~al.} 2004, in
  preparation

\bibitem[{{Marigo} {et~al.}(1996){Marigo}, {Bressan}, \& {Chiosi}}]{marigo96}
{Marigo}, P., {Bressan}, A., \& {Chiosi}, C. 1996, \aap, 313, 545

\bibitem[{{Matsuura} {et~al.}(2002){Matsuura}, {Yamamura}, {Cami}, {Onaka}, \&
  {Murakami}}]{matsuura02}
{Matsuura}, M., {Yamamura}, I., {Cami}, J., {Onaka}, T., \& {Murakami}, H.
  2002, \aap, 383, 972

\bibitem[{{Messineo} {et~al.}(2004{\natexlab{a}}){Messineo}, {Habing},
  {Menten}, {Omont}, \& {Sjouwerman}}]{messineo03_3}
{Messineo}, M., {Habing}, H.~J., {Menten}, K.~M., {Omont}, A., \& {Sjouwerman},
  L.~O. 2004{\natexlab{a}}, \aap\ (Chapter\,IV)

\bibitem[{{Messineo} {et~al.}(2004{\natexlab{b}}){Messineo}, {Habing},
  {Menten}, {Omont}, \& {Sjouwerman}}]{messineo03_2}
---. 2004{\natexlab{b}}, \aap, 418, 103 (Chapter\,III)

\bibitem[{{Messineo} {et~al.}(2002){Messineo}, {Habing}, {Sjouwerman}, {Omont},
  \& {Menten}}]{messineo02}
{Messineo}, M., {Habing}, H.~J., {Sjouwerman}, L.~O., {Omont}, A., \& {Menten},
  K.~M. 2002, \aap, 393, 115 (Chapter\,II)

\bibitem[{{Nagata} {et~al.}(1993){Nagata}, {Hyland}, {Straw}, {Sato}, \&
  {Kawara}}]{nagata93}
{Nagata}, T., {Hyland}, A.~R., {Straw}, S.~M., {Sato}, S., \& {Kawara}, K.
  1993, \apj, 406, 501

\bibitem[{{Nakada} {et~al.}(1991){Nakada}, {Onaka}, {Yamamura}, {Deguchi},
  {Hashimoto}, {Izumiura}, \& {Sekiguchi}}]{nakada91}
{Nakada}, Y., {Onaka}, T., {Yamamura}, I., {et~al.} 1991, \nat, 353, 140

\bibitem[{{Ng} \& {Bertelli}(1996)}]{ng96}
{Ng}, Y.~K. \& {Bertelli}, G. 1996, \aap, 315, 116

\bibitem[{{Nyman} {et~al.}(1993){Nyman}, {Hall}, \& {Le Bertre}}]{nyman93}
{Nyman}, L.-A., {Hall}, P.~J., \& {Le Bertre}, T. 1993, \aap, 280, 551

\bibitem[{{Olivier} {et~al.}(2001){Olivier}, {Whitelock}, \&
  {Marang}}]{olivier01}
{Olivier}, E.~A., {Whitelock}, P., \& {Marang}, F. 2001, \mnras, 326, 490

\bibitem[{{Omont} {et~al.}(2003){Omont}, {Gilmore}, {Alard}, {Aracil},
  {August}, {Baliyan}, {Beaulieu}, {B{\' e}gon}, {Bertou}, {Blommaert},
  {Borsenberger}, {Burgdorf}, {Caillaud}, {Cesarsky}, {Chitre}, {Copet}, {de
  Batz}, {Egan}, {Egret}, {Epchtein}, {Felli}, {Fouqu{\' e}}, {Ganesh},
  {Genzel}, {Glass}, {Gredel}, {Groenewegen}, {Guglielmo}, {Habing},
  {Hennebelle}, {Jiang}, {Joshi}, {Kimeswenger}, {Messineo}, {Miville-Desch{\^
  e}nes}, {Moneti}, {Morris}, {Ojha}, {Ortiz}, {Ott}, {Parthasarathy}, {P{\'
  e}rault}, {Price}, {Robin}, {Schultheis}, {Schuller}, {Simon}, {Soive},
  {Testi}, {Teyssier}, {Tiph{\` e}ne}, {Unavane}, {van Loon}, \&
  {Wyse}}]{omont03}
{Omont}, A., {Gilmore}, G.~F., {Alard}, C., {et~al.} 2003, \aap, 403, 975

\bibitem[{{Ortiz} {et~al.}(2002){Ortiz}, {Blommaert}, {Copet}, {Ganesh},
  {Habing}, {Messineo}, {Omont}, {Schultheis}, \& {Schuller}}]{ortiz02}
{Ortiz}, R., {Blommaert}, J.~A.~D.~L., {Copet}, E., {et~al.} 2002, \aap, 388,
  279

\bibitem[{{Price} {et~al.}(2001){Price}, {Egan}, {Carey}, {Mizuno}, \&
  {Kuchar}}]{price01}
{Price}, S.~D., {Egan}, M.~P., {Carey}, S.~J., {Mizuno}, D.~R., \& {Kuchar},
  T.~A. 2001, \aj, 121, 2819

\bibitem[{{Renzini} \& {Voli}(1981)}]{renzini81}
{Renzini}, A. \& {Voli}, M. 1981, \aap, 94, 175

\bibitem[{{Schuller}(2002)}]{schuller02}
{Schuller}, F. 2002, PhD Thesis: Universit\'e Pierre et Marie Curie, Paris 6.

\bibitem[{{Schuller} {et~al.}(2003){Schuller}, {Ganesh}, {Messineo}, {Moneti},
  {Blommaert}, {Alard}, {Aracil}, {Miville-Desch{\^ e}nes}, {Omont},
  {Schultheis}, {Simon}, {Soive}, \& {Testi}}]{schuller03}
{Schuller}, F., {Ganesh}, S., {Messineo}, M., {et~al.} 2003, \aap, 403, 955

\bibitem[{{Schultheis} {et~al.}(2003){Schultheis}, {Lan{\c c}on}, {Omont},
  {Schuller}, \& {Ojha}}]{schultheis03}
{Schultheis}, M., {Lan{\c c}on}, A., {Omont}, A., {Schuller}, F., \& {Ojha},
  D.~K. 2003, \aap, 405, 531

\bibitem[{{Sevenster}(1999)}]{sevenster99}
{Sevenster}, M.~N. 1999, \mnras, 310, 629

\bibitem[{{Sevenster} {et~al.}(1997{\natexlab{a}}){Sevenster}, {Chapman},
  {Habing}, {Killeen}, \& {Lindqvist}}]{sevenster97a}
{Sevenster}, M.~N., {Chapman}, J.~M., {Habing}, H.~J., {Killeen}, N.~E.~B., \&
  {Lindqvist}, M. 1997{\natexlab{a}}, \aaps, 122, 79

\bibitem[{{Sevenster} {et~al.}(1997{\natexlab{b}}){Sevenster}, {Chapman},
  {Habing}, {Killeen}, \& {Lindqvist}}]{sevenster97b}
---. 1997{\natexlab{b}}, \aaps, 124, 509

\bibitem[{{Sevenster} {et~al.}(2001){Sevenster}, {van Langevelde}, {Moody},
  {Chapman}, {Habing}, \& {Killeen}}]{sevenster01}
{Sevenster}, M.~N., {van Langevelde}, H.~J., {Moody}, R.~A., {et~al.} 2001,
  \aap, 366, 481

\bibitem[{{Sjouwerman} {et~al.}(1999){Sjouwerman}, {Habing}, {Lindqvist}, {van
  Langevelde}, \& A.}]{sjouwerman99}
{Sjouwerman}, L.~O., {Habing}, H.~J., {Lindqvist}, M., {van Langevelde}, H.~J.,
  \& A., W. 1999, "The Central Parsecs of the Galaxy" ASP Conf. Series 186 (p.
  379)

\bibitem[{{Sjouwerman} {et~al.}(1998){Sjouwerman}, {van Langevelde},
  {Winnberg}, \& {Habing}}]{sjouwerman98}
{Sjouwerman}, L.~O., {van Langevelde}, H.~J., {Winnberg}, A., \& {Habing},
  H.~J. 1998, \aaps, 128, 35

\bibitem[{{van Loon} {et~al.}(2003){van Loon}, {Gilmore}, {Omont}, {Blommaert},
  {Glass}, {Messineo}, {Schuller}, {Schultheis}, {Yamamura}, \&
  {Zhao}}]{vanloon03}
{van Loon}, J.~T., {Gilmore}, G.~F., {Omont}, A., {et~al.} 2003, \mnras, 338,
  857

\bibitem[{{Vassiliadis} \& {Wood}(1993)}]{vassiliadis93}
{Vassiliadis}, E. \& {Wood}, P.~R. 1993, \apj, 413, 641

\bibitem[{{Whitelock} {et~al.}(1991){Whitelock}, {Feast}, \&
  {Catchpole}}]{whitelock91}
{Whitelock}, P., {Feast}, M., \& {Catchpole}, R. 1991, \mnras, 248, 276

\bibitem[{{Whitelock} {et~al.}(2000){Whitelock}, {Marang}, \&
  {Feast}}]{whitelock-hyp}
{Whitelock}, P., {Marang}, F., \& {Feast}, M. 2000, \mnras, 319, 728

\bibitem[{{Whitelock} {et~al.}(1994){Whitelock}, {Menzies}, {Feast}, {Marang},
  {Carter}, {Roberts}, {Catchpole}, \& {Chapman}}]{whitelock94}
{Whitelock}, P., {Menzies}, J., {Feast}, M., {et~al.} 1994, \mnras, 267, 711

\bibitem[{{Whitelock} {et~al.}(2003){Whitelock}, {Feast}, {van Loon}, \&
  {Zijlstra}}]{whitelock03}
{Whitelock}, P.~A., {Feast}, M.~W., {van Loon}, J.~T., \& {Zijlstra}, A.~A.
  2003, \mnras, 342, 86

\bibitem[{{Winfrey} {et~al.}(1994){Winfrey}, {Barnbaum}, {Morris}, \&
  {Omont}}]{winfrey94}
{Winfrey}, S., {Barnbaum}, C., {Morris}, M., \& {Omont}, A. 1994, Bulletin of
  the American Astronomical Society, 26, 1382

\bibitem[{{Wood} {et~al.}(1998){Wood}, {Habing}, \& {McGregor}}]{wood98}
{Wood}, P.~R., {Habing}, H.~J., \& {McGregor}, P.~J. 1998, \aap, 336, 925

\bibitem[{{Zoccali} {et~al.}(2003){Zoccali}, {Renzini}, {Ortolani}, {Greggio},
  {Saviane}, {Cassisi}, {Rejkuba}, {Barbuy}, {Rich}, \& {Bica}}]{zoccali03}
{Zoccali}, M., {Renzini}, A., {Ortolani}, S., {et~al.} 2003, \aap, 399, 931

\end{thebibliography}

\addcontentsline{toc}{section}{Bibliography}

\vspace{-0.5cm}
\begin{thebibliography}{33}
\expandafter\ifx\csname natexlab\endcsname\relax\def\natexlab#1{#1}\fi

\bibitem[{{Athanassoula} \& {Bureau}(1999)}]{athanassoula99}
{Athanassoula}, E. \& {Bureau}, M. 1999, \apj, 522, 699

\bibitem[{{Binney} {et~al.}(1997){Binney}, {Gerhard}, \& {Spergel}}]{binney97}
{Binney}, J., {Gerhard}, O., \& {Spergel}, D. 1997, \mnras, 288, 365

\bibitem[{{Binney} {et~al.}(1991){Binney}, {Gerhard}, {Stark}, {Bally}, \&
  {Uchida}}]{binney91}
{Binney}, J., {Gerhard}, O.~E., {Stark}, A.~A., {Bally}, J., \& {Uchida}, K.~I.
  1991, \mnras, 252, 210

\bibitem[{{Binney} \& {Tremaine}(1987)}]{binney87}
{Binney}, J. \& {Tremaine}, S. 1987, {Galactic dynamics} (Princeton, NJ,
  Princeton University Press, 1987, 747 p.)

\bibitem[{{Bissantz} {et~al.}(2003){Bissantz}, {Englmaier}, \&
  {Gerhard}}]{bissantz03}
{Bissantz}, N., {Englmaier}, P., \& {Gerhard}, O. 2003, \mnras, 340, 949

\bibitem[{{Blitz} \& {Spergel}(1991)}]{blitz91}
{Blitz}, L. \& {Spergel}, D.~N. 1991, \apj, 379, 631

\bibitem[{{Combes}(1991)}]{combes91}
{Combes}, F. 1991, \araa, 29, 195

\bibitem[{{Dame} {et~al.}(2001){Dame}, {Hartmann}, \& {Thaddeus}}]{dame01}
{Dame}, T.~M., {Hartmann}, D., \& {Thaddeus}, P. 2001, \apj, 547, 792

\bibitem[{{de Vaucouleurs}(1964)}]{devaucouleurs64}
{de Vaucouleurs}, G. 1964, in IAU Symp. 20: The Galaxy and the Magellanic
  Clouds, 195--+

\bibitem[{{Debattista} {et~al.}(2002){Debattista}, {Gerhard}, \&
  {Sevenster}}]{debattista02}
{Debattista}, V.~P., {Gerhard}, O., \& {Sevenster}, M.~N. 2002, \mnras, 334,
  355

\bibitem[{{Englmaier} \& {Gerhard}(1999)}]{englmaier99}
{Englmaier}, P. \& {Gerhard}, O. 1999, \mnras, 304, 512

\bibitem[{{Fux}(1997)}]{fux97}
{Fux}, R. 1997, \aap, 327, 983

\bibitem[{{H{\" a}fner} {et~al.}(2000){H{\" a}fner}, {Evans}, {Dehnen}, \&
  {Binney}}]{hafner00}
{H{\" a}fner}, R., {Evans}, N.~W., {Dehnen}, W., \& {Binney}, J. 2000, \mnras,
  314, 433

\bibitem[{{Lindqvist} {et~al.}(1992){Lindqvist}, {Habing}, \&
  {Winnberg}}]{lindqvist92}
{Lindqvist}, M., {Habing}, H.~J., \& {Winnberg}, A. 1992, \aap, 259, 118

\bibitem[{{Messineo} {et~al.}(2004{\natexlab{a}}){Messineo}, {Habing},
  {Menten}, {Omont}, \& {Sjouwerman}}]{messineo03_4}
{Messineo}, M., {Habing}, H.~J., {Menten}, K.~M., {Omont}, A., \& {Sjouwerman},
  L.~O. 2004{\natexlab{a}}, \aap\ in preparation (Chapter\,V)

\bibitem[{{Messineo} {et~al.}(2004{\natexlab{b}}){Messineo}, {Habing},
  {Menten}, {Omont}, \& {Sjouwerman}}]{messineo04}
{Messineo}, M., {Habing}, H.~J., {Menten}, K.~M., {Omont}, A., \& {Sjouwerman},
  L.~O. 2004{\natexlab{b}}, \aap, 418, 103, Chapter\,III

\bibitem[{{Messineo} {et~al.}(2002){Messineo}, {Habing}, {Sjouwerman}, {Omont},
  \& {Menten}}]{messineo02}
{Messineo}, M., {Habing}, H.~J., {Sjouwerman}, L.~O., {Omont}, A., \& {Menten},
  K.~M. 2002, \aap, 393, 115, Chapter\,II

\bibitem[{{Morris} \& {Serabyn}(1996)}]{morris96}
{Morris}, M. \& {Serabyn}, E. 1996, \araa, 34, 645

\bibitem[{{Nakada} {et~al.}(1991){Nakada}, {Onaka}, {Yamamura}, {Deguchi},
  {Hashimoto}, {Izumiura}, \& {Sekiguchi}}]{nakada91}
{Nakada}, Y., {Onaka}, T., {Yamamura}, I., {et~al.} 1991, \nat, 353, 140

\bibitem[{{Paczynski} {et~al.}(1994){Paczynski}, {Stanek}, {Udalski},
  {Szymanski}, {Kaluzny}, {Kubiak}, {Mateo}, \& {Krzeminski}}]{paczynski94}
{Paczynski}, B., {Stanek}, K.~Z., {Udalski}, A., {et~al.} 1994, \apjl, 435,
  L113

\bibitem[{{Press} {et~al.}(1992){Press}, {Teukolsky}, {Vetterling}, \&
  {Flannery}}]{press92}
{Press}, W.~H., {Teukolsky}, S.~A., {Vetterling}, W.~T., \& {Flannery}, B.~P.
  1992, {Numerical recipes in FORTRAN. The art of scientific computing}
  (Cambridge: University Press, |c1992, 2nd ed.)

\bibitem[{{Sevenster} {et~al.}(1999){Sevenster}, {Saha}, {Valls-Gabaud}, \&
  {Fux}}]{sevenster99a}
{Sevenster}, M., {Saha}, P., {Valls-Gabaud}, D., \& {Fux}, R. 1999, \mnras,
  307, 584

\bibitem[{{Sevenster}(1999{\natexlab{a}})}]{sevenster99b}
{Sevenster}, M.~N. 1999{\natexlab{a}}, \mnras, 310, 629

\bibitem[{{Sevenster}(1999{\natexlab{b}})}]{sevenster99}
{Sevenster}, M.~N. 1999{\natexlab{b}}, \mnras, 310, 629

\bibitem[{{Sevenster}(2002)}]{sevenster02}
{Sevenster}, M.~N. 2002, \aj, 123, 2772

\bibitem[{{Sevenster} {et~al.}(1997{\natexlab{a}}){Sevenster}, {Chapman},
  {Habing}, {Killeen}, \& {Lindqvist}}]{sevenster97a}
{Sevenster}, M.~N., {Chapman}, J.~M., {Habing}, H.~J., {Killeen}, N.~E.~B., \&
  {Lindqvist}, M. 1997{\natexlab{a}}, \aaps, 122, 79

\bibitem[{{Sevenster} {et~al.}(1997{\natexlab{b}}){Sevenster}, {Chapman},
  {Habing}, {Killeen}, \& {Lindqvist}}]{sevenster97b}
{Sevenster}, M.~N., {Chapman}, J.~M., {Habing}, H.~J., {Killeen}, N.~E.~B., \&
  {Lindqvist}, M. 1997{\natexlab{b}}, \aaps, 124, 509

\bibitem[{{Sevenster} {et~al.}(2000){Sevenster}, {Dejonghe}, {Van Caelenberg},
  \& {Habing}}]{sevenster00}
{Sevenster}, M.~N., {Dejonghe}, H., {Van Caelenberg}, K., \& {Habing}, H.~J.
  2000, \aap, 355, 537

\bibitem[{{Sevenster} {et~al.}(2001){Sevenster}, {van Langevelde}, {Moody},
  {Chapman}, {Habing}, \& {Killeen}}]{sevenster01}
{Sevenster}, M.~N., {van Langevelde}, H.~J., {Moody}, R.~A., {et~al.} 2001,
  \aap, 366, 481

\bibitem[{{Sjouwerman} {et~al.}(1998){Sjouwerman}, {van Langevelde},
  {Winnberg}, \& {Habing}}]{sjouwerman98}
{Sjouwerman}, L.~O., {van Langevelde}, H.~J., {Winnberg}, A., \& {Habing},
  H.~J. 1998, \aaps, 128, 35

\bibitem[{{Weiland} {et~al.}(1994){Weiland}, {Arendt}, {Berriman}, {Dwek},
  {Freudenreich}, {Hauser}, {Kelsall}, {Lisse}, {Mitra}, {Moseley}, {Odegard},
  {Silverberg}, {Sodroski}, {Spiesman}, \& {Stemwedel}}]{weiland94}
{Weiland}, J.~L., {Arendt}, R.~G., {Berriman}, G.~B., {et~al.} 1994, \apjl,
  425, L81

\bibitem[{{Whitelock} \& {Catchpole}(1992)}]{whitelock92}
{Whitelock}, P. \& {Catchpole}, R. 1992, in ASSL Vol. 180: The center, bulge,
  and disk of the Milky Way, 103

\bibitem[{{Zhao}(1996)}]{zhao96}
{Zhao}, H.~S. 1996, \mnras, 283, 149

\end{thebibliography}

\begin{thebibliography}{5}
\expandafter\ifx\csname natexlab\endcsname\relax\def\natexlab#1{#1}\fi

\bibitem[{{Bellazzini} {et~al.}(2002){Bellazzini}, {Fusi Pecci}, {Montegriffo},
  {Messineo}, {Monaco}, \& {Rood}}]{bellazzini02}
{Bellazzini}, M., {Fusi Pecci}, F., {Montegriffo}, P., {et~al.} 2002, \aj, 123,
  2541

\bibitem[{{Blommaert} {et~al.}(2003){Blommaert}, {Siebenmorgen}, {Coulais},
  {Metcalfe}, {Miville-Deschenes}, {Okumura}, {Ott}, {Pollock}, {Sauvage}, \&
  {Starck}}]{blommaert03}
{Blommaert}, J.~A.~D.~L., {Siebenmorgen}, R., {Coulais}, A., {et~al.}, eds.
  2003, {The ISO Handbook, Volume II - CAM - The ISO Camera}

\bibitem[{{Omont} {et~al.}(2003){Omont}, {Gilmore}, {Alard}, {Aracil},
  {August}, {Baliyan}, {Beaulieu}, {B{\' e}gon}, {Bertou}, {Blommaert},
  {Borsenberger}, {Burgdorf}, {Caillaud}, {Cesarsky}, {Chitre}, {Copet}, {de
  Batz}, {Egan}, {Egret}, {Epchtein}, {Felli}, {Fouqu{\' e}}, {Ganesh},
  {Genzel}, {Glass}, {Gredel}, {Groenewegen}, {Guglielmo}, {Habing},
  {Hennebelle}, {Jiang}, {Joshi}, {Kimeswenger}, {Messineo}, {Miville-Desch{\^
  e}nes}, {Moneti}, {Morris}, {Ojha}, {Ortiz}, {Ott}, {Parthasarathy}, {P{\'
  e}rault}, {Price}, {Robin}, {Schultheis}, {Schuller}, {Simon}, {Soive},
  {Testi}, {Teyssier}, {Tiph{\` e}ne}, {Unavane}, {van Loon}, \&
  {Wyse}}]{omont03}
{Omont}, A., {Gilmore}, G.~F., {Alard}, C., {et~al.} 2003, \aap, 403, 975

\bibitem[{{Schuller} {et~al.}(2003){Schuller}, {Ganesh}, {Messineo}, {Moneti},
  {Blommaert}, {Alard}, {Aracil}, {Miville-Desch{\^ e}nes}, {Omont},
  {Schultheis}, {Simon}, {Soive}, \& {Testi}}]{schuller03}
{Schuller}, F., {Ganesh}, S., {Messineo}, M., {et~al.} 2003, \aap, 403, 955

\bibitem[{{Simon}(2004)}]{simon04}
{Simon}, G. 2004, in preparation

\end{thebibliography}

\appendix
\section{SAAO and 2MASS colours and magnitudes}
Transformation equations between the colours and magnitudes measured
in the SAAO \citet{carter90} and 2MASS photometric systems have been
derived by \citet{carpenter01} using a list of mostly blue 94
photometric standards. Figure 12 in \citet{carpenter01} shows that the
differences between  magnitudes and colours obtained with the two
systems are smaller than 0.15 mag.

Considering  that Mira stars have typically a pulsation amplitude
in the near-infrared of 1-2 mag and that  2MASS data are from a
single-epoch observation randomly taken with respect to the stellar
phase, the system transformations have only a secondary effect in the
total colour and magnitude uncertainty, when comparing  data from
2MASS with data taken with the SAAO telescope.

Since Mira stars are cold objects, molecular absorption bands
characterise their infrared spectra and we must exclude that a
combination of molecular bands and filter transmissions could generate
a different colour transformation for these special class of objects.
To address that we looked for 2MASS counterparts of the 104 outer
Bulge Mira stars monitored by \citet{whitelock94}. As demonstrated in
\citetalias{messineo03_2}, Mira stars are among the brightest objects
detected in the $K_{\rm s}$ band and therefore the identification of
their 2MASS counterparts is straightforward.  A number of 101 2MASS
counterparts were found within 60\arcsec\ (mostly within 10\arcsec)
from the IRAS position.  We excluded three sources because they had
not unique counterparts (IRAS 17287$-$1955, IRAS 17030$-$2801, IRAS
18264$-$2720).

The differences between the mean magnitudes obtained with SAAO
observations \citep{whitelock91} and the single-epoch 2MASS data have
a dispersion of up to 0.8 mag.

\begin{figure}[t!]
\begin{centering}
\resizebox{0.8\hsize}{!}{\includegraphics{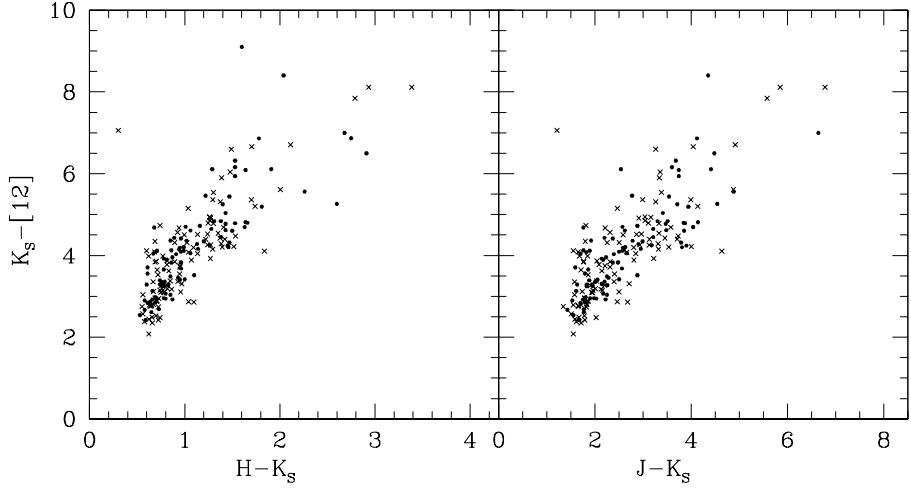}}
\caption{\label{fig:saao2mass.ps} Colour-colour plots of Mira stars.
The dots indicate average colours of Mira stars obtained from SAAO
observations \citep{whitelock91}. Crosses indicate colours from
single-epoch 2MASS data for the same sample of Mira stars
\citep{whitelock91}. }
\end{centering}
\end{figure}

We obtain the following mean differences:

$$ {K_{\rm s}}_{(\rm 2MASS)} - K_{(\rm SAAO)} =  -0.15 \pm 0.06 {\rm ~mag};$$
$$(J-K_{\rm s})_{(\rm 2MASS)} - (J-K)_{(\rm SAAO)} = -0.14 \pm 0.05 {\rm ~mag};$$
$$ (H-K_{\rm s})_{(\rm 2MASS)} - (H-K)_{(\rm SAAO)} = -0.06 \pm 0.03 {\rm ~mag}.$$

To verify whether the colour-colour relations found in Sect.\
\ref{miras}, using data in the SAAO photometric system, hold also when
using  2MASS photometry, in Fig.\ \ref{fig:saao2mass.ps} we plot
both  SAAO data and  2MASS data for the same sample of Mira
stars \citep{whitelock91}. No systematic trend is present.

\section{IRAS and MSX filters}
Most of the past work has been carried out using the IRAS photometry,
and therefore the currently available colour-colour relations of Mira
stars use mid-infrared data from the IRAS catalogue.  A comparison of
mid-infrared filters is therefore mandatory to translate  old
findings into  new MSX and ISOGAL colours.

\begin{figure}[t!]
\begin{centering}
\resizebox{0.5\hsize}{!}{\includegraphics{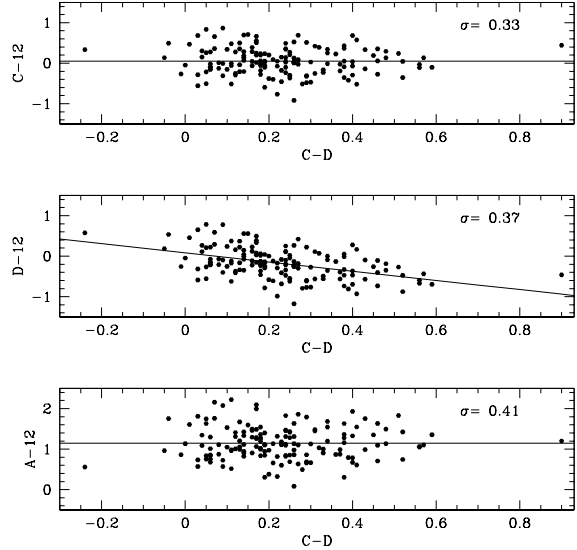}}
\caption{\label{fig:cc2.ps} Colour-Colour diagrams of our SiO targets.
The IRAS 12$\mu$m magnitude is defined as $[12]=-2.5 \log F_{12}[{\rm
Jy}]/28.3$. The continuous lines are our best fits.}
\end{centering}
\end{figure}

Figure \ref{fig:cc2.ps} shows the difference between MSX magnitudes
and IRAS 12 $\mu$m magnitude for the SiO targets.  Note that the $D$
filter excludes the silicate feature around 9.7 $\mu$m, while $A$ and
$C$ filters include part of it (see Fig. \ref{fig:filter.ps}).
Therefore the ($D-[12])$ colour shows a dependence on the ($C-D$)
colour, which increases when the silicate feature at 9.7 $\mu$m starts
to be self-absorbed.  The ($A-[12]$) and ($C-[12]$) colours do not
show any trend with the ($C-D$) colour.  Due to both uncertainties of
photometric measurements and source variability the scatter is large,
however we derived relations between the $A$, $C$, and $D$ and the
[12] magnitudes, as follow:\\ $A-[12]=1.15\pm0.03$ mag\\
$C-[12]=0.05\pm0.03$ mag\\$D-[12]=0.08(\pm0.18) -1.13(\pm0.05)(C-D)$ mag. \tappendix

\addcontentsline{toc}{section}{Bibliography}

\setlength{\bibsep}{0.5mm}

   \chapter{
   86 GHz SiO maser survey of late-type stars in the Inner Galaxy
   IV. Bolometric magnitudes }
\chaptermark{Bolometric magnitudes }
 
\begin{authorline}
        M.\ Messineo, H.\ J.\ Habing, K.\ M.\ Menten,
        A.\ Omont, L.\ O.\ Sjouwerman 
        \journal{Astronomy and Astrophysics (2004), in preparation}
\end{authorline}
 
\begin{abstract} 
We present a study of DENIS, 2MASS, ISOGAL and MSX
photometry for a sample of evolved late--type stars in the inner
Galaxy, which we previously searched for 86 GHz SiO maser emission
\citep{messineo02}.  Bolometric magnitudes are computed for each SiO
star by direct integration of the observed energy distribution, and
bolometric corrections as a function of colours are derived.  Adopting
a distance of 8 kpc the SiO stars within 5\degr\ from the Galactic
Centre show a distribution of bolometric magnitudes that peaks at
$M_{\mathrm{bol}} = -5.1$ mag, i.e., very similar to the OH/IR stars
close to the Galactic centre.  From their bolometric luminosities and
interstellar extinction we find that 11\% of the SiO stars are likely
to be foreground to the bulge. Furthermore the small velocity
dispersions of those foreground stars suggest a disk component.  The
15 known large amplitude variables included in our sample fall above
the Mira period--luminosity relation of \citet{glass95}, which
suggests a steepening of the period--luminosity relation for periods
larger than 450 days, as also seen in the Magellanic Clouds.  From
this period--luminosity relation and from their colours, the envelopes
of SiO stars appear less evolved than those of OH/IR stars, which have
thicker shells due to higher mass--loss rates.
\end{abstract}

%
\section{Introduction}

The early 1960's observations of non-circular gas motions in the
inner Galaxy produced evidence for the existence of a Galactic
bar \citep{devaucouleurs64}.  Its existence is also supported by an
apparent asymmetry of the integrated light as seen in COBE maps
\citep{blitz91} and of the stellar counts \citep{nakada91}. Thus,
stellar kinematic studies provided less stringent constraints, due to
the small number of measured stellar radial velocities. OH and SiO
maser emission lines from the envelopes of evolved late-type stars can
measure stellar line-of-sight velocities with an accuracy of a few
\kms\ throughout the Galaxy. Maser emission thereby provides a ready
means to measure line-of-sight velocities in the Galactic plane even
where the optical interstellar extinction is high.

\citet{lewis89} analysed the colours and maser emission of IRAS
sources,  suggesting a chronological sequence of increasing
mass-loss from SiO, to H$_2$O and OH maser emission. This sequence
links AGB stars through the Mira and OH/IR stages with Planetary
Nebulae.  The presence of particular maser lines apparently depends on
the envelope's mid-infrared opacity: a higher mass-loss rate makes a
more opaque dust shell, which better shields molecules against
photodissociation.  However, parameters other than mass-loss, such as
initial stellar masses and chemical abundances, probably also play
an important role \citep{habing96}.

A number of maser surveys have been carried out to measure stellar
line-of-sight velocities toward the inner Galaxy, between 30\degr\ and
$-30$\degr\ in longitude,
\citep[e.g.][]{baud79,lindqvist92,blommaert94,sevenster01,
sevenster97a,sevenster97b,sjouwerman98,izumiura99,deguchi00a,
deguchi00b}.  These surveys mostly detected visually obscured OH/IR
stars, i.e., AGB stars with 1612 MHz OH maser emission and high
mass-loss rates.

Our survey of 86 GHz SiO maser emission towards infrared-selected
Mira-like stars in the inner Galaxy, mostly at $b<0.5$\degr,
\citep[][ hereafter Chapter\,II]{messineo02}
\defcitealias{messineo03_2}{Chapter\,III} led to the determination of
255 new line-of-sight velocities.  The sample of targets (hereafter
``SiO targets'') was selected to be complementary to previous OH/IR
surveys, so the sources with the reddest mid- and near-infrared
colours were excluded.

In studying Galactic structure and kinematics it is important to
combine the kinematic information and the stellar properties, such as
luminosities, to explore any differences in the velocity fields of
different tracers. In particular, it must be clarified whether OH/IR
and SiO masing stars trace the same dynamic population, so that they
can simply be combined to study the kinematics of the inner Galaxy.

The infrared photometry of our 441 SiO targets, derived from the large
surveys  DENIS \citep{epchtein94}, 2MASS \citep{2massES},
ISOGAL \citep{omont03,schuller03} and MSX \citep{egan99,price01}, was
given by \citet[][ hereafter Chapter\,III]
{messineo03_2}. \defcitealias{messineo03_2}{Chapter\,III} Corrections for
interstellar extinction were discussed by \citet[][ hereafter
Chapter\,IV] {messineo03_3}. \defcitealias{messineo03_3}{Chapter\,IV}
\defcitealias{messineo02}{Chapter\,II} The present paper derives the
bolometric magnitude of each target star, which we then compare with
those of a sample of OH/IR stars.

Since most of the SiO targets are variable stars
\citepalias{messineo03_2}, their characterisation would require
long-term, multi-frequency flux monitoring programs, which are not
available yet.  A first, although necessarily approximate discussion
of their luminosities is possible even with single epoch observations,
making use of the most recent infrared surveys.

The individual source numbers (e.g \#99) are taken from Table 2 (86
GHz SiO maser detections) and Table 3 (non-detections) of
\citetalias{messineo02}.

\section{Apparent bolometric magnitudes}\label{ambol}
\label{mbol}

The photometric measurements of each SiO target
\citepalias{messineo03_3} were corrected for interstellar extinction
using the mean \ks--band extinction found from the surrounding field
stars. For the wavelength-dependence of the extinction, we assume a
power law $A_\lambda/A_K = (\lambda/2.12\mu m)^{-1.9}$, so that for
the individual bands we find effective band extinctions {\it (A$_I$,
A$_J$, A$_H$) } = (6.78, 2.86, 1.66) \Aks.  For the mid-infrared range
we use the extinction law of Lutz \citepalias[the extinction ratios
are given in Table 2 of ][]{messineo03_3}.

The bolometric stellar magnitudes, $m_{\mathrm{bol}}$, were computed
by integrating over frequency $\nu$ using linear interpolations between
the dereddened flux densities, $F_\nu(\nu)$.  At the low frequency end
we extrapolated to $F_{\nu=0}=0$, and at the upper end we extrapolated
the two highest frequency data--points, provided the flux decreases, to
zero intensity.  \citet{loup03} presented a comparison of different
methods commonly used to compute bolometric magnitudes.  Using model
spectra of O-rich AGB stars \citep[from][]{groenewegen93}, they
showed that the integration method we adopt yields bolometric
magnitudes which should on average be accurate within 0.3 mag.
Since the DENIS $I,J,$\ks\ as well as the 2MASS $J,H,$\ks\ observations
were each taken simultaneously, but both sets at different times, we
computed the bolometric magnitudes separately using either the DENIS
or 2MASS data.  The SiO targets are mostly variable stars
\citepalias{messineo03_2}, so a comparison of both datasets yields
some information on their average variability.



For all but two of the SiO target stars the low-frequency
extrapolation is insignificant since it contains a negligible fraction
of the total flux. Only for sources \#76 and \#347 the low-frequency
extrapolation contributes more than 20\% to the total integrated flux.

Given that usually (\ks$-[15])_0 < 5$ mag, the main uncertainty in
computing the bolometric magnitudes of the SiO target stars arises
from the extrapolation at high frequencies. For  20  SiO targets a
linear extrapolation was not possible because the flux density rises
at high frequencies. For another 100 targets the blue extrapolation
using 2MASS data or  DENIS data contributes more than 20\%
to the bolometric flux. 

For about half of our SiO stars (233 stars) we could compute
bolometric magnitudes through a direct integration using either the
DENIS or 2MASS data, with the flux in the extrapolated regions
contributing less than 20\% to the total. Unless otherwise stated, we
adopt the average of these two integrations to attenuate the effect of
variability.

For the other half of our sample, a bolometric magnitude could be
integrated from only one dataset, since either the other data set is
incomplete or the blue extrapolation is too uncertain in that the
extrapolated spectral region contributes more than 20\% to the total
flux.  For this half of the sample, when both the DENIS and 2MASS \ks\
measurements were available, we estimate bolometric magnitudes using
the bolometric corrections described in  Appendix A and the average
of the two \ks--band flux densities. The average flux densities at 15
\um, when both ISOGAL and MSX 15$\mu$m measurement were available, was
used.

\begin{figure*}[t!]
\begin{center}
\resizebox{0.7\hsize}{!}{\includegraphics{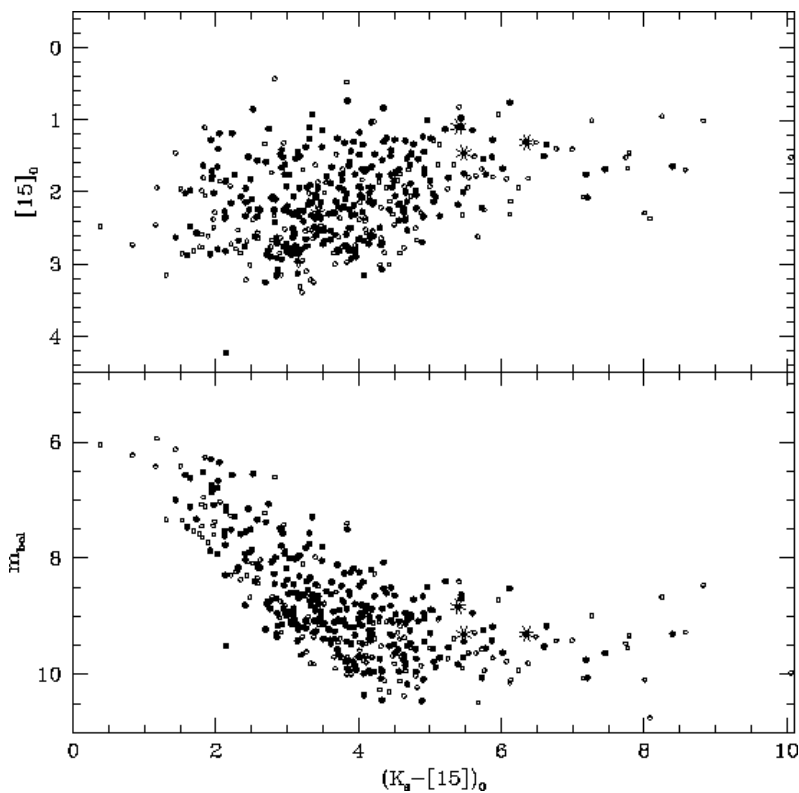}}
\caption{\label{fig:mb.ps} {\bf Lower panel:} Apparent bolometric
magnitudes, $m_{\mathrm{bol}}$, versus (\ks$-[15])_0$. Filled circles
indicate objects with detected SiO maser emission, and open circles
objects with no SiO detection.  The four starred symbols show stars
with observed OH emission \citepalias{messineo03_2}.  When both DENIS
and 2MASS \ks\ measurements are available, or when both ISOGAL and MSX
15$\mu$m measurement are available, the average flux density is
adopted.  {\bf Upper panel:} Apparent $[15]_0$ magnitudes versus the
(\ks$-[15])_0$ colour.  Symbols are as in the lower panel.}
\end{center}
\end{figure*}

The apparent bolometric magnitudes are plotted against the
(\ks$-[15])_0$ colour in Fig. \ref{fig:mb.ps}. There is an misleading
correlation because the range in $[15]_0$ is small compared with the
range in $K_{\rm s0}$, and $m_{\mathrm{bol}}$ is dominated by the
near-infrared flux.

There is a tail of stars at faint apparent bolometric magnitudes with
a wider colour spread and redder (\ks$-[15])_0$ colours (Fig.\
\ref{fig:mb.ps}) \citepalias[which is the consequence of selection of MSX
sources,][]{messineo03_2}.  These redder colours are indicative of
higher mass-loss rates.

\subsection{Variability}

Bolometric magnitudes of known Mira variables are found to vary up
to 2 mag from minimum to maximum  light
\citep[e.g.][]{whitelock-hyp,whitelock91}.

Since we are using single epoch observations taken at random phase,
the obtained bolometric magnitudes are also at a random phase.

\begin{figure}[ht!]
\resizebox{\hsize}{!}{\includegraphics{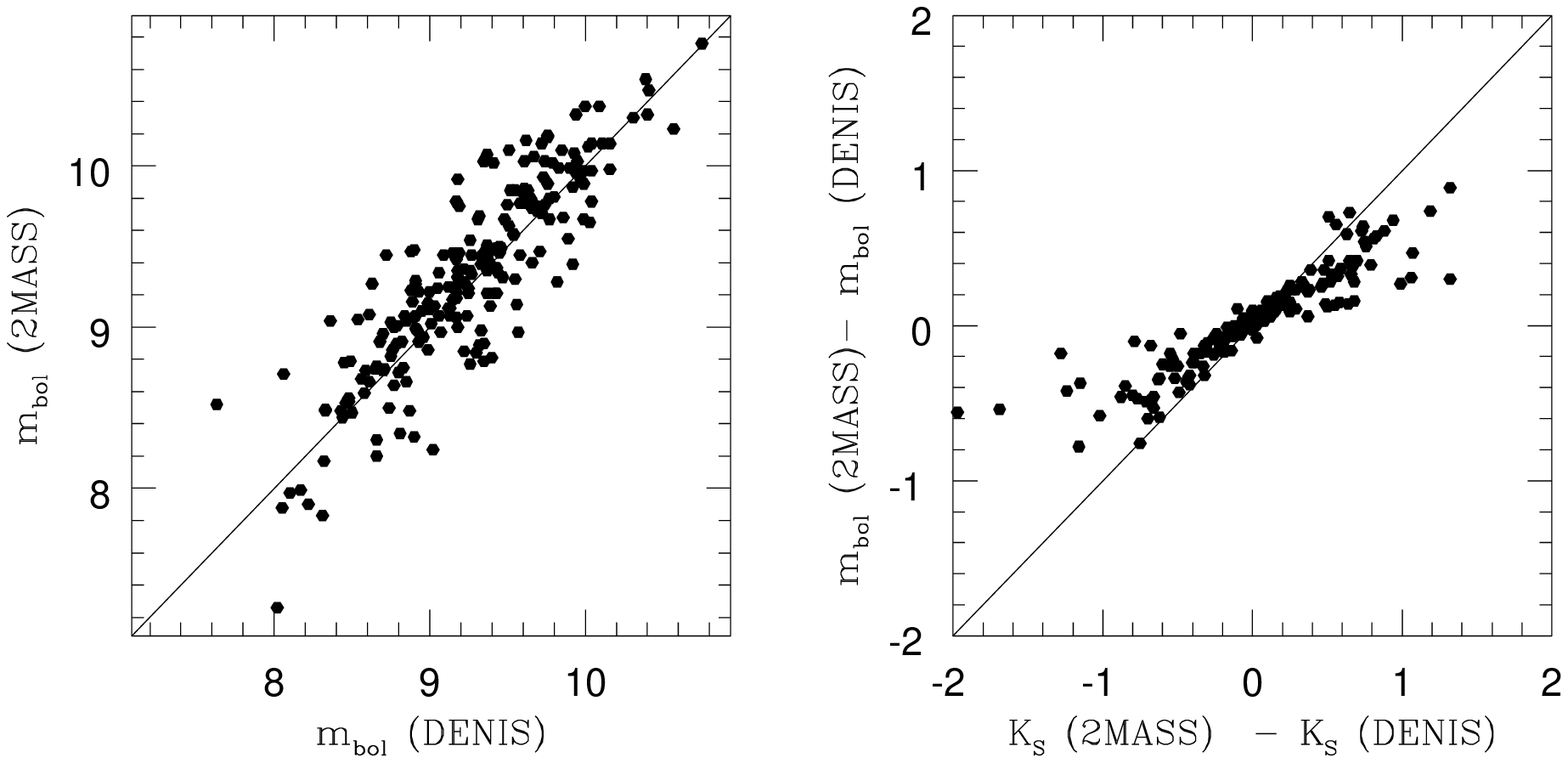}}
\caption{\label{fig:diffe.ps} {\bf Left panel:} Apparent bolometric
magnitudes derived using 2MASS data versus apparent bolometric
magnitudes using DENIS data.  {\bf Right panel:} Differences in
$m_{\mathrm{bol}}$ versus the corresponding differences between 
DENIS-\ks\ and  2MASS-\ks.}
\end{figure}

Figure \ref{fig:diffe.ps} shows that there is an rms scatter of
$\sim0.35$ mag between $m_{\mathrm{bol}}^{2MASS}$ and
$m_{\mathrm{bol}}^{DENIS}$, which is mainly due to variability
\citepalias{messineo03_2}.  There is a tight correlation between
$m_{\mathrm{bol}}$ and \ks\ since both quantities are related by the
bolometric correction $BC_{K_{\rm s}}$ (Appendix A). If we assume that
all stars are sinusoidally variable with maximum-to-minimum amplitudes
$A_i$, and that 2MASS and DENIS measurements are unrelated in time,
then the rms of the amplitude distribution is equal to the rms of the
differences between two measurements at random phase, i.e.\ 0.35 mag.

The observed scatter is only a lower limit to the stellar flux
variability, because it only accounts for the near-infrared variation.
However, since the dominant part of the stellar energy distribution of
a Mira star is at near-infrared frequencies and since its pulsation
amplitude decreases at lower frequencies, this variation is nearly
that of the bolometric flux.

\section{ SiO targets with
$\vert l \vert < 5^\circ$} \label{lum} 
\subsection{Luminosities}
\begin{figure}[t!]
\begin{centering}
\resizebox{0.7\hsize}{!}{\includegraphics{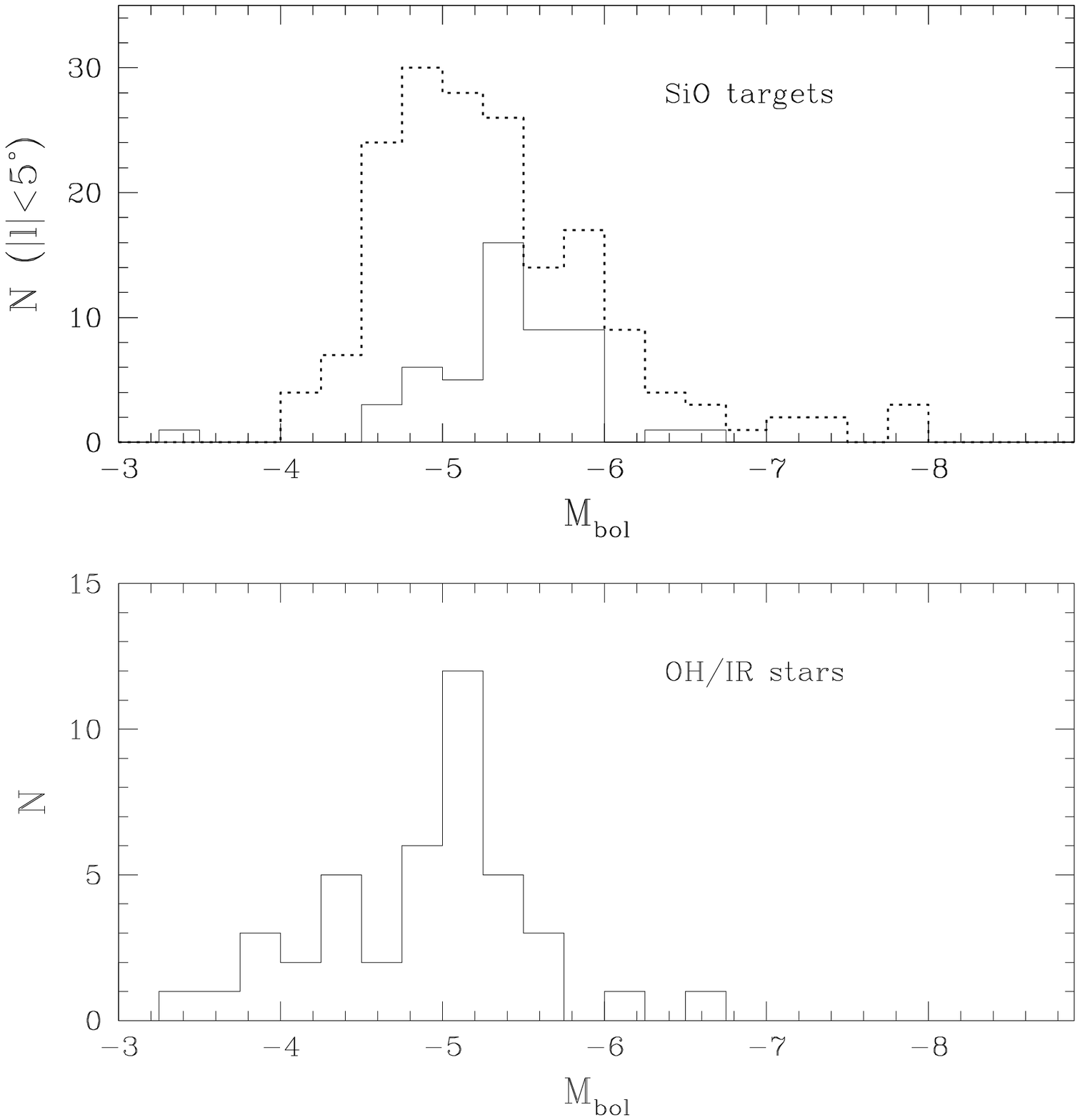}}
\caption{\label{fig:mbas.ps} Distribution of  absolute bolometric
magnitudes for an adopted distance of 8 kpc. SiO targets identified as
possible foreground objects on the basis of their extinction
\citepalias{messineo03_3} are omitted from the plot.  {\bf Upper panel:}
The dotted line shows the distribution of all SiO targets with $\vert
l \vert < 5^\circ$.  The continuous line shows the distribution of SiO
stars with (\Aks $> 2.0$ mag); most of them belong to the Nuclear Disk
(see Sect.\ \ref{ND}).  {\bf Lower panel:} For comparison the
distribution of  absolute magnitudes of a sample of OH/IR stars
within 1\degr\ from the Galactic centre \citep{ortiz02} is shown.}
\end{centering}
\end{figure}

We derive the ``absolute'' bolometric magnitudes, $M_{\mathrm{bol}}$,
of 252 sources within the central 5\degr\ of the Galactic Centre
adopting a distance of 8 kpc. Their ``absolute'' bolometric
magnitudes, as shown in Fig.\ \ref{fig:mbas.ps}, ranges from $-4$ to
$-8$ mag, with a peak around $-5.0$ mag.  Our SiO targets appear to
have mostly luminosities above that of the tip of the red giant branch
at $M_{\mathrm{bol}} \sim -3.6$  \citep{ferraro00}, where the
helium flash stops the ascent of the star on the red giant
branch. This is typical for long period variable (LPV) stars in old
metal-rich globular clusters.

Only two SiO targets (\#77 and \#347) have $M_{\mathrm{bol}}$ fainter than
$-4.0$ mag; those stars could be in the early AGB phase or at a larger
distance, or could have lower initial masses than the bulk of the SiO
targets.

Another 33 (11\%) of all SiO targets within 5 degrees from the
Galactic Centre have $M_{\mathrm{bol}}<-6.5$ mag, and even brighter
than $-7.2$ mag, which is the classical AGB limit \citep{iben83}.  The
maximum average luminosity observed for Galactic Mira stars
corresponds to $M_{\mathrm{bol}}$$=-5.5$ mag
\citep{whitelock91,glass95,olivier01}. Since the \ks--band luminosity
of a Mira star varies by up to about 2 magnitudes, for a single epoch
measurement at random phase, we expect a distribution extending to
$M_{\mathrm{bol}}=-6.5$ mag.  The 33 targets brighter than this limit,
could be either foreground stars, massive, young AGB stars, or evolved
massive stars (red supergiants) \citep{nagata93,schuller02}.

Near-infrared spectroscopy could reveal their true nature
\citep[e.g.][]{schultheis03}.  By combining extinction information and
luminosity Messineo et al. found that 20 of the 33 bright sources
are likely to be foreground stars \citepalias{messineo03_3}.  Their
total extinction (interstellar plus circumstellar), which was
calculated by assuming a photospheric colour for the central star,
appears to be lower than the average interstellar extinction of stars
in their respective surrounding fields.

We find five sources (\#31, \#75, \#92, \#128, \#294) with
$M_{\mathrm{bol}} < -7.2$ mag that cannot be explained as foreground
on the basis of their extinction.  One of them (\#92) is classified
as a possible red supergiant stars by \citet{nagata93}.  The bolometric
magnitude given by Nagata et al. is in good agreement with our
measurement. It is not unlikely that some of our other bright sources
are also red supergiants.

\subsection{Initial masses and ages}

Estimates of the initial mass of pulsating AGB stars is problematic
due to uncertainties in the pulsation mode, effective temperature
(i.e., stellar radii), and of the mass-loss history
\citep[e.g.][]{vassiliadis93,marigo96}.  A degeneracy between age and
metallicity further complicates the analysis of the distribution of
stellar colours and luminosities
\citep[e.g.][]{frogel87,whitelock91,vanloon03}.

The maximum luminosity reached at the end of the AGB phase strongly
depends on the metallicity and initial mass of the star: a lower
metallicity leads to a higher luminosity for a given initial mass.

Assuming solar metallicity and using the relation from \citet{marigo96}
between the initial mass and maximum luminosity reached before the
onset of the superwind \citep{vassiliadis93,renzini81}, we derive a
distribution of initial masses for our SiO targets that ranges from
1.0 to 4.0 M$_\odot$. This range is perhaps wider than the real due to
the variability and our use of single epoch observations. For a solar
metallicity, the majority of the SiO targets ($M_{\mathrm{bol}}=-4.5$
to $-5.5$) has ages ranging from 0.8 to 5 Gyr \citep{bertelli00}.
However, these are lower limits since stars with higher than the
assumed solar metallicity would stay longer on the main sequence and
would therefore be older when they reach the AGB.

A reliable empirical calibration of  mass and age is possible
only for Mira stars in globular clusters.  Bulge Mira stars are
more massive and/or metal rich than Mira stars in Galactic globular
clusters \citep[e.g.][]{whitelock91}, as follows from their longer
pulsation period (up to 800 days vs. 200-300 days in globular
clusters) and higher luminosity. The most luminous AGB star in
globular clusters has $M_{\mathrm{bol}}=-4.8$ mag \citep{guarnieri97},
while in the bulge they can reach $-5.5$ mag \citep{whitelock91,
glass95}.

The presence of an intermediate age population ($2.2 M_\odot <$
initial mass $< 8.0$$ M_\odot$) in the inner Galaxy was suggested by,
e.g., \citet[][]{cole02,vanloon03}.  However, most of these studies
are based on photometric observations of only giant stars. Because of
differential reddening and the degeneracy between metallicity and age
those studies could not securely confirm the presence of such a young
population.

To infer the age of a stellar population it is crucial to detect the
main-sequence turn-off.  Using deep near-infrared photometry
\citet{zoccali03} studied the stellar population of a bulge field at
latitude $b=6$\degr, from main--sequence stars to the AGB tip.  They
estimated an age of 10 Gyr for the field population and did not
observe any other turn--off that could suggest the existence of an
intermediate age population.  Their number of luminous AGB stars
agrees with that predicted for a population of that age.  However,
Zoccali's field is at $b=6$ \degr\ and it may not be representative of
the in--plane population of the inner Galaxy.

\citet{feltzing00} studied stars in the Baade's window field using
visual HST data and also concluded that  bulge stars are generally
old, although the existence of a young and metal--rich population
cannot be ruled out.

Evidence for a younger component comes from the distribution of OH/IR
stars, in the central bulge that have a vertical scale height $<100$
pc \citep{sevenster99}. They could belong to a distinct in--plane
component.  Furthermore, low latitude OH/IR stars stars and SiO
targets do not trace the axisymmetric and old (8-10 Gyr) bulge
population studied by \citet{ng96}, but the longitude--velocity
diagram of maser stars shows evidence for a Galactic bar
\citep{sevenster99,messineo02}.  The masing population could be
related to a more recent star formation event, e.g., triggered by the
formation of the Galactic bar \citep[e.g.][]{sevenster99,cole02,
sjouwerman99}.

New near--infrared photometric observations down to the main-sequence
turn--off for inner Galactic fields at low latitude are 
important for improving our understanding of Galactic stellar
population history, and can verify the possible existence of an
intermediate age population.

Near--infrared spectroscopy would yield estimates of the 
metallicity of masing stars. The  SiO targets,
being bright at near--infrared wavelengths, are ideal targets
for such a spectroscopic program.

\subsection{Red Supergiant stars ?} \label{RSG}
Red supergiant stars are massive stars ($>9$ M$_\odot$), which are
burning helium (or carbon) in non-degenerate cores. They are often
located in OB associations.  The luminosity distribution of the SiO
targets within 5\degr\ from the Galactic Centre, shown in Fig.\
\ref{fig:mbas.ps}, suggests that a few red supergiant stars are also
included, as supported by other indications described below.

IRAS counterparts of our SiO targets are mostly located in region
IIIa of the IRAS two--colour diagram \citepalias{messineo03_2}.
Previous studies of IRAS sources have shown that most of the sources
 in region IIIa are O--rich AGB stars, but  26\% may be red
supergiant stars \citep{josselin96,winfrey94}.

Weak SiO maser emission is found in semiregular AGB and red
supergiant stars \citep{alcolea90}, but red supergiants with
large amplitudes are strong SiO emitters and their SiO maser intensity
is comparable to that of Mira stars \citep{alcolea90}.  The central
star has an extended shell, generated by the strong pulsations as in
LPVs, where the SiO maser activity takes place.


The  line profile of the SiO masers  normally has several
components a few \kms\ wide over a range from 10 \kms\ (giant) to 20-40
\kms\ (supergiant) \citep{alcolea99,lebertre90}.  Our entire sample
has an average SiO maser line width of 4.6 \kms\ with a scatter
of 2.3 \kms. There are 12 sources with line width $>9$ \kms: \#31,
\#71, \#87, \#113, \#117, \#129, \#135, \#173, \#190, \#203, \#223,
\#232. These spectra look similar to the broad and multiple peaked
spectra typical of red supergiants \citep{haikala90,cho98}.  Seven of
those 12 are located within 5\degr\ of the Galactic centre, and 4
(\#31, \#87, \#113, \#129) are also very luminous ($M_{\rm bol}<-6.5$
mag).
The velocity--longitude distribution of bright ($M_{\rm bol}<-6.5$
mag) stars detected in our SiO maser survey appears distinct from that
of the nuclear disk component or from bulge stars.  The linear
dependence of $v$ and $l$ suggests (see Sect.\ \ref{lvmbol}) that it
is probably a Galactic disk component.

\section{Comparison with OH/IR stars}

\begin{figure*}[t!]
\begin{center}
\resizebox{0.7\hsize}{!}{\includegraphics{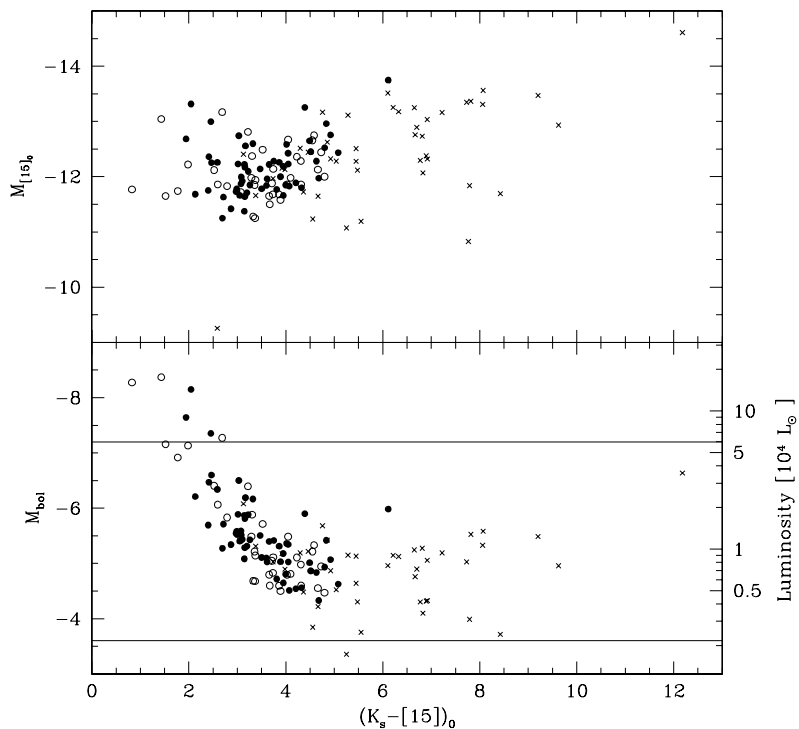}}
\caption{\label{fig:mbas2.ps} {\bf Lower panel:} ``Absolute''
bolometric magnitudes, $M_{bol}$, versus dereddened (\ks$-[15])_0$,
adopting a distance of 8 kpc. When both DENIS and 2MASS \ks\
measurements, or both ISOGAL and MSX 15$\mu$m measurements are
available, the average flux density is adopted.  Circles show SiO
targets from the ISOGAL fields of \citet{ortiz02}, they cover about
$\sim 0.84$\degr.  Filled and open symbols indicate SiO maser
detections and non--detections, respectively.  For comparison, crosses
show OH/IR stars within 1\degr\ from the Galactic Centre with a known
period and ISOGAL counterpart \citep{ortiz02}. The line at
$M_{\mathrm{bol}} = -3.6$ indicates the location of the tip of the red
giant branch and the line at $M_{\mathrm{bol}} = -7.2$ the AGB
limit. {\bf Upper panel:} ``Absolute'' $[15]_0$ magnitudes,
$M_{[15]}$, versus dereddened (\ks$-[15])_0$.  Symbols are as in the
lower panel.}
\end{center}
\end{figure*}

For a comparison with our SiO targets, we also consider a sample of
OH/IR stars (AGB stars with 1612 MHz OH maser emission) within 1\degr\
from the Galactic Centre \citep[from Table 3 of][]{ortiz02}, a region
fully mapped at 1612 MHz
\citep{lindqvist92,sjouwerman98,sevenster97a}. Ortiz et al.\ found
ISOGAL counterparts for all OH sources in ISOGAL fields, counterparts
which are typically bright at 15 $\mu$m and have very red
(\ks$-[15])_0$ colours (reaching 12 mag; see upper panel of Fig.\
\ref{fig:mbas2.ps}), which is indicative of  high mass loss rates.

To directly compare luminosities of OH/IR stars with those of SiO
targets, bolometric magnitudes must be calculated in the same way.  We
therefore calculated the interstellar extinction \Aks\ toward each
OH/IR star using 2MASS stars within 2\arcmin\ from the position of the
OH/IR star, following the procedure described in
\citetalias{messineo03_3}. Extinctions, \Aks, are found systematically
larger (by up to 0.6 mag) than those adopted by \citet{ortiz02}. This
is due to our use of the ($H$, \ks) data, which are less sensitive to
extinction and therefore less affected by a low--extinction bias than
($J$, $H$) or ($J$, \ks) data used by \citet{ortiz02}.  We adopted the
near--infrared mean flux densities of the OH/IR stars given by
\citet{wood98}.  When using Mathis' extinction law \citep[for
comparison with][]{ortiz02} we obtained bolometric magnitudes
consistent with those of \citet{ortiz02}: since the spectral energy
distribution of an OH/IR star peaks long-ward of 3\um\ (due to the
presence of a thick circumstellar envelope), its bolometric magnitude
is less sensitive to near--infrared interstellar extinction
corrections than for SiO targets.  However, when we adopt the
mid--infrared extinction law suggested by Lutz et al.\ \citepalias[as for
the SiO targets,][]{messineo03_3} we obtain bolometric magnitudes for
the OH/IR stars brighter by up to 0.6 mag.  Whenever a bolometric
magnitude determination via direct integration (Sect. \ref{mbol}) was
not possible, it was estimated using the bolometric correction
described in the Appendix.

Figure \ref{fig:mbas.ps} shows a comparison between the bolometric
magnitude of our SiO targets (Sect.\ \ref{ambol}) with that of the
OH/IR stars.

The peak of the OH/IR magnitude distribution is at $\sim-5.1$ mag, and
translates to an initial stellar mass of about 1.8 $M_\odot$ and an
age of about 2 Gyr, using the models of \citet{bertelli00} with solar
metallicity. The peak for OH/IR stars appears consistent with the
median of the bolometric magnitudes of SiO targets in the central
5\degr, although the distribution of the latter is broadened by
variability and the use of single epoch observations.

For the 96 SiO targets (58 detections) from the same ISOGAL fields as
the OH/IR stars of \citet{ortiz02}, Fig.\ {\ref{fig:mbas2.ps} shows the
absolute bolometric magnitude of both  OH/IR and SiO stars plotted
against their (\ks$-[15])_0$ colour. Distributions appear to
differ in that  SiO targets have bluer  colours
\citepalias[due to our selection,][]{messineo03_2}.  This indicates that
OH/IR stars have thicker envelopes due to higher mass--loss rates.

Note that here we are comparing only our SiO targets with known OH/IR
stars.  We imposed colour constraints for our target selection and we
tried to be complementary to previous OH/IR studies by discarding the
reddest mid- and near--infrared colour sources.  It is quite well
possible that there are SiO masing stars with colours much redder than
what we considered, though according to \citet{nyman93} there should
be a cut--off in the 86 GHz SiO maser intensity for very optically
thick circumstellar envelopes.

In contrast to SiO targets, there are no OH/IR stars brighter than
$M_{\rm bol} = -6.5$ (for a distance of 8 kpc) in our sample. The
luminous SiO targets tend to be bluer than fainter ones, suggesting a
thinner circumstellar envelope, which could also explain the lack of
1612 MHz OH maser line emission.  Therefore, our sample includes
luminous SiO masing AGB stars or red supergiants, with moderate
mass--loss rates, not masing in OH.  If the mass--loss rate increases
with time \citep[e.g.][]{lewis89} these objects could later also show
1612 MHz OH maser emission, which is more likely for thicker envelopes
\citep{sevenster01,habing96}.

In the chronological sequence of circumstellar masers presented by
\citet{lewis89}, as first suggested by \citet{bedijn87}, circumstellar
envelopes of AGB stars gradually evolve under slowly increasing rates
of mass loss. Since the SiO masing shell is near the photosphere, the
first maser emission to appear is that from SiO, while OH maser
emission appears as the shell grows.  The SiO maser emission is then
the first to disappear when the mass loss declines.

The distributions of bolometric magnitudes of both SiO and OH/IR stars,
as seen in Fig.\ \ref{fig:mbas.ps}, peak at about the same value, which
suggests that both populations trace the same epoch of star formation,
but SiO masing stars have less evolved circumstellar envelopes
\citepalias[see also][]{messineo03_2}.  This is supported also by the
different distribution of SiO and OH/IR stars in the
period--luminosity diagram (Sect.\ \ref{lp}).

\section{Period--Luminosity relation}\label{lp}

Our SiO targets include 15 LPVs from the sample of \citet{glass01},
\citetalias[see also Sect. 4.2 in][]{messineo02}.  Figure
\ref{fig:glass.ps} plots their bolometric magnitudes against their
periods, showing that they fall above the period--luminosity
distribution of  Mira stars in the Sgr--I field
\citep{glass95}. However, in the Sgr--I field Mira stars with period
longer than 450 days are rare. Our SiO stars may support a steepening
of the period--luminosity relation for period longer than 450 days, as
observed in the Magellanic Clouds \citep{hughes90}.

\begin{figure}[t]
\begin{centering}
\resizebox{0.7\hsize}{!}{\includegraphics{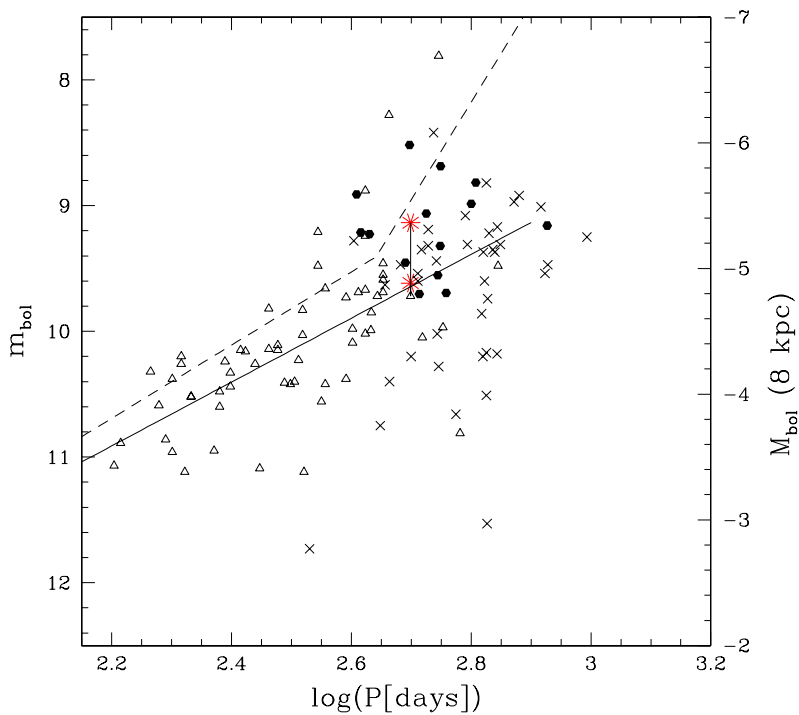}}
\caption{\label{fig:glass.ps} The period--luminosity relation of LPV
stars.  Dots represent 15 LPV stars in our SiO maser sample
located within 24\arcmin\ from the Galactic Centre \citep{glass01}.
They have a bright ISOGAL counterpart, but no known OH maser emission.
Triangles show Mira stars located in the Sgr--I field
\citep{glass95}, for which we plot the mean bolometric magnitudes.
For \#153, the only SiO target located in the Sgr--I field, we plot
both  magnitudes (stars) obtained using  DENIS and 2MASS data,
and connect them to the mean $m_{\mathrm {bol}}$ value found by
\citet{glass95} with a vertical line.  Crosses show OH/IR stars
with ISOGAL counterpart \citep{ortiz02}.  The continuous line
represents the relationship found by \citet{glass95}, while the dashed
line is a fit for LMC Mira stars \citep{hughes90}, assuming a distance
modulus of 18.55 mag for the LMC and 14.5 mag for the Galactic
Centre. }
\end{centering}
\end{figure}

The standard deviation in the period--luminosity relation of Mira
stars in Sgr--I is 0.36 mag \citep{glass95}, which partly arises from
the spread in distance, and partly from the uncertainty in the
apparent bolometric magnitude.  A spread of $\pm2$ kpc in distance at
a distance of 8 kpc causes a spread in $M_{\mathrm {bol}}$ of
$\pm0.55$ mag. Bolometric magnitudes given by \citet{glass95} were
obtained adopting the mean values of monitored infrared flux
densities.  The period--luminosity relation obtained using mean
magnitudes of Mira stars in the Magellanic Clouds is less sensitive to
distance uncertainties, resulting in a scatter of only 0.15 mag
\citep{feast89}.

Our sample has one SiO star located in the Sgr--I field, \#153.  Its
$m_{\mathrm {bol}}$ calculated using  DENIS data is 9.62 mag,
identical to the mean value obtained by \citet{glass01}.  When using
2MASS data we find $m_{\mathrm {bol}} = 9.14$ mag.  Considering
the pulsation amplitude $\Delta K = 1.38$ mag \citep{glass95}, both
measurements are consistent.

The 15 LPV/SiO targets in the Galactic Centre from \citet{glass01}
suffer from strong extinction (\Aks$>2$ mag) and using the kinematic
information for the 11 which were detected in our SiO maser survey, we
conclude that they are members of the central nuclear disk (see Sect.\
\ref{ND}). They are therefore at about the distance of the Galactic
centre.  Assuming an isotropic distribution of stars in the nuclear
disk, a range of longitudes of $\pm1.5$\degr\ translates to a distance
range of $\pm200$ pc, and a corresponding distance modulus range of
$\pm0.05$ mag.  Their dispersion on the period--luminosity plane is
mostly due to the uncertainty in the bolometric magnitude ($\sim$0.35
mag, see Sect.\ \ref{ambol}) and in the extinction
\citepalias{messineo03_3}.

Since we selected bright ISOGAL and MSX sources \citepalias{messineo02} and
since  mid--infrared flux density correlates with  period
\citep[see the 12$\mu$m period--luminosity relation in][]{whitelock91},
it is not surprising that the LPVs we observed have  periods 
above 400 days (Fig.\ \ref{fig:glass.ps}).  \citet{imai02} searched for
SiO maser emission toward the entire sample of Mira stars near the
Galactic centre detected in $K$--band by \citet{glass01} and confirmed
the expected increase of the detection rate with increasing
period. While for periods below 300 days the SiO maser detection rate
is below 20\%, above 400 days it rises to 60\% .  OH/IR stars also have
typical periods above 400 days (Fig.\ \ref{fig:glass.ps}).
Thereby, SiO and OH maser emission mainly trace long period AGB
variables.

Unlike Mira stars, OH/IR stars do not follow a period--luminosity
relation in the Galactic Centre region \citep[][ and references
therein]{ortiz02}.  Most of them for a given magnitude have a longer period
than that predicted by the Mira period--luminosity relation, so
OH/IR stars and SiO targets are differently distributed. 

Pulsation models predict that due to a dramatic increase of the mass
loss rate (superwind phase), a Mira star significantly stretches its
pulsation period, while its luminosity remains almost constant
\citep{vassiliadis93}, and becomes an OH/IR star. OH/IR stars have then
longer periods for a given luminosity than those found in Mira stars.

\section{Stars in the the Nuclear Disk and a fourth dimension: extinction}
\label{ND}
\begin{figure}[t]
\begin{centering}
\resizebox{0.7\hsize}{!}{\includegraphics{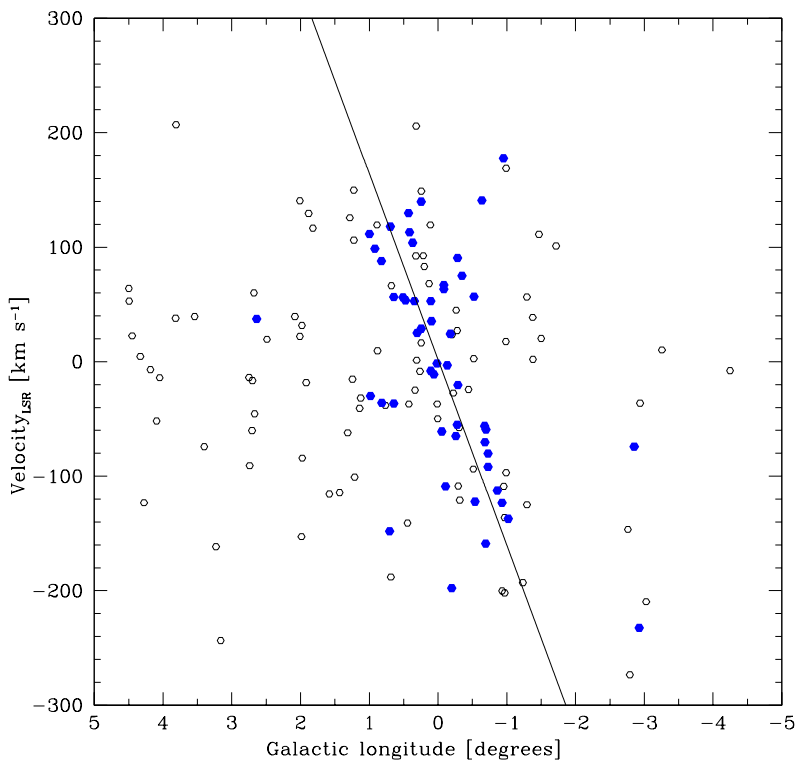}}
\caption{\label{fig:ND.ps} Stellar longitude--velocity diagram of our
86 GHz SiO masers.  Filled circles indicate sources with
interstellar extinction \Aks\ above 2 mag. Most of these
belong to the fast rotating nuclear disk.  The continuous line
indicates our best fit to the nuclear disk component.
 }
\end{centering}
\end{figure}

The line of sight extinction is a useful fourth dimension along with
position ($l,b$) and velocity to characterise various Galactic
components.

Around zero longitude our stellar longitude--velocity ($l-v$) diagram
reveals a stellar nuclear disk, which follows the high velocity
gaseous {\bf nuclear disk} \citep[e.g.][]{burton78,binney91}.

There is a unique correspondence between interstellar extinction and
velocity.  When we select SiO targets at interstellar extinction \Aks$
>$ 2 mag (Fig.\ \ref{fig:ND.ps}), we find that most of these stars
kinematically trace the nuclear disk.  Their line--of--sight velocities
range from $\sim+150$ to $\sim-200$ km s$^{-1}$, similar to the gaseous
nuclear disk line--of--sight velocities \citep[cf. the $^{13}$CO ($l-v$)
diagram in Fig.\ 4 of][]{bally88}. Therefore, the extinction enables us
to identify individual stars belonging to the nuclear disk.

The stellar nuclear disk rotates rapidly around the Galactic Centre:
our best--fit yields a gradient of 162($\pm 40)$ km s$^{-1}$ per degree
longitude. This slope is consistent with a value of $180(\pm15)$ km
s$^{-1}$ found for  OH/IR stars with high circumstellar expansion
velocity located within 1\degr\ of the the Galactic Centre
\citep{lindqvist92}.

To investigate whether the 56 SiO targets with \Aks$ >2$ mag also show
distinct physical properties, their bolometric magnitude distribution
is shown in Fig.\ \ref{fig:mbas.ps}.  The distribution peaks at $-5.4$
mag, a peak slightly more luminous than the average luminosity of SiO
targets within 5\degr. However, their luminosity is consistent with
those of variable AGB stars.  There is only one SiO target with
$M_{\rm bol} < -6.5$ mag, \#99, which may be a young star. Its 
location ($l=0.8$\degr; $v= -36.0$ \kms; $2\sigma$ away from our best
fit) indicates that it is probably unrelated to the nuclear disk
in origin and kinematics.

The 11 LPVs from \citet{glass01} that we detected at 86 GHz (see also
Sect.\ \ref{lp}) all belong to the nuclear disk population, as their
positions ($|l|<1.5,|b|<0.5$), their interstellar extinctions
(\Aks$>2.0$ mag) and their line--of--sight velocities confirm.

\section{Bolometric magnitudes and the $l-v$ diagram}

Different Galactic components give a different signature in the $(l -
v)$ diagram. However, a stellar $(l - v)$ diagram alone is not
sufficient to locate individual stellar components, mainly due to the
velocity dispersion of the stars, that smooths  various
features. The inclusion of additional distance information  will
notably improve the understanding of the $(l - v)$ diagram.  The
stellar bolometric magnitude may serve as a first approximation of
distance.

In section \ref{lum}, we have already noticed the presence of a group
of stars brighter than the typical luminosity of a Mira star at its
maximum ($M_{\rm bol}= -6.5$ mag) when we assume a distance of 8
kpc. In Figure \ref{fig:lvmbol.ps} we illustrate the $l-v$ diagram of
SiO targets with $M_{\rm bol}$ fainter and brighter than $-6.5$ mag.
The bright sample appears to be distributed differently from the faint
sample, since it has a low velocity dispersion ($\sim 30$ \kms),
independently of longitude.  The stars that were classified as
foreground stars on the basis of extinction consideration
\citepalias[crosses in Fig.\ \ref{fig:lvmbol.ps}; see
also][]{messineo03_3}, almost all belong to the bright group. This
indicates that the bright stars are closer on average to the Sun than
fainter ones and that their luminosities are overestimated by
assuming the distance of the Galactic centre.  But then how nearby are
those stars? Do they trace a specific Galactic component?

\label{lvmbol}
\begin{figure}[t]
\begin{centering}
\resizebox{0.6\hsize}{!}{\includegraphics{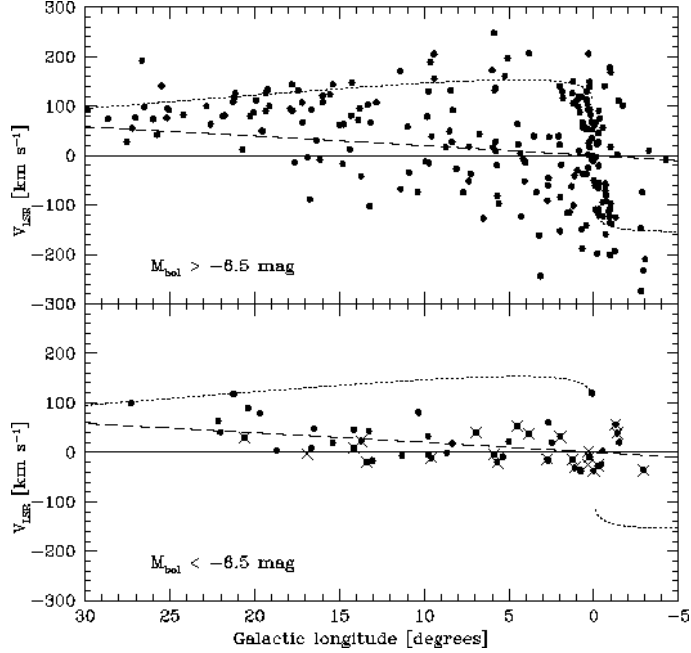}}
\caption{\label{fig:lvmbol.ps} Longitude--velocity diagram of SiO
targets.  {\bf Upper panel:} SiO targets with $M_{\rm bol} > -6.5$ mag
(assuming a distance modulus for the Galactic centre of 14.5 mag). The
dotted line shows the curve defined by the line-of-sight velocities of
tangent points to circular orbits, as predicted by Eq.\ 5.1.  The
dashed line indicates a circular orbit at a galactocentric distance of
5 kpc.  {\bf Bottom panel:} Dots indicate SiO targets with $M_{\rm
bol} < -6.5$ mag. Crosses indicate SiO targets classified as
foreground on the basis of extinction \citepalias{messineo03_3}.}
\end{centering}
\end{figure}

An estimate of their distance can be inferred by assuming circular
orbits. However, this assumption clearly does not hold for the central
kiloparsecs, due to both the presence of a bar and the fact that bulge
stars might have a velocity dispersion larger than the rotation
velocity. Hence, we calculated Galactocentric distances, under the
assumption of circular orbits, for SiO targets at longitudes
$>10$\degr.  We adopted the relation between the longitude, radial
velocity and distance \citep{burton88}:

\begin{equation}
v = R_\odot [\frac{V(R)}{R}-\frac{V(R_\odot)}{R_\odot}] {\rm sin}(l),
\end{equation}

\noindent 
where $V(R)$ is the circular rotation velocity at Galactocentric
distance $R$, the Galactocentric distance of the Sun is taken to be
$R_\odot=8.0$ kpc, and the circular velocity of the Sun
$V(R_\odot)=200$ \kms.  Adopting $V(R)=V(R_\odot)(R/R_\odot)^{0.1}$
\kms\ \citep{binney91}, and inverting equation (1), it follows that:

\begin{equation}
R = \frac{8.0}{(1+\frac{v}{(200 \times {\rm sin}(l)})^{1.11}} ({\rm
in~kpc}),
\end{equation}

\begin{figure}[t]
\begin{centering}
\resizebox{0.9\hsize}{!}{\includegraphics{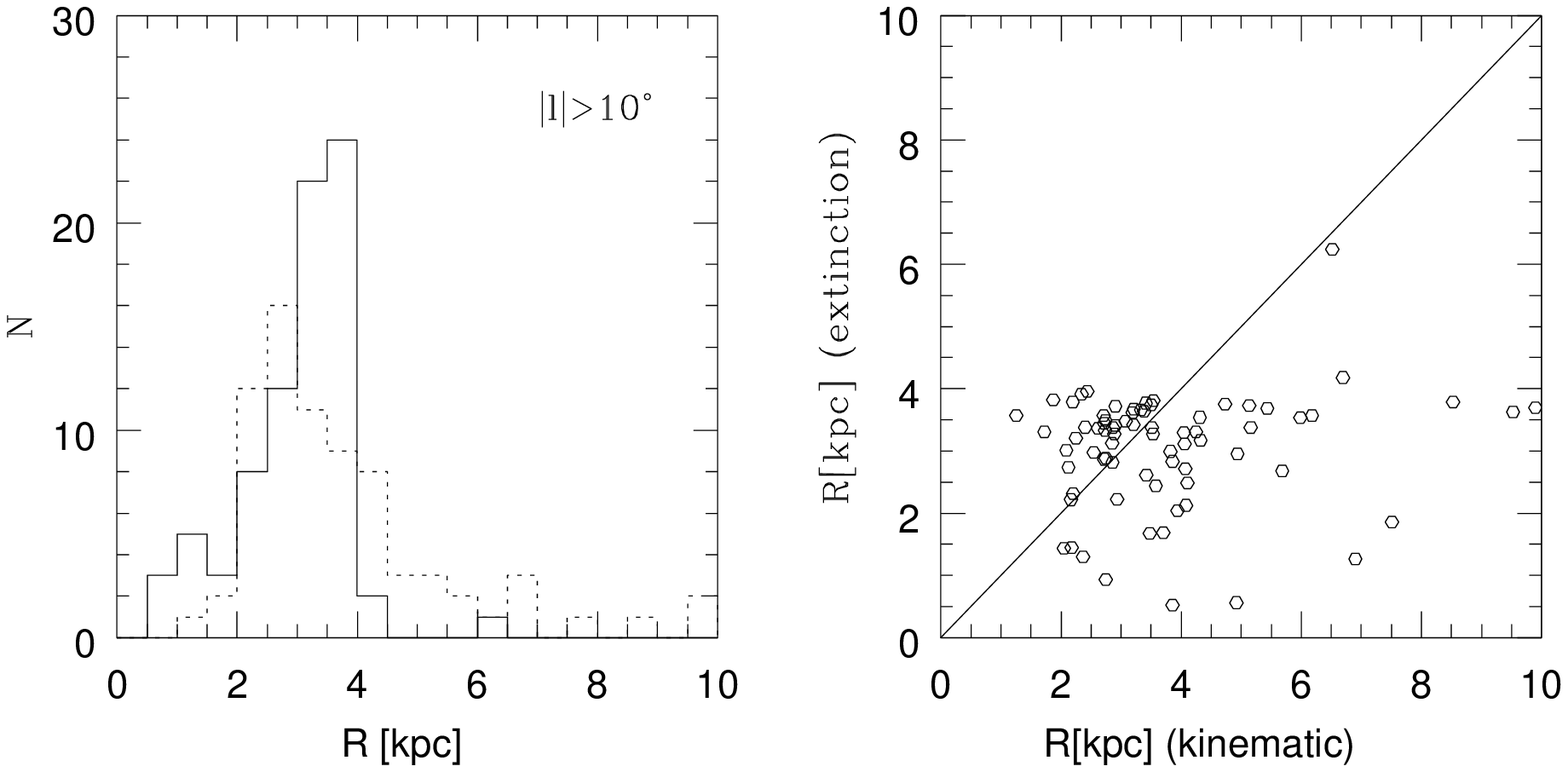}}
\caption{\label{fig:lopez.ps}{\bf Left side}:The continuous histogram
shows the distribution of Galactocentric distances inferred for SiO
targets at longitudes $l>10$\degr\ using interstellar extinction
estimates towards the line of sight of each SiO target and a model of
Galactic dust distribution \citep{drimmel03}. The dotted histogram
shows the distribution of Galactocentric distances inferred for SiO
targets at longitudes $l>10$\degr\ assuming circular orbits. {\bf
Right side}: Distance estimates from interstellar extinction versus
kinematic distances. }
\end{centering}
\end{figure}

Bright sources are found to have an almost homogeneous distribution in
Galactocentric distance, from 0 to 15 kpc.  Fainter stars (Fig.\
\ref{fig:lopez.ps}) peak at about 3 kpc from the Galactic centre,
i.e.\ at about the corotation radius of $\sim3.4$ kpc
\citep[e.g.][]{englmaier99}.

Since interstellar extinction increases with distance, another
independent way to estimate distances is to use a model of dust
distribution in the inner Galaxy. Using the implementation of
\citet{drimmel03} based on COBE/DIRBE data, we calculated for each
line of sight toward a SiO target the interstellar extinction as a
function of distance.  By interpolating the models at the values of derived
interstellar extinction \citepalias{messineo03_3}, we have estimated
distances to each SiO target.

The model of Drimmel et al.\ does not include the nuclear disk
molecular complex and therefore it fails in reproducing the distance of
nuclear disk stars. It predicts that all SiO targets are within a
Galactocentric distance of 5 kpc and confirms a peak around
corotation.

We used two independent methods to estimate distances, one based on
kinematics and another based on extinction. Although the uncertainty
are large (see right panel of Fig.\ \ref{fig:lopez.ps}), from both
methods we find that most of SiO targets have a Galactocentric
distance within 5 kpc, with a peak around corotation.  The bright
stars, however, do not seem confined at a specific Galactocentric
distance. 

It is difficult to argue about the nature of each single SiO target.
The difference in distance modulus of a star at the Galactic centre
and at a Galactocentric distance of 5 kpc is 2.1 mag, thus we might
have overestimated the luminosity of the SiO target stars. The true
luminosity of the ``bright'' stars could be below the AGB limit
($M_{\rm bol}< -7.2$ mag) and therefore they could be all AGB stars,
rather than red supergiants.  However, the luminosity range of these
two classes of stars overlaps, and some of them could still be red
supergiants as the SiO maser line widths and the low velocity
dispersion suggest. Near--infrared spectroscopy is needed to
distinguish between AGB and red supergiants. Furthermore, a
near-infrared monitoring program would provide pulsation periods and
therefore another independent estimate of the distance of each SiO
target.

We are currently working on a kinematic study of the SiO targets based
on a more realistic Galactic gravitational potential, which will be
presented in a forthcoming paper.

\section{Conclusion}\label{conclusion}

Bolometric magnitudes for  SiO target stars were computed from 
available flux density measurements. Adopting a distance of 8 kpc for
all stars within 5\degr\ from the Galactic Centre we find a peak in the
luminosity distribution  at M$_{\mathrm {bol}}= -5.1$ mag,
which coincides with the peak shown by OH/IR stars in the Galactic
Centre \citep[e.g.][]{ortiz02}.  Assuming a solar metallicity this peak
magnitude corresponds to stars with an initial mass of 1.8 M$_\odot$
and an age of about 2 Gyr.

Our  sample includes 15 LPVs from the sample of \citet{glass01}.
We found that these fall above the period--luminosity relation found by
\citet{glass95}, thus differing from OH/IR stars. This supports a
chronological sequence for circumstellar masers proposed by
\citet{lewis89}. We also find evidence for a steepening of the bulge
period--luminosity relation for periods larger than 450 days, as found
for Magellanic Cloud LPVs.

In contrast to OH/IR stars \citep{ortiz02}, SiO targets contain a
 significant fraction (11\%) of blue and luminous stars (M$_{\mathrm
 {bol}} < -6.5$ mag). These stars are most likely massive AGB stars,
 or red supergiant stars, foreground with respect to the bulge.  Their
 longitude--velocity distribution has a low velocity dispersion, which
 identifies a component, distinct from the bulge, perhaps related to
 the disk or to the molecular ring. A full dynamical analysis of SiO
 stars will be presented in a subsequent paper.

We found a unique relation between interstellar extinction and
kinematic properties of SiO stars. Those with \Aks$>2.0$ mag
belong to the stellar Nuclear Disk.

\begin{acknowledgements}
We thank C. Loup and M.--R. Cioni for making their work on bolometric
magnitudes available prior publication.  MM is grateful to J. Meisner
for help with statistical computations and to Glenn van de Ven for his
enthusiastic and great discussions on stellar dynamics. The DENIS
project was carried out in the context of EARA, the European
Association for Research in Astronomy.  This publication makes use of
data products from the IRAS data base server,
from the Two Micron All Sky Survey, 
%
from the Midcourse Space Experiment, 
and from the SIMBAD data base.
The work of MM is funded by the Netherlands Research School for
Astronomy (NOVA) through a {\it netwerk 2, Ph.D. stipend}.
\end{acknowledgements}

\appendix
\section{Bolometric corrections}
The bolometric correction for a given photometric band is defined as
$BC_{\lambda}=m_{\mathrm {bol}}-m_{\lambda}$.  Bolometric corrections
for late--type stars as a function of spectral type or colour are used
to derive stellar luminosities.  However, there are uncertainties in
their estimates and applicability.  Stars of different temperature,
metallicity and mass--loss rate have a different energy distribution
and therefore a different bolometric corrections. Variability is also
a complication for a proper determination of the bolometric
correction. Long period variables are characterised by extended
atmospheres and their energy distribution differs from other static
late--type giants because of water absorption seen from the $J$--band
to the $L$--band \citep{matsuura02, frogel87}. The amount of
absorption varies from star to star and with stellar phase.  At the
present time there are a number of already completed near--infrared
monitoring programs of large amplitude AGB stars, in the solar
neighbourhood, in the South Galactic Cap and in the Magellanic Clouds,
which provide accurate average bolometric magnitudes of oxygen--rich
large amplitude variables
\citep{whitelock03,olivier01,whitelock-hyp,whitelock94}.  We use these
data to derive various bolometric corrections for these type of
objects and compare these with results for stars in the inner Galaxy.

\subsection {Bolometric corrections of monitored LPV stars}
In order to have a large sample of LPV stars, well studied and
covering a wide range of colours, we assembled photometric catalogues
of LPV stars from \citet{olivier01} and
\citet{whitelock03,whitelock-hyp,whitelock94}.  All near--infrared
magnitudes are in the SAAO system \citep{carter90}.  Stellar
fluxes are corrected for reddening only in the work of
\citet{olivier01}.  However, the effect of interstellar extinction
is negligible because these stars are nearby or outside of the
Galactic plane, and because the bulk of their energy is emitted at
infrared wavelengths.  Bolometric magnitudes are reported by the
authors and, with the exception of the Hipparcos sample, were derived for
each star by integrating under a spline curve fitted to the mean
near--infrared ($J H K L$) flux densities and the IRAS 12 and 25
$\mu$m flux densities as a function of frequency, and using two
extrapolations for longer and shorter wavelengths as described in
\citet{whitelock94}.  For the sample of low mass--loss Mira stars
observed by Hipparcos \citep{whitelock-hyp} bolometric magnitudes were
calculated by fitting blackbody curves to the ($J H K L$) flux
densities as a function of frequency.  The blackbody fitting is a good
approximation of these low mass--loss stars and for them blackbody and
spline--fit bolometric magnitudes agree to better than 0.2 mag
\citep{whitelock03}.
%
Figure \ref{fig:lpvbc.ps} shows various bolometric corrections versus
colours. The IRAS 12 $\mu$m magnitude, [12], is obtained adopting a
zero point of 28.3 Jy \citep{iras}.  $BC_{K}$ is well correlated with
$K-[12]$ rather than $J-K$.  The least--squares polynomial fits to
the bolometric correction $BC_{K}$ and $BC_{[12]}$, which give a good
match to the data ($\sigma=0.08$ mag) over the range $1.0<
{({K}-[12])} <7.0$, shown by the continuous lines in
Fig. \ref{fig:lpvbc.ps}, are given by:
$$BC_{K} = 2.219+0.7351{({K}-[12])} -0.1299  {({K}-[12])}^2$$
$$BC_{[12]} = 2.219+1.735{({K}-[12])}-0.1299  {({K}-[12])}^2.$$
\begin{figure}
\begin{centering}
\resizebox{\hsize}{!}{\includegraphics{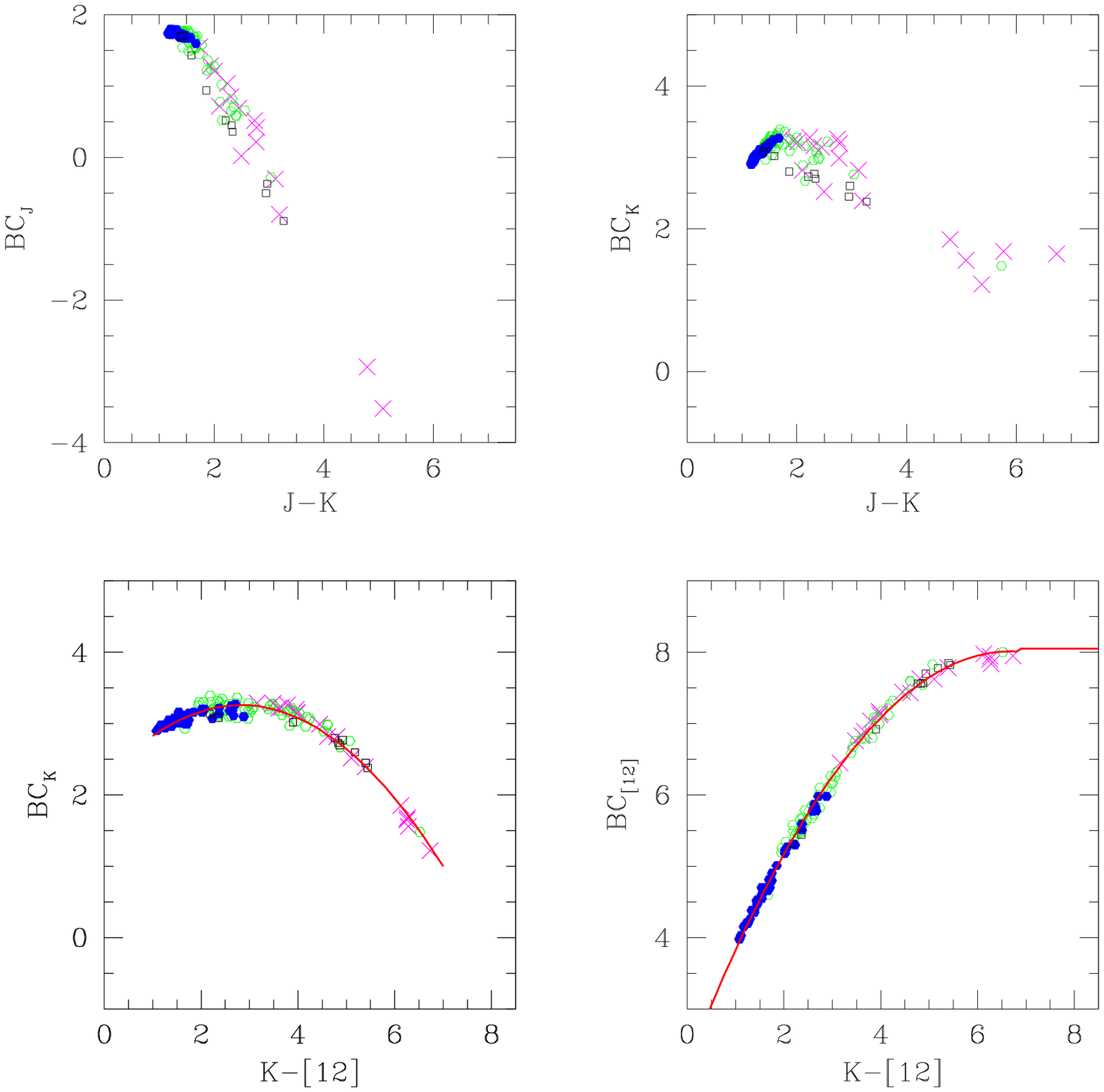}}
\caption{\label{fig:lpvbc.ps} Monitored LPV stars - Bolometric
corrections as a function of various colours. Filled circles represent
LPV stars in the solar vicinity detected by Hipparcos
\citep{whitelock-hyp}; open circles represent LPV stars in the
Southern Galactic Cap \citep{whitelock94}; crosses indicate local high
mass--loss LPV stars \citep{olivier01}; squares represent obscured LPV
stars in the Large Magellanic Cloud \citep{whitelock03}.  The IRAS
12$\mu$m magnitude is defined as $[12]=-2.5 \log F_{12}[{\rm
Jy}]/28.3$.}
\end{centering}
\end{figure}

These two curves are complementary to each other. Practically for
${K}-[12]$ smaller than 4 mag it is better to use $BC_{K}$ due to
the flatness of the curve.  For ${K}-[12]$ redder than 4 mag 
$BC_{K}$ decreases steeply with increasing colour, while $BC_{[12]}$
is more stable.  The narrowness of  ($BC_{K}$,${K}-[12]$) and
($BC_{[12]}$, ${K}-[12]$) sequences also suggests that they do not
depend on metallicity; in fact, the Magellanic Clouds objects also
follow the same relation. Instead, the ($BC_{K}$, $J-K$) sequence
depends on metallicity as found by \citet[e.g.][]{frogel87} comparing
local stars and bulge stars.

\begin{figure}[h!]
\begin{centering}
\resizebox{0.56\hsize}{!}{\includegraphics{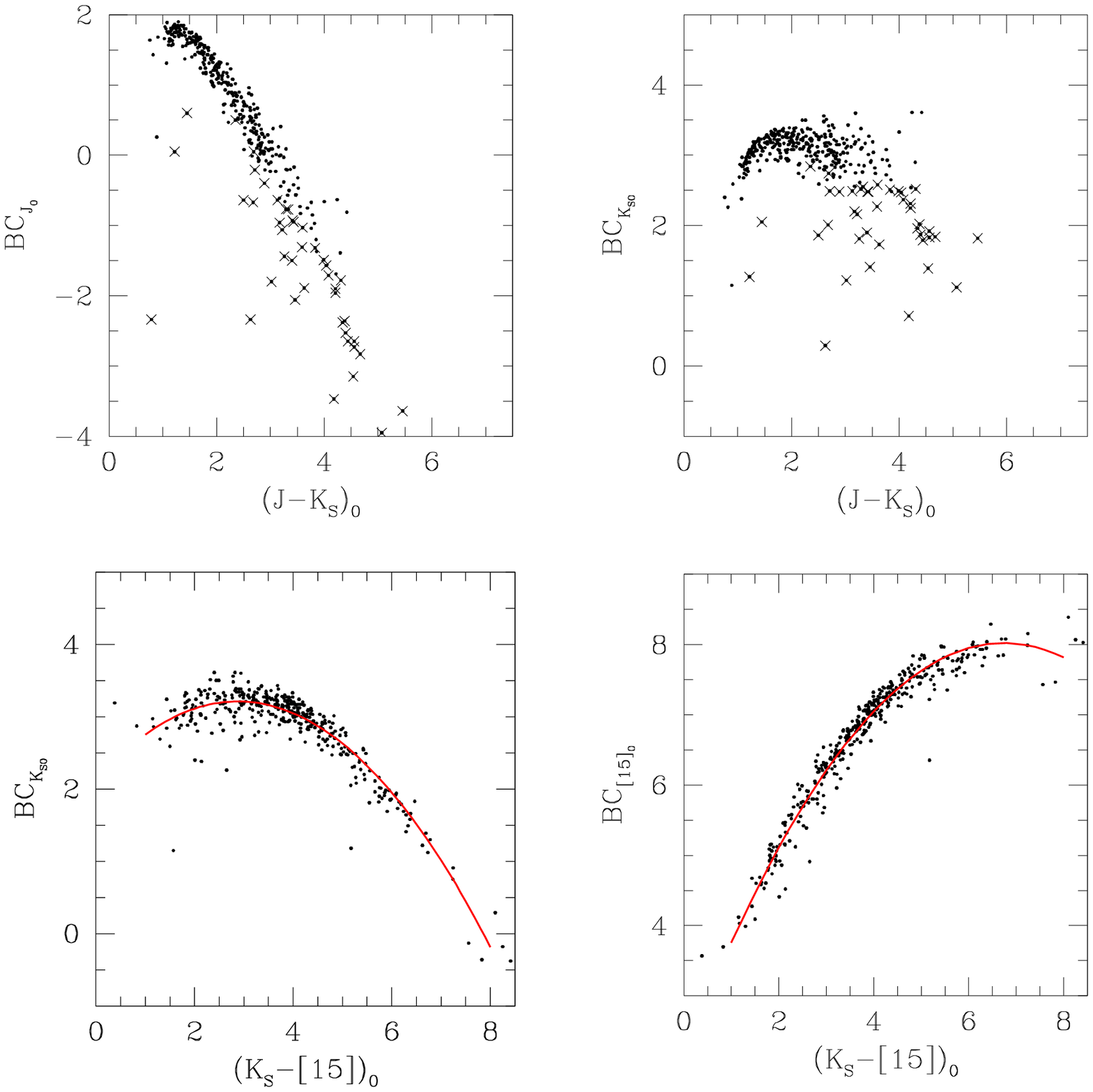}}
\resizebox{0.56\hsize}{!}{\includegraphics{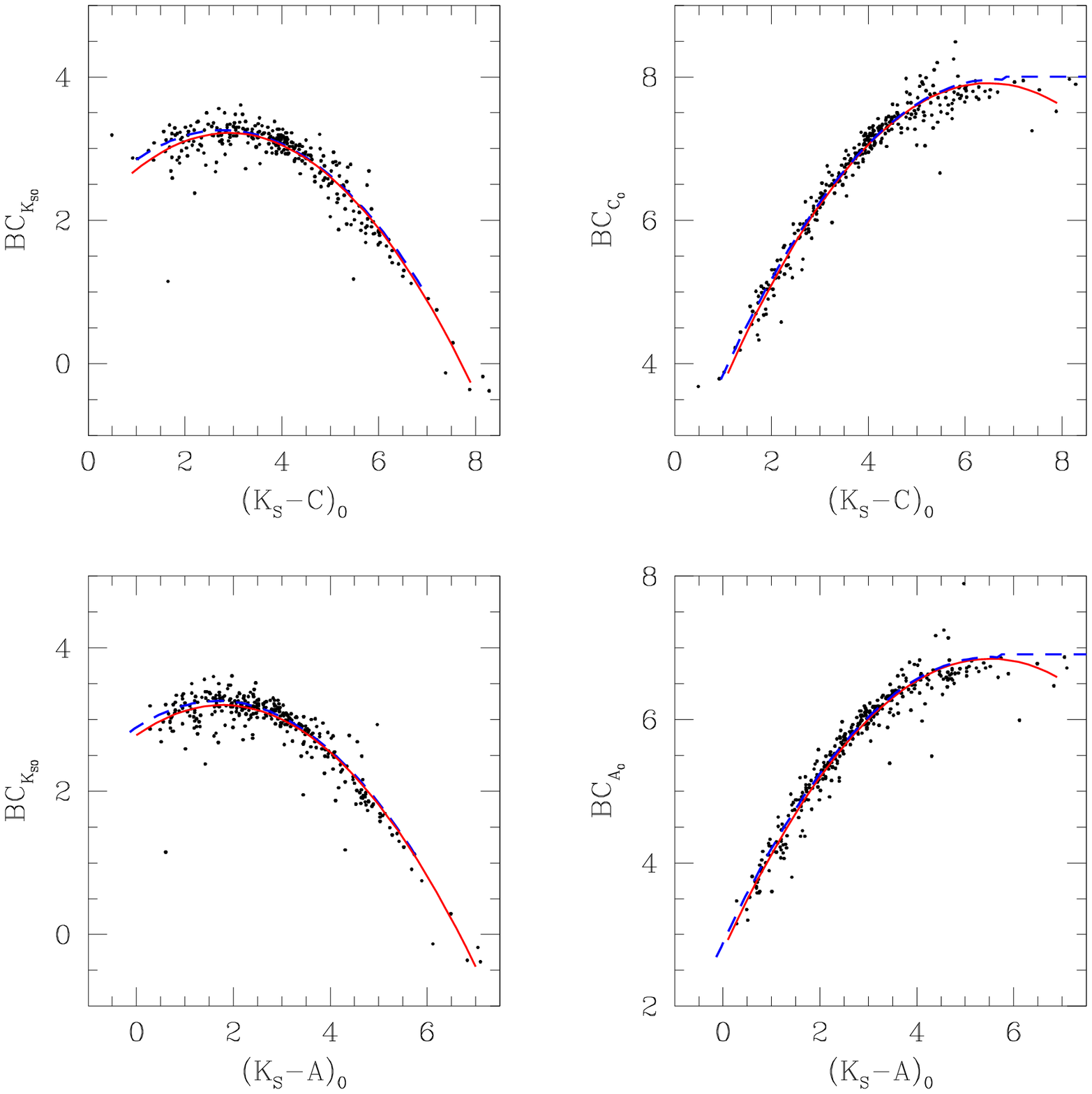}}
\caption{\label{fig:bc.ps} SiO targets - Bolometric corrections versus
colours. Crosses are sources with (\ks$-[15])_0$ redder than 5 mag.
Only sources with bolometric magnitudes well determined are plotted,
i.e. when the extrapolation to shorter wavelengths has a negligible
contribute to the integrated bolometric magnitude ($<20$\%).  The
2MASS dataset is plotted.  DENIS data are plotted only for sources
with poor 2MASS photometry.  When both ISOGAL and MSX 15\um\
measurement are available, then the average flux density is adopted.
Continuous lines indicate our best fits.  Dashed lines show
the relation found for  monitored LPV stars, adapted using the
equation $A-[12]=1.15, C-[12]=0.05$ mag \citepalias{messineo03_3}.}
\end{centering}
\end{figure}

\subsection{Bolometric corrections for stars in the inner Galaxy}
For the SiO targets we derived bolometric magnitudes as described in
Sect.\ \ref{mbol}.  Bolometric corrections for the \ks\ and 15 \um\
band are computed from the $m_{\mathrm {bol}}$ and plotted against the
dereddened (\ks$-[15])_0$ colour in Fig.\ \ref{fig:bc.ps}.
Identical results are found using separately 2MASS $m_{\mathrm
{bol}}$ and \ks\ or DENIS $m_{\mathrm {bol}}$ and \ks\ and/or ISOGAL
[15] or MSX $D$ (15\um) band magnitudes. The plotted continuous line
is a least--squares polynomial fit, which gives a good match to the
data ($\sigma=0.25$ mag) over the range $1.0< $(\ks$-D)_0 <8.0$:
$$BC_{K_{\rm s}} =2.138+0.745{(K_{\rm s}-[15])_0}-0.1294{(K_{\rm
s}-[15])_0}^2.$$

Analogously, using the MSX $A$ (8\um) band, the following fit is
obtained as a function of  (\ks$-A)_0$ colour in the range $0.0 <
$(\ks$-A)_0 > 7.0$ ($\sigma = 0.25$ mag):
$$BC_{K_{\rm s}} =2.780+0.475{(K_{\rm s}-A)_0}-0.1337{(K_{\rm
s}-A)_0}^2.$$

Eventually, using the MSX $C$ (12\um) band, the following fit is
obtained as a function of (\ks$-C)_0$ colour in the range $1.0 <
($\ks$-C)_0 < 8.0$ ($\sigma = 0.25$ mag):
$$BC_{K_{\rm s}} =2.037+0.813{(K_{\rm s}-C)_0}-0.1398{(K_{\rm
s}-C)_0}^2.$$
	
The fit to our SiO targets for (\ks$-C)_0 < 2$ mag gives a bolometric
correction $BC_{K_{S_0}}$ smaller (up to 0.1 mag) than those derived
for Hipparcos Mira stars, which have a similar ($K-[12]$)
colour. However, within the uncertainty of the fits the two curves derived
are consistent. 
\tappendix  
\addcontentsline{toc}{section}{Bibliography}

\setlength{\bibsep}{0.5mm}


\chapter{Considerations on the dynamics of maser stars in our Galaxy.}
\chaptermark{Considerations on the kinematics of maser stars}
 
\begin{authorline}
A preliminary report on an
ongoing study by \\
H.J.\ Habing, M.\ Messineo, G.\ van de Ven, M.\ Sevenster and K.H.\ Kuijken\\
\end{authorline}

\section{Introduction}
\defcitealias{messineo02}{Chapter\,II}
\defcitealias{messineo04}{Chapter\,III}
\defcitealias{messineo03_4}{Chapter\,V}

In this thesis we present results from a survey of SiO maser stars
undertaken for the explicit reason to investigate stellar kinematics
and dynamics in the inner regions of the Milky Way. This survey is a
continuation of the thesis project of M. Sevenster in which she looked
at 1612 MHz (18cm) for OH/IR stars with the VLA in the northern
hemisphere and with the ATCA in the southern hemisphere
\citep{sevenster97a,sevenster97b,sevenster01}. The results of
Sevenster's survey have been presented and analysed in several papers
\citep{sevenster99a,sevenster99b, debattista02}.

In discussing the velocities of the SiO maser stars we came across a
question that until now has not been studied in depth: can the \los\
velocities (\vlos)\footnote[1]{We will use the term "line-of-sight
velocity" instead of the more common term "radial velocity" to avoid
confusion: in the kinematic and dynamical discussion "radial velocity"
will be used for the motion along the radius vector from the Galactic
Centre} of SiO and OH maser stars in the forbidden quadrants of the
longitude-velocity, $(l-v)$, diagram be explained by a rotating bar,
and if so, can they be used to constrain the parameters of this bar?
A full answer requires two consecutive steps: a) to find a potential
and orbits in this potential that will fill the required areas of the
$(l-v$) diagram; 2) to find a physical explanation why stars fill
these orbits.  To answer this question we have started numerical
calculations of the orbits in a gravitational potential and compared
the predicted longitudes and line-of-sight velocities with those of
the observations.  Here we describe the first preliminary results of
this ongoing study.

\section{Available evidence for a Galactic bar}
To date, there is clear evidence that the Galactic gravitational field
has a weak bar. This was first proposed by \citet{devaucouleurs64}
based on the analysis of HI gas velocities and it was later confirmed
by, e.g., \citet{binney91} and \citet{bissantz03}.  Other evidence
comes from the asymmetry around $l=0$\degr\ seen in the COBE data
\citep[e.g.][]{blitz91,weiland94,binney97}, star counts
\citep{nakada91,whitelock92}, and microlensing studies
\citep{paczynski94}.

Stellar maser emission provides a unique tool for studying stellar
kinematics.  Unbiased samples of stellar \los\ velocities in the inner
Galaxy are obtained through stellar OH and SiO maser searches, and
these are not affected by interstellar extinction. In this small
chapter we focus on the kinematics and spatial distribution of maser
stars, whose properties have been discussed in the previous chapters
or in the existing literature.

\subsection{Asymmetry in the longitude distribution of maser stars}
The data-set resulting from the OH/IR maser surveys by
\citet{sevenster97a,sevenster97b,sevenster01} uniformly covers the
entire longitude range from $-45$\degr\ to $+45$\degr. It can
therefore be used very well to study a possible symmetry in star
counts around $l=0$\degr.  By plotting $l/|l|$ against $|l|$ in a
cumulative fashion, a deviation from (axial) symmetry in the inner
Galaxy shows up as a non--horizontal section.  In part of the sample
$(-10^\circ < l < 10^\circ)$ an asymmetry was found that could be
explained most naturally by a triaxial (m=2) component in the inner
Galaxy, rather than a m=1 asymmetry \citep{sevenster99}. Here we
present the same figure of the cumulative distribution (Fig.\
\ref{fig:maartje.ps}), but for a larger sample of OH/IR stars. The
distribution is given for OH/IR stars as well as for MSX sources with
AGB colours \citep[as defined in][]{sevenster02} in the same region
$(|b|<3$\degr). For OH/IR stars, there is an over-density at negative
longitudes close to $l=0$\degr. At larger absolute longitudes, the
over-density is at positive longitudes; the slope of the curve is
positive. Around $l=40$\degr, the curve seems to level out, but this
cannot be assessed in more detail as the sample doesn't go out far
enough.

All these aspects are explained by a bar-like density distribution
sampled out to distances well beyond the centre of the Galaxy, up to
the far end of the bar.  For the MSX sources, the initial negative
slope is not seen and the curve starts to rise at lower
longitudes. This may be explained by the same bar-like distribution
sampled out to smaller distances \citep[for a more detailed discussion
see][]{sevenster99}.


Different models used to describe the density distribution of the bar
lead to different values of the semi-major axis ($a$) of the bar and
the viewing angle ($\phi$), the angle between the \los\ and the major
axis. However, they do not vary independently, and possible models
seem to range very roughly from $\phi=20$\degr\ and $a=3$ kpc to
$\phi=50$\degr\ and $a=2$ kpc; this relation is not necessarily
linear. From measurements of the pattern speed
\citep[e.g.][]{debattista02} we only have an upper limit for the
semi--major axis of about 3 kpc, so it will probably be hard to
constrain the viewing angle as described here. However, if some
parameters or even the functional form of the bar density are known
from other arguments, this will limit the degrees of freedom
considerably.

\begin{figure}[h!]
\begin{center}
\rotatebox{-90}{\resizebox{0.8\hsize}{!}{\includegraphics{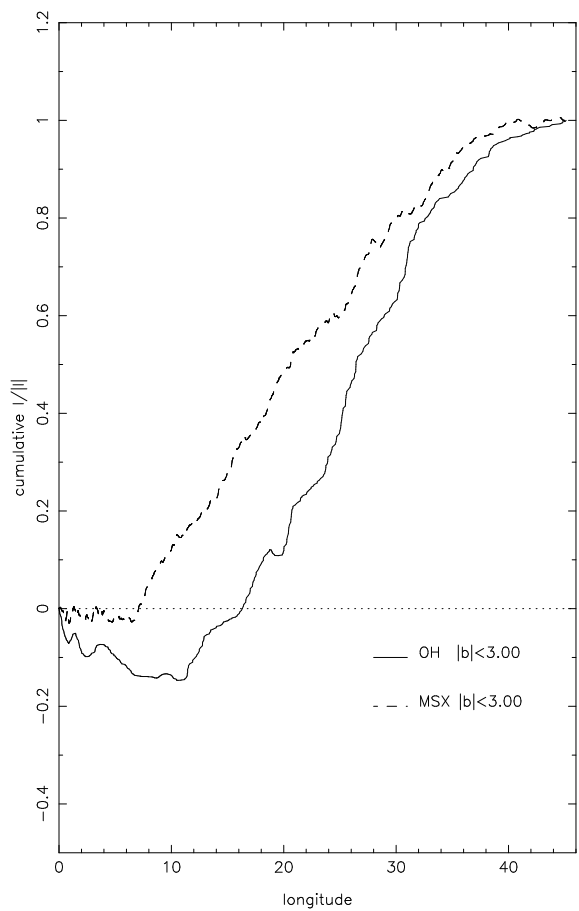}}}
\end{center} \hfill
\caption{\label{fig:maartje.ps} On the horizontal axis the absolute
longitude, $|l|$, is displayed and on the vertical axis the cumulative
sum of $l/|l|$.}
\end{figure}

\subsection{Longitude--velocity diagram}
We will use three observational $(l-v)$ diagrams. In the top panel of
Fig.\ \ref{fig:stars.ps} the CO line spectrum from \citet{dame01} is
shown; the middle panel shows the OH/IR stars observed by Sevenster
and collaborators and the lower panel shows the SiO masers studied in
this thesis.

\begin{figure}[h!]
\begin{center}
\resizebox{0.7\hsize}{!}{\includegraphics{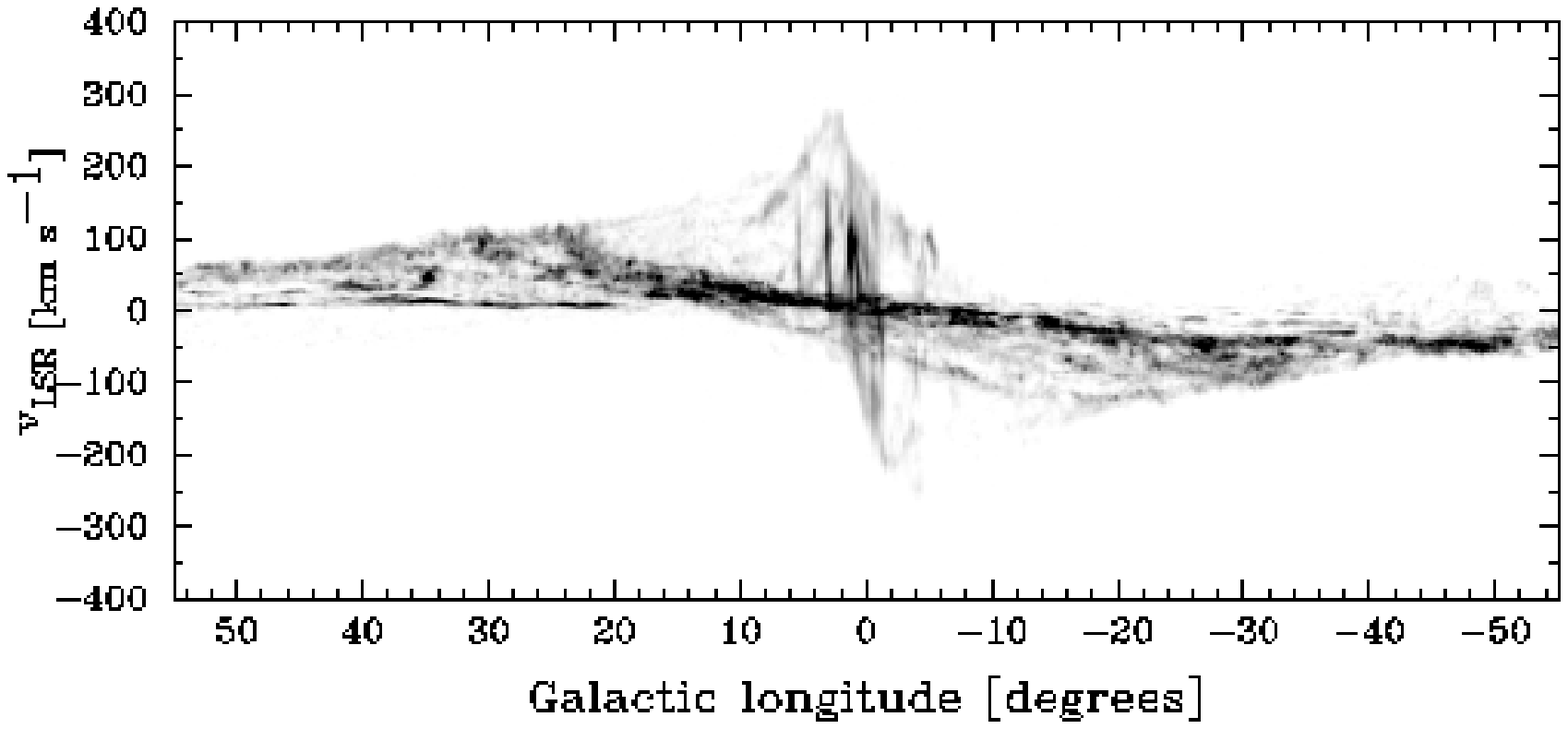}}
\resizebox{0.7\hsize}{!}{\includegraphics{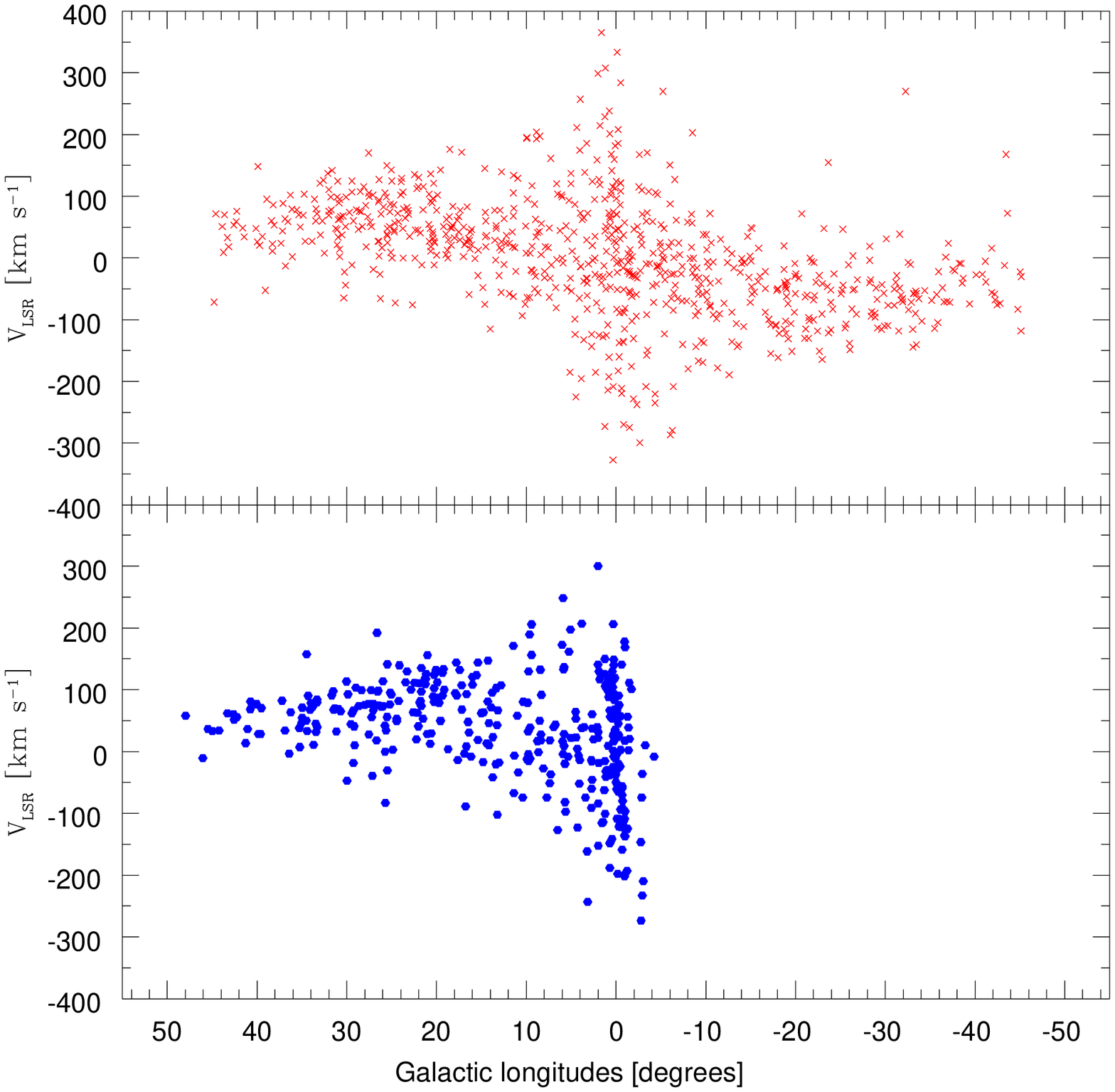}}
\end{center} \hfill
\caption{\label{fig:stars.ps} Longitude--velocity diagrams.  In the
upper panel the CO distribution from \citet{dame01} is shown; the
middle panel shows the sample of 766 OH/IR stars by
\citet{sevenster97a,sevenster97b,sevenster01}; the lower panel shows
the distribution of the 271 SiO masing stars from
\citetalias{messineo02} plus another $\sim 90$ unpublished SiO maser
detections.}
\end{figure}

If a cloud is located at the tangent point to the line of sight of a
circular orbit around the Galactic centre (GC), its velocity vector
will point entirely along the \los; the velocity at the tangent point
is the highest velocity seen along a given \los.  It will be called
the terminal velocity, $v_{\rm term}$. An analytic relation exists between
terminal velocity and longitude: $v_{\rm term}= v_{\rm los}(R_{\rm tang}) -
V_{\odot} {\rm sin}(l)$. This is plotted in Fig.\
\ref{fig:chap6_2.ps}, as two continuous curves, adopting a constant
tangential velocity, $V_\odot$ of 220 \kms, all the way to the GC.
The terminal velocity relation gives a good approximation to the
highest velocity of the CO gas outside longitudes of about $\pm
20$\degr, confirming that the gas in the Galactic plane moves largely
in circular orbits around the GC.  Most of the stars appear also
constrained by the same curve. However, since stars have higher
velocity dispersions than the gas, some stars cross the edge defined
by the gas terminal velocity by up to 60 \kms.

There is a lower density gas region at positive velocities between
longitudes 20\degr\ and almost 0\degr\ and similarly at negative
velocities and longitudes between 0\degr\ and $-20$\degr.  Gas
simulations show that a barred potential explains the gas distribution
well, including the low--density regions \citep{athanassoula99}. Gas
on intersecting orbits will collide with itself and a steady flow
is not possible. The gas looses angular momentum and it will flow into
smaller orbits at lower energies.  Holes seem to appear also in the
stellar $(l-v)$ diagram of the maser stars, but they can not be
explained in the same way.

In the region close to the GC, not only the holes in the gas $(l-v)$
distribution, but also the presence of stars at forbidden velocities
in the two quadrants ($l>0$ and \vlos$< 0$) and ($l<0$ and \vlos $>
0$) show that the assumption of circular orbits breaks down.  In
figure \ref{fig:chap6_2.ps}, where maser stars are over--plotted on
the CO gas, the curve $-V_\odot\sin (l)$ is drawn. At
positive (negative) longitudes the stars above (below) this curve
could move on circular \textit{prograde} orbits. This apparently
applies to all maser stars in the longitude range $345^\circ\, > l
>\,15^\circ$.  However, forbidden velocities at longitudes
$|l|<15$\degr\ are a clear sign of non--circular orbits. Both the OH
and the SiO masers populate these regions within pretty much the same
boundaries.

\begin{figure}[h]
\begin{center}
\resizebox{0.7\hsize}{!}{\includegraphics{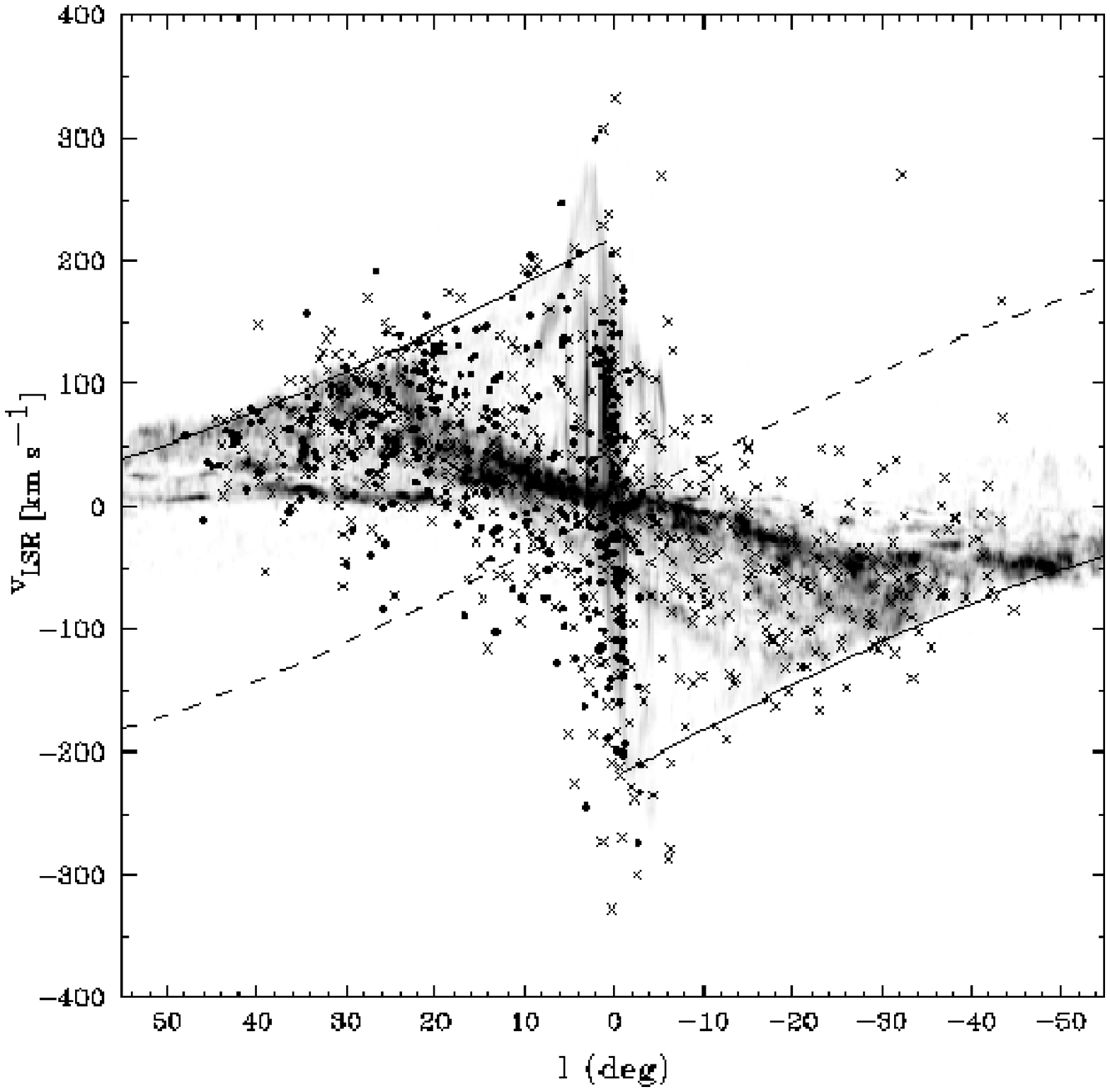}}
\end{center} \hfill
\caption{\label{fig:chap6_2.ps} This $(l-v)$ diagram shows the maser
stars on the CO map.  Symbols are as in Fig.\
\ref{fig:stars.ps}. Continuous lines show the expected terminal
velocities under the assumption of circular orbits. The dashed line is
the curve $-V_\odot\sin (l)$, which corresponds to the opposite of the
velocity of the Sun along the \los. At positive (negative) longitudes
the stars on circular orbits above (below) this curve are moving
prograde in the Galactic standard of rest. The remaining regions are
then the forbidden regions.}
\end{figure}

\subsection{Nuclear Disk}
The distribution of gas in the inner one degree from the Galactic
centre, the circumnuclear zone (CMZ) or nuclear ring, is well
described by a disk with a radius of about 200 pc radius
\citep{morris96,combes91}.  This disc appears in the $(l-v)$ diagram
as a distinct feature: a strong correlation between longitude and
velocity, at longitudes $-1.4<l<1.5$\degr, with maximum velocities of
about 200 \kms.  Since the gas is a collisional medium, intersecting
orbits are forbidden to gas. Dynamical models in a barred potential
predict 2 kinds of orbits: X1, along the major axis of the bar, and X2
perpendicular to the X1. When the X1 start to be self--intersecting
the gas moves inward in to the lower energy X2 orbits.  The transition
between the cusped X1 orbit and the X2 appear as a shock region where
atomic gas is possibly converted in molecular gas.  In principle,
stars can populate the intersecting X1 orbits not accessible to gas
clouds.
  
A strong correlation between longitude and velocity is seen also in
the maser stars.  It was first seen in OH/IR stars within 1\degr\ from
the GC \citep{lindqvist92,sjouwerman98} and clearly appears in our SiO
maser stars (Fig.\ \ref{fig:stars.ps}).  A linear regression fit using
the OH/IR stars gives a slope of $180$ \kms\ pc$^{-1}$
\citep{lindqvist92}, consistent with that derived for SiO targets
\citepalias{messineo03_4}.

As seen in \citetalias{messineo03_4}, nuclear disk stars are highly
obscured (\Aks\ $ > 2$ mag). The use of extinction estimates enabled
us to exclude possible foreground objects and to select individual
stellar members of the nuclear disk.  Furthermore, since SiO targets
are very bright at near--infrared wavelengths \citepalias{messineo04},
they are ideal for follow--up spectroscopic studies of the
nuclear--disk population.

\section{A simple dynamical model}
In an axisymmetric potential, the angular momentum of each star,
$L=r^2\,\partial\phi/\partial t$ is conserved and thus the stars will
keep the same direction (i.e. $\partial\phi/\partial t$ will not
change sign) when a star moves along its orbit. Clockwise moving stars
(i.e. with positive \los\ velocity) seen at a positive longitude will
move away from us (after correction for LSR motion). At negative
longitudes these clockwise rotating stars are all coming toward us
(i.e. \vlos $< 0$). In the $(l-v)$--diagram counter rotating stars
will appear \textit{only} in the "forbidden quadrants" ($l>0$ and
\vlos$< 0$) and ($l<0$ and \vlos $> 0$).

Kinematic deviations from what would be expected in an axisymmetric
potential were found by \citet{sevenster00}, when modelling the
underlying distribution function (DF). The observed \los\ dispersions
for instance could not be explained by an axisymmetric model with
$\sigma_R$ equal to $\sigma_z$. Moreover, to explain the stars in the
forbidden quadrants, isotropic components had to be invoked.

A barred potential explains the Galactic kinematics well.  Several
N--body dynamical models of the barred Milky Way exist
\citep[e.g.][]{fux97}.  They can be compared with observations
qualitatively, but do not allow a real fit to the data. The latter is
possible with the few dynamical models that are built by superposition
of either (analytical) DF components \citep[]{sevenster00} or of
numerically integrated orbits \citep[e.g.][]{hafner00,zhao96}. The
latter method is also known as Schwarzschild's method and is more
general than the DF method, as it does not require a priori
assumptions about the form of the DF, which is even more complicated
by the fact that $E_J$ is the only known integral of motion.

For the population of maser stars, no realistic dynamical model with a
barred potential has yet been constructed. Therefore we started to
calculate orbits in a barred potential with the ultimate goal to
predict the observed distribution of maser points in the $(l-v)$
diagram. This is work under way, but the first results are
promising. In the following we briefly sketch the various assumptions
and steps made for our calculation and some first results.

\subsection{Geometry}
We adopt a Cartesian coordinate system $(x,y,z)$, corotating with the
bar--like bulge at a (clockwise) angular speed of $\Omega_b = 60$
\kmskpc \citep{debattista02}. The origin of the coordinate system is
at the Galactic centre, the $x$--axis is aligned with the major axis
of the bar--like bulge and the $y$--axis with its minor axis.  The Sun
is assumed to lie at a distance $R_\odot$ from the Galactic centre, in
the Galactic plane. The Sun-GC line makes an angle $\phi$ w.r.t. the
long--axis of the bulge. We fix the Sun's distance to the Galactic
centre at $R_\odot=8$ kpc and its (clockwise) azimuthal velocity to
$V_\odot=200$ \kms. For a given mass model, this leaves two free
parameters: the angular speed $\Omega_b$ of the bar-like bulge and the
angle $\phi$ of the Sun w.r.t. to long--axis of this bulge.

\subsection{Equations of motion}
We calculate numerically the orbits of stars in a frame of reference
that is rotating in the Galactic plane at an angular speed,
$\vec{\Omega_b}=(0,0,\Omega_b)$, solving the equation:

\begin{equation}
  \label{eq:eqofmotion}
  \ddot{\mathbf{r}} = -\nabla\Phi_{\mathrm{eff}} -
  2\left(\mathbf{\Omega_b}\times\dot{\mathbf{r}}\right).
\end{equation}

In the right--hand side of this equation the first term is the
acceleration induced by the effective gravitational potential; the
second term is the Coriolis acceleration. The Coriolis acceleration
introduces a dependence of the acceleration in the $x$--direction on
the velocity in the $y$--direction and vice versa. For further explanations
see \citet{binney87}. 

\subsection{Gravitational potential}
As a first qualitative study of the stellar dynamics in our Milky
Way we consider the non--axisymmetric planar logarithmic potential
 \citep{binney87} 

\begin{equation}
  \label{eq:logpot}
  \Phi(x,y) = \frac12 v_0^2 \ln \left(R_C^2+x^2+
  \frac{y^2}{q^2}\right),
\end{equation}

with constant velocity $v_0$, core radius $R_C$ and axial ratio $q \le
1$.  Near the centre ($R\equiv\sqrt{x^2+y^2} \ll R_C$) the logarithmic
potential approximates that of a two-dimensional harmonic oscillator,
such that the corresponding central density is nearly
homogeneous. Going outwards the rotation curve rapidly flattens to
$v_c \sim v_0$.  The constant axial ratio implies that the influence
of the non--axisymmetry is similar at all radii. Although at larger
radii this is not realistic for our Galaxy, the orbits calculated in
this rotating potential are still representative as they become nearly
circular beyond the corotation radius.

\subsection{First results}
Taking the above logarithmic potential with $v_0=200$ \kms, $R_C=0.14$
kpc and $q=0.9$, we calculated a set of closed orbits by numerically
solving Eq. (\ref{eq:eqofmotion}) with $\Omega_b=60$ \kmskpc.  We used a
fifth order Runge-Kutta algorithm as described in "Numerical Recipes"
\citep{press92}. A representative example of these orbits in the $XY$-plane
and in the ($l-v$) diagram, is shown in Fig.\ \ref{fig:orbit.ps}.

We have just started these simulations and we need to further compare
our model to observations in a quantitative way.  However, as already
proposed by \citet{binney91} for the gas, it clearly appears from the
simple superposition of the orbits in the $(l-v)$ diagram that orbits
from the X1 and X2 families can explain the observed forbidden stellar
velocities.

\begin{figure}[h!]
\begin{center}
\resizebox{0.49\hsize}{!}{\includegraphics{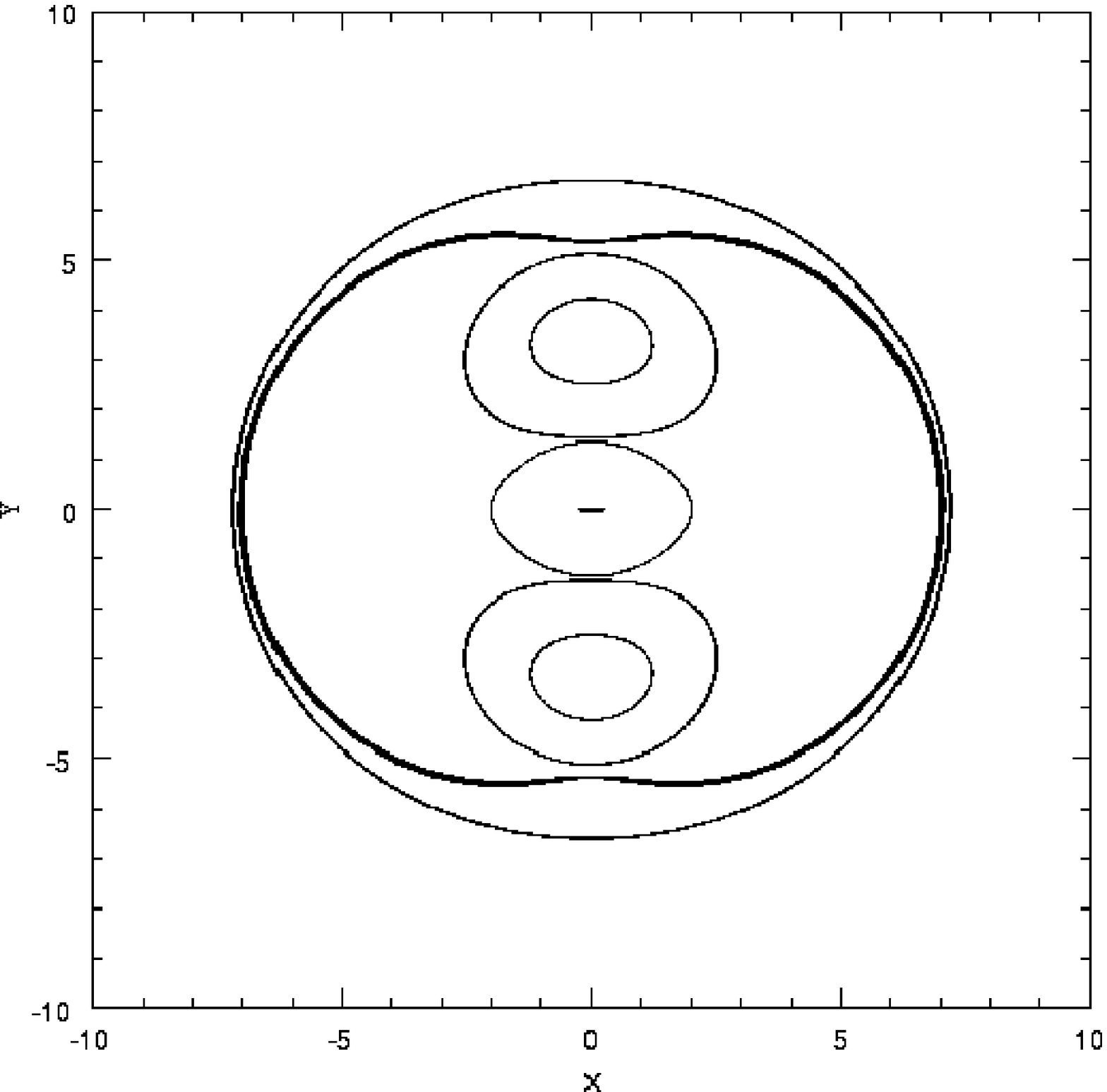}}
\resizebox{0.49\hsize}{!}{\includegraphics{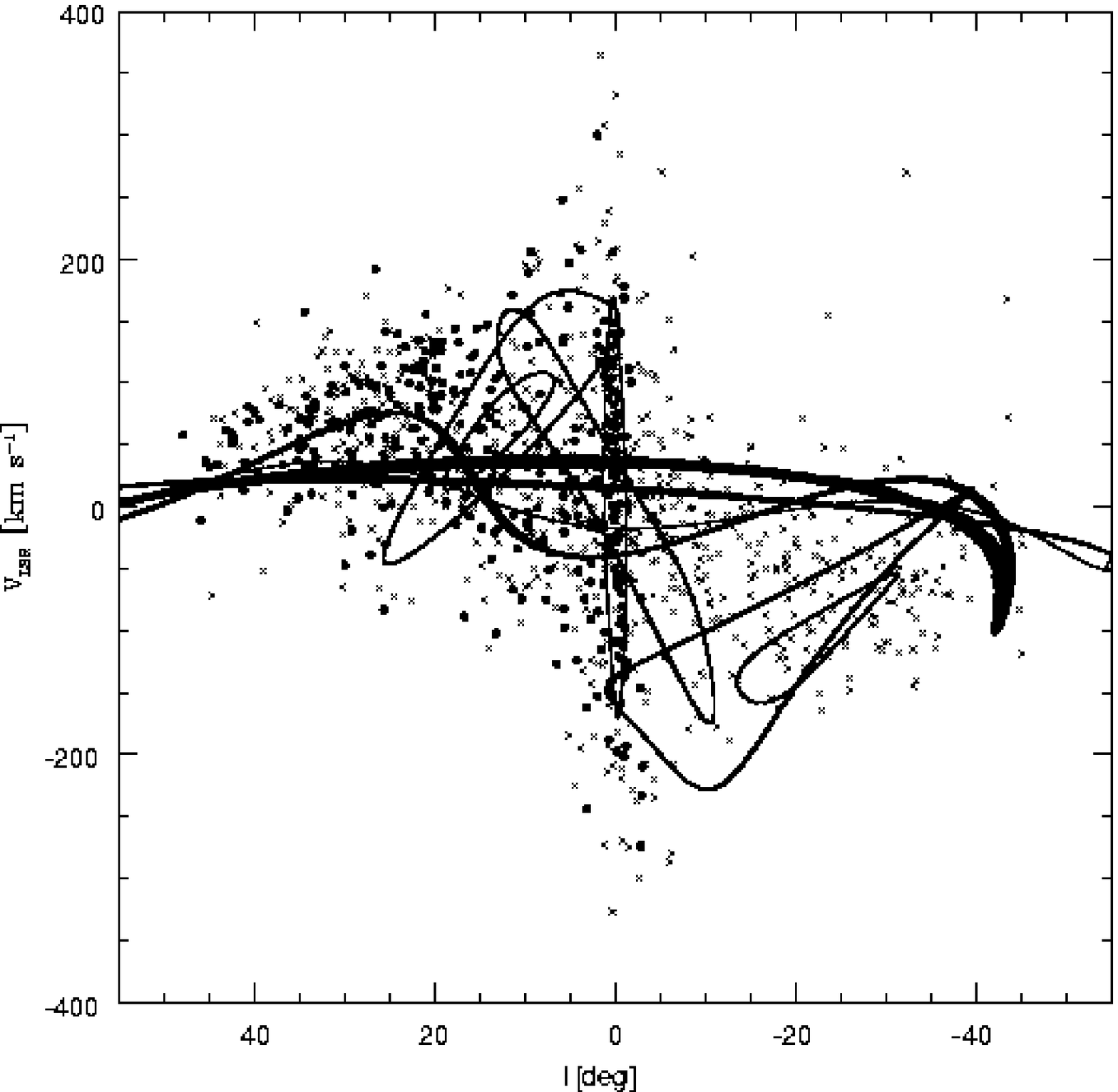}}
\end{center} \hfill
\caption{\label{fig:orbit.ps} Left panel: $(x,y)$ plot of an example
of closed orbits calculated in a rotating logarithmic potential (see
text).  Right panel: the same orbits projected in a ($l-v$) diagram
assuming $\phi = 45$\degr. }
\end{figure}

\section{Summary and future plans}

In summary, the SiO and OH maser stars have a similar distribution in
the velocity-longitude diagram.  Their forbidden velocities are
difficult to understand in an axisymmetric potential, but they can be
understood in a rotating barred potential.

The disk maser stars beyond longitudes $|l|>15$\degr\ are probably
moving on loop orbits. Outside of the bar region the potential must be
close to axisymmetry as observed in the terminal velocities of the
gas.
 
For stars within longitudes $|l|<15$\degr, their forbidden velocities
can be explained by X1/X2 orbits.

The most realistic potential currently available is that obtained by
the Basel group \citep{englmaier99,bissantz03}. This is based on a
mass model of the Milky Way derived from the dust corrected COBE
maps. We plan to use the Basel potential to calculate a library of
orbits and to fit these orbits to the available kinematics of maser
stars. The fit can be done by maximising the likelihood of the \los\
velocity distribution at the (discrete) observations.

\begin{acknowledgements} HH is grateful to the A\&A Office in Paris
for its kind support.  Dr. Englmaier and Dr. Gerhard kindly provided
us with their model of the Galactic potential.  
\end{acknowledgements}

\setlength{\bibsep}{0.5mm}


\chapter{The ISOGAL survey and the completeness analysis}
\chaptermark{The ISOGAL survey and the completeness analysis}

\section{Introduction} 

The ISOGAL project is an ISO infrared survey of specific regions in
the Galactic Plane, which were selected to provide information on
Galactic structure, the stellar populations and mass-loss, and the
recent star formation history of the inner disk and Bulge of the
Galaxy.  Several (about 25) scientific papers have been published
based on the ISOGAL data. They present studies of the Galactic
structure, an analysis of the complete AGB population, and studies of
infrared dark clouds and young stellar objects \citep{omont03}.

The survey was performed at 7 and 15~$\mu$m with ISOCAM, covering 16
square degrees with a spatial resolution of 3-6\arcsec and a
sensitivity of 10-20 mJy, two orders of magnitude deeper than IRAS at
12\um, and a factor 10 deeper than the MSX A band (8\um).

The 7 and 15 \um\ ISOGAL data were combined with the $I$, $J$, $K_{s}$
(effective wavelengths 0.79, 1.22 and 2.14~$\mu$m) ground-based data
from the DENIS survey, resulting in a 5-wavelength catalogue of point
sources. The combination of mid- and near-infrared measurements
permits a determination of the nature of the individual sources and of
the interstellar extinction towards them.

A complete overview of the first scientific results from ISOGAL data
is given by \citet{omont03}, while the description of the point source
catalogue is given in an Explanatory Supplement \citep{schuller03}.

In the present chapter I will briefly describe the survey and than I
will move to the description of the ISOGAL completeness analysis
(Sect.\ 1.4 below), to which I contributed significantly.

\section{Observations}

\begin{figure*}[htbp]
\begin{center}
\resizebox{12cm}{!}{ \rotatebox{0}{\includegraphics{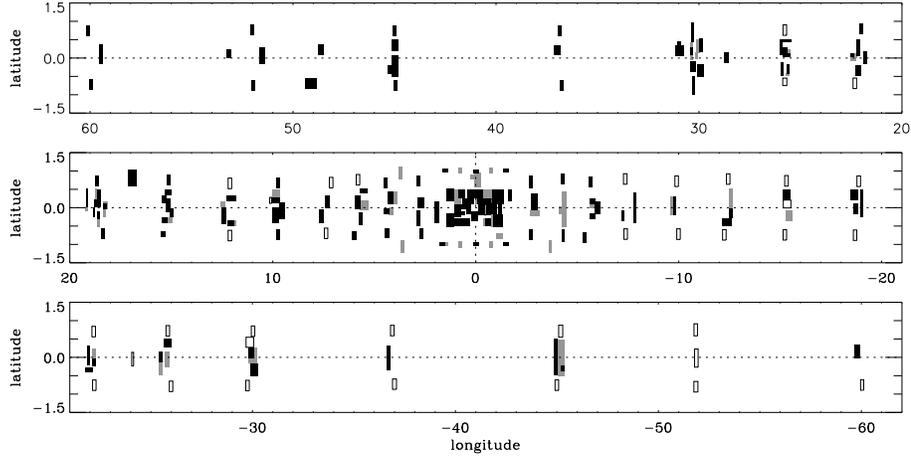}}}
\caption{Galactic map of ISOGAL fields with $\vert \ell
\vert$~$<$~60$^\circ$. Black, grey and open boxes show fields which
have been observed at both 7 \& 15~$\mu$m, at 7~$\mu$m only and at
15~$\mu$m only, respectively. Twenty--one additional northern fields
(not displayed) were also observed, at $\ell$~$\approx$~+68$^\circ$,
+75$^\circ$, +90$^\circ$, +98$^\circ$, +105$^\circ$, +110$^\circ$,
+134$^\circ$, +136$^\circ$, \& +138$^\circ$. Figure adapted from
\citet{schuller03}.}
\end{center}
\label{figure1}
\end{figure*}

The ISOGAL observational program --250 hours of observing time-- is one of
the largest ISO programs.  The ISOCAM observations were taken from
January 1996 to April 1998, i.e., over the whole ISO mission.

The observed fields ($\sim$16 deg$^2$) are distributed along the inner
Galactic Disk, mostly within $\vert \ell \vert$~$<$~30$^\circ$, $\vert
{\it b} \vert$~$<$~1$^\circ$, as shown in Figure 1.

Detailed information on the observation parameters and on the field
positions is available in the ISOGAL Explanatory Supplement
\citep{schuller03} and on the ISOGAL web server {\tt
www-isogal.iap.fr/}.

Most of the observations were performed with the broad filters $LW2$
and $LW3$ and a pixel scale of 6\arcsec. A few regions around the
Galactic centre were observed with the narrow filters $LW5$ or $LW6$,
and $LW9$ and a pixel scale of 3\arcsec, to reduce the effects of
bright sources that would saturate the detector (thus moving the
saturation limit from $Flux_{12}>6$ Jy to $Flux_{12}>20$ Jy). A list
of the ISOCAM filters used for the ISOGAL survey is given in Table
\ref{table:filters}.

\begin{table}[htbp]\normalsize
\begin{minipage}{\linewidth}
\caption[]{\label{table:filters}ISOCAM filters used for ISOGAL:
reference wavelengths and bandwidths, zero point magnitudes and flux
densities, and total observed area.  Table adapted from
\citet{schuller03}.}
\begin{center}
\begin{tabular}{|c|c|c|cc|c|}
\hline
Filter & $\lambda_{ref}$ & $\Delta\lambda$ &
ZP\footnote{The magnitude of a source with a flux
density $F_{\nu}$ expressed in $mJy$ is given by 
$mag=ZP-2.5 \times log (F_{\nu}$)} & $F_{mag=0}$ & Area \\
& [$\mu m$] & [$\mu m$] & [mag] & [Jy] & [deg$^2$] \\
\hline
LW2 &  6.7 & 3.5 & 12.39 & 90.36 & 9.17 \\
LW5 &  6.8 & 0.5 & 12.28 & 81.66 & 0.64 \\
LW6 &  7.7 & 1.5 & 12.02 & 64.27 & 2.97 \\
\hline
LW3 & 14.3 & 6.0 & 10.74 & 19.77 & 9.92 \\
LW9 & 14.9 & 2.0 & 10.62 & 17.70 & 3.53 \\
\hline
\end{tabular}
\end{center}
\end{minipage}
\end{table}

The observations, performed in raster mode ($\sim$0.1 deg$^2$), were
oriented in Galactic coordinates. At each raster position 19 basic
ISOCAM frames (32~x~32 pixels) were taken, resulting in a total
integration time of 21 s per raster position. The raster steps were
typically 90\arcsec\ in one direction and 150\arcsec\ in the other
direction, and each sky point was observed for a maximum of 4 times,
with an average of 1.5.  Only 384 of the 463 raster positions were
used for the production of the first ISOGAL point source catalogue,
because only one raster was used in case of overlapping areas to avoid
redundancy. The total number of ISOGAL fields (rectangular area of the
sky whose edges are aligned with the galactic axes and observed by
ISOGAL) is 263. They can be divided in 43 fields (FA) only observed at
7 \um, 57 fields (FB) only observed at 15 \um, and 163 fields (FC)
observed at both 7 and 15\um.

Systematic cross-identification with the near-infrared $I,J,K_{\rm
s}$ sources of the DENIS survey was an integral part of the ISOGAL
program and special DENIS observations were performed for this purpose
\citep{simon04}. DENIS data were available for 95\% of the fields
surveyed with ISOCAM.

\section{Data processing and analysis}  
Data reduction was performed with standard procedures of the CAM
Interactive Analysis Package (CIA version 3). A sophisticated pipeline
was developed for the ISOGAL data \citep{schuller03}, which involves
more steps than the standard treatment of ISOCAM data \citep[see
ISOCAM Handbook,][]{blommaert03}. This was necessary because of the
extreme conditions of the ISOGAL observations.  In addition to the
usual problems, i.e. glitches, dead pixels and the time-dependent
behaviour of the detectors, one needs also to consider bright
background emission, crowding, high spatial density of bright sources
(which causes pixel-memory effects), and the short integration time.

The point sources were extracted using a dedicated PSF fitting routine
\citep{schuller03}.

The completeness of point source extraction has been systematically
addressed through retrieval of artificial sources, which is described
in the next section.

\section{\label{artif}Artificial sources}
Synthetically reproducing the complete process of photometric
measurements is the only way to properly characterise all undesired
effects associated with observations in a crowded field.

An artificial star experiment consists of adding artificial stars to
 ISOGAL images, and re-extracting all point sources with the same
 pipeline as the one used to generate the ISOGAL catalogue.  The
 analysis of  input magnitudes of  artificial sources with those
 in output enables one to characterise the effects of crowding on the
 photometric quality and the completeness of the extracted point
 source catalogue.

Artificial star experiments were conducted on ISOGAL images following
a procedure similar to that applied by \citet{bellazzini02}:
\begin{itemize}
\item The magnitude of artificial stars was randomly extracted from
the observed luminosity function.
\item 
The goal of the procedure is to study the effects of
crowding. Therefore, it is of primary importance that the artificial
stars do not interfere with each other.  The interference between
artificial stars would, in fact, change the actual crowding and affect
the results of the artificial experiment study.  To avoid this serious
bias, one can divide the image (or raster in the ISOGAL case) into
grids of cells of known width (20 pixels). One artificial star is
randomly positioned in each cell, avoiding the border of the cell in
order to control the minimum distance between adjacent artificial
stars.
\item The stars were simulated using the point-spread-function (PSF)
determined directly from the average ISOGAL data corresponding to the
observational setup (filter, pixel scale).    A new image was then
built by adding the artificial stars and their Poisson photon noise
into the original raster image.
\item The measurements process was performed in 
the same way as the original measures.
\item The output magnitudes were recorded, as well as the positions of the lost
stars.
\item To generate a significant number of artificial stars, for each
image, the whole process was repeated between 100 and $\sim$300 times,
depending on the source density and image size.  A total of
5$\times$10$^{3}$ to 4$\times$10$^{4}$ sources were generated.
\end{itemize}

\begin{figure*}[htbp]
\begin{center}
\resizebox{12cm}{!}{ \rotatebox{0}{\includegraphics{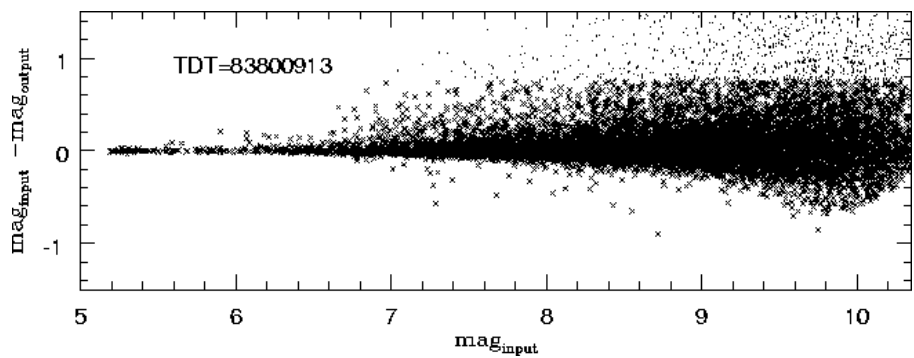}}}
\caption{\label{figure:diffmag.ps} Differences between the input and
output magnitudes of the 19250 artificial stars vs. input magnitudes
for the raster TDT=83600913. If the difference is larger then 0.75 mag
the artificial star is considered lost.}
\end{center}
\label{figure1}
\end{figure*}

As an example, figure \ref{figure:diffmag.ps} shows the differences
between input and output magnitudes, $mag_{input}-mag_{output}$,
obtained for the raster TDT=83600913 (filter=$LW2$, pixel scale
6\arcsec), which is centred at longitudes 1.37\degr\ and latitude
$-2.63$\degr, covers a region of $0.22\times0.24$\degr, and has a
density of 6660 sources per deg$^{-2}$. A number of 55 artificial star were
added each run and a total of 350 runs were performed.

The distribution of the difference is not symmetric. There is a strong
concentration at zero, but also a positive tail of stars.  This means
that the output magnitudes are brighter then the input ones and this is due to
the blending between artificial stars and real stars.  When an
artificial star is blended with a fainter source its output magnitude
will be brighter than its input magnitude because of the flux
contributed by the fainter blended source. Artificial sources having
an output magnitude  $+$0.75 mag brighter than their input magnitude 
were considered lost. In fact, if at the position of the artificial
star one measures a point source more than 0.75 mag brighter than the
magnitude of the input artificial star, this means that the artificial
star falls on a brighter real source and in this case the star
actually recovered is the real one.

Artificial star simulations were conducted on 35 images (total area
$\sim$2~deg$^2$) selected to have all possible observational setups
and different crowding levels (the source density ranges from 0.0017
to 0.03 sources per pixel). Artificial star experiments were used to
evaluate both random and systematic photometric errors due to
crowding, as well as the completeness level of the extraction as a
function of source density.

\begin{figure*}[htbp]
\begin{center}
\resizebox{12cm}{!}{\rotatebox{0}{\includegraphics{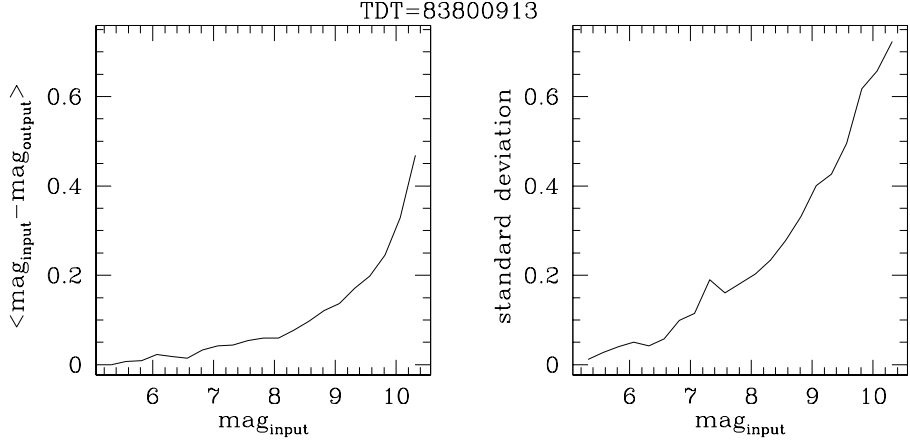}}}
\caption{\label{fig:std.ps}{\bf Left-panel:} Mean differences between
the input and recovered magnitudes per bin of input magnitude,
relative to the 19250 artificial stars simulated for the raster
TDT=83600913.  {\bf Right-panel:} Standard deviation of the
differences between the input and recovered magnitudes as a function
of the input magnitude.}
\end{center}
\label{figure1}
\end{figure*}

Output magnitudes were generally found to be brighter than input
magnitudes. This bias is very small for bright stars, but can reach
0.3 magnitude for the faintest ones in the densest fields, where the
probability of blending with real stars is higher (see
Fig. \ref{fig:std.ps}).

The completeness of the extraction was quantified by analysing for
each simulation the fraction of retrieved simulated sources as a
function of input magnitude. A smooth curve appears which drops at
faint magnitudes. The magnitude where this fraction becomes less than
50\% depends on the density of the field.  The point source catalogues
extracted from the various ISOGAL rasters were found at least 50\%
complete down to the faintest end in fields with low stellar density,
but this was not the case in denser regions.

\begin{table*} 
\begin{center}
\caption[]{Sensitivities\footnotemark[1] at 7 and 15~$\mu$m for
typical ISOGAL conditions \label{table:sens}.  Table adapted from
\citet{omont03}.}
\vspace{0cm}
\begin{tabular}{@{\extracolsep{-.09in}}l c c c c c c c c}\\
\hline
Region\footnotemark[2] & Source & Background & Pixel & Filter & \multicolumn{2}{c}{7
$\mu m$} & \multicolumn{2}{c}{15~$\mu m$}  \\ 
 & density &       &  &   & mag & flux (mJy) & mag & flux (mJy)  \\
\hline
A & low       & weak   & 6\arcsec & broad  &10& 9&8.7& 7 \\
B & high      &moderate& 6\arcsec & broad  & 9&22&8  &12 \\
C & very high & strong & 3\arcsec & narrow & 8.4&35&7  &30 \\
D & high   &very strong& 6\arcsec & narrow & 7.7 & 55 & 6.5  &  45 \\
\hline
\label{table1}
\end{tabular}
\end{center}
\vspace*{-0.5cm}
\begin{small}
\noindent 
\footnotemark[1] Sensitivity limits of ISOGAL sources published in PSC
{\it Version 1}, corresponding approximately to detection completeness
of 50\% (Schuller et al. 2003).\\ 
\footnotemark[2] Typical regions:\\
A Lowest density Bulge fields, $\vert${\it b}$\vert$ $\ge$ 2$^\circ$\\
B Standard Disk fields, $\vert${\it b}$\vert$ $<$ 0.5$^\circ$, $\vert$$\ell$$\vert$ $\le$ 30$^\circ$\\
C Central Bulge/Disk fields, $\vert${\it b}$\vert$ $<$ 0.3$^\circ$, $\vert$$\ell$$\vert$ $\le$ 1$^\circ$\\
D Most active star formation regions such as M16, W51. 
\end{small}  
\end{table*}

For each observational setup (combination of pixel scale and filter) a
relation between the estimated 50\% completeness limit and the field
source density was derived and used to define the limiting magnitude
of each ISOGAL observation, corresponding to the faintest sources that
were included in the published catalogue.  The completeness limit
depends on the source density, on the intensity and the structure of
the local diffuse background, and on the filter.  The
sensitivities reached at 7 and 15~$\mu$m for standard ISOGAL
conditions are summarised in Table \ref{table:sens}.

The distribution of limiting magnitudes, for all ISOGAL observations
is shown in Fig.~\ref{distrib_mag_lim}. Since most observations were
done with the broad $LW2$ and $LW3$ filters, these histograms show
that the typical reached sensitivity is around 20~mJy at 7~$\mu$m and
12~mJy at 15~$\mu$m.

\begin{figure}[htbp] \begin{center} \resizebox{8.cm}{!}{
\rotatebox{0}{\includegraphics{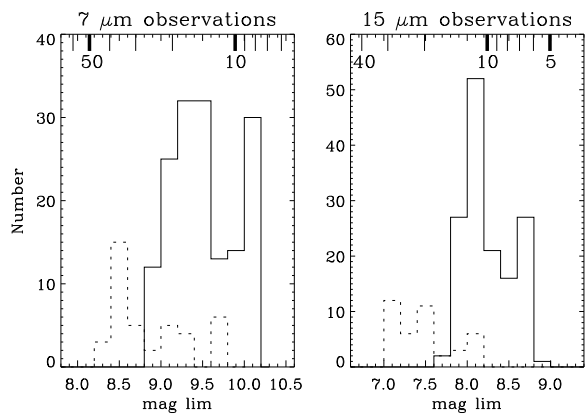}}} 
\caption{Distribution of the magnitudes at which the catalogues become
incomplete at the 50\% level for the broad filters LW2 and LW3 (full
lines), and for the narrow filters (dotted lines).  The logarithmic
scales at the top of each panel show the corresponding flux densities
in mJy for LW2 and LW3. A small correction has to be applied for the
corresponding flux densities with narrow filters. Figure adapted from
\citet{schuller03}.} \label{distrib_mag_lim} \end{center}
\end{figure}

About $\approx$25\% of extracted point sources fall below these
magnitude limits.  Analysing the quality flags of discarded sources,
the photometric cut is far more severe for moderate quality sources
than for good quality ones.

The completeness findings have been complemented and checked by the
results of several repeated observations (in one case with 3$\arcsec$
pixels, rather than the typical 6$\arcsec$ pixels, and hence with
greatly reduced crowding), and by comparison with DENIS (or 7~$\mu$m)
red giant source counts.

\section{Concluding remarks} 
In summary the ISOGAL PSC (version 1.0) contains 106 150 sources.  It
gives I, J, K$_{\rm s}$, [7], [15] magnitudes, at five wavelengths
(0.8, 1.25, 2.15, 7 \& 15~$\mu$m); DENIS associations (I,J,K$_{\rm
s}$) are given when available.  About half of the sources have
7-15~$\mu$m associations and 78\% have DENIS associations.  Quality
flags are provided for each source at each wavelength, as well as for
source associations, and only sources with a reasonable quality and
with a magnitude above the 50\% completeness limit are included in the
catalogue \citep{schuller03}.

\begin{acknowledgements}   I am grateful to all members of the ISOGAL
consortium, in particular to Prof. Alain Omont from the Institute
d'Astrophisique de Paris, P.I. of the survey, and to Frederic Schuller
for their collaboration.  ``Grazie molto'' to Paolo Montegriffo for
the fruitful discussions on the completeness
analysis. \end{acknowledgements}

\setlength{\bibsep}{0.5mm}

\addcontentsline{toc}{section}{Bibliography}

\begin{nedervat}  
Op een heldere avond kunnen we aan de hemel een witte, op sommige
plekken onderbroken band van licht tegenkomen. Wat we zien zijn
miljoenen sterren die samen de schijf van ons eigen sterrenstelsel, de
Melkweg, vormen. De Melkweg is opgebouwd uit een enorme bolvormige
halo die grotendeels uit donkere materie bestaat, verder een grote,
vlakke schijf van gas en sterren, en in het midden van de Melkweg
bevindt zich een verdikking in de schijf, de lens (zie
fig. 1). Sterren in de schijf van de Melkweg draaien rond het centrum,
waarvan we nu weten dat het een zwart gat bevat. Hoe dichter een ster
bij het centrum staat, des te korter is de tijd die het de ster kost
om {\'e}{\'e}n keer rond te gaan. De zon bevindt zich op een
afstand van ongeveer 26.000 lichtjaren van het centrum van de Melkweg
en beweegt met een snelheid van zo'n 220 km per seconde rond het
centrum. Zelfs met deze snelheid kost het ons zonnestelsel rond de 226
miljoen jaar om {\'e}{\'e}n keer rond het centrum van de Melkweg te
gaan.\\
Het bestuderen van onze Melkweg is belangrijk voor ons
begrip van de vorm, het ontstaan en de ontwikkeling van
sterrenstelsels in het algemeen. De structuur en dynamica van het gas
en de sterren die deel uit maken van de Melkweg kunnen door hun
relatieve nabijheid (vergeleken met andere sterrenstelsels) in groot
detail worden bestudeerd. Maar omdat wij ons ook zelf in de Melkweg
bevinden krijgen we te maken met projectie-effecten die het ons lastig
maken om de grote schaal structuur van de Melkweg te achterhalen.

\vskip 0.5cm
\noindent {\bf De Melkweg: een spiraalstelsel met een balk in het centrum}
\vskip 0.2cm

De sterren en het gas in de schijf van de Melkweg liggen gegroepeerd
in vier spiraalarmen. Wijzelf bevinden ons aan de rand van de Orion
spiraalarm, tussen de Sagittarius arm en de Perseus arm in.\\

Over het algemeen is de lens in het centrale deel van een
sterrenstelsel ofwel bolvormig, net als een bal, of uitgerekt tot een
balk, waarbij de spiraalarmen aan de uiteinden van de balk
ontstaan. De beweging van zowel gas als sterren geeft aan dat zich in
het centrum van de Melkweg een balk bevindt. Ofschoon we tegenwoordig
overtuigd zijn van het bestaan van deze balk in het centrum van de
Melkweg, weten we nog nauwelijks iets af van zijn eigenschappen, zoals
zijn lengte en dikte, of de hoek waaronder we de balk zien.

\begin{figure}[t]
\begin{center}
\resizebox{0.6\hsize}{!}{\includegraphics{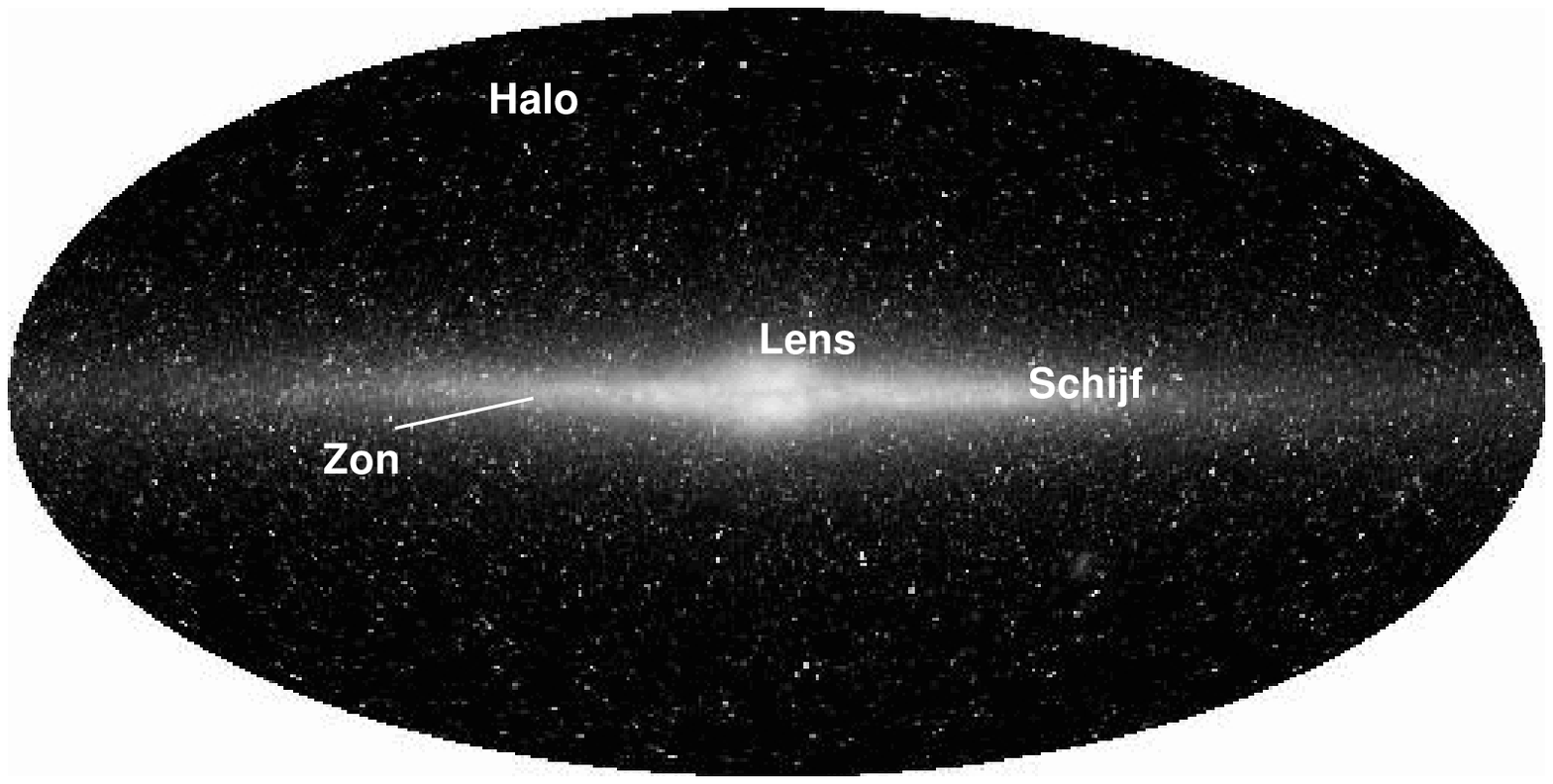}} 
\caption{\label{fig:mw.ps} Deze afbeelding gemaakt met de COBE
satelliet laat zien hoe onze Melkweg eruit ziet in het nabij-infrarode
deel van het spectrum, waar het meeste licht afkomstig is van
ge{\"e}volueerde sterren genaamd rode reuzen. De extinctie door
interstellair stof is veel zwakker op infrarode golflengten dan op
golflengten in het zichtbare deel van het spectrum. Door nu juist op
deze golflengten te kijken kunnen we ons dus een veel beter beeld
vormen van de Melkweg.  }
\end{center} \hfill
\end{figure}

\vskip 0.5cm
\noindent {\bf Interstellaire extinctie}
\vskip 0.2cm

Een van de grootste problemen die we tegenkomen bij het bestuderen van
de structuur van onze Melkweg is de verduistering door interstellair
stof die sterker wordt naarmate we dichter bij het centrum van de
Melkweg komen. Interstellair stof dat zich langs de gezichtslijn
bevindt absorbeert en verstrooit het licht afkomstig van de sterren
die we waarnemen. Dit effect genaamd extinctie (of 'uitdoving') zwakt
de helderheid van de ster af en het maakt het spectrum van het
sterlicht roder. Het sterspectrum laat zien hoe het licht afkomstig
van de ster verdeeld is over de verschillende golflengten. Voor licht
in het zichtbare deel van het spectrum is de extinctie zeer hoog, maar
voor langere golflengten, zoals in het nabije- en mid-infrarode deel
van het spectrum, is de extinctie minder sterk (zie fig. 2). Op
infrarode golflengten kunnen we dus een veel duidelijker beeld krijgen
van onze Melkweg. In de afgelopen tien jaar zijn verschillende
onderzoeken (surveys) uitgevoerd op infrarode golflengten om de
structuur en de vormingsgeschiedenis van onze Melkweg te
achterhalen. Deze data bevat een schat aan informatie over de opbouw
van sterpopulaties die er nog op wacht om volledig geanalyseerd te
worden.\\

Dit promotie-onderzoek gaat over de bepaling van de structuur en
samenstelling van de binnenste delen van onze Melkweg, daarbij gebruik
makend van ge{\"e}volueerde sterren (sterren die al het grootste deel
van hun leven achter de rug hebben) om het zwaartekrachtsveld in dit
deel van de Melkweg in kaart te brengen. Het zwaartekrachtsveld kan
ons namelijk meer vertellen over de verdeling van massa in de Melkweg.
\begin{figure}[t]
\begin{center}
\resizebox{0.32\hsize}{!}{\includegraphics{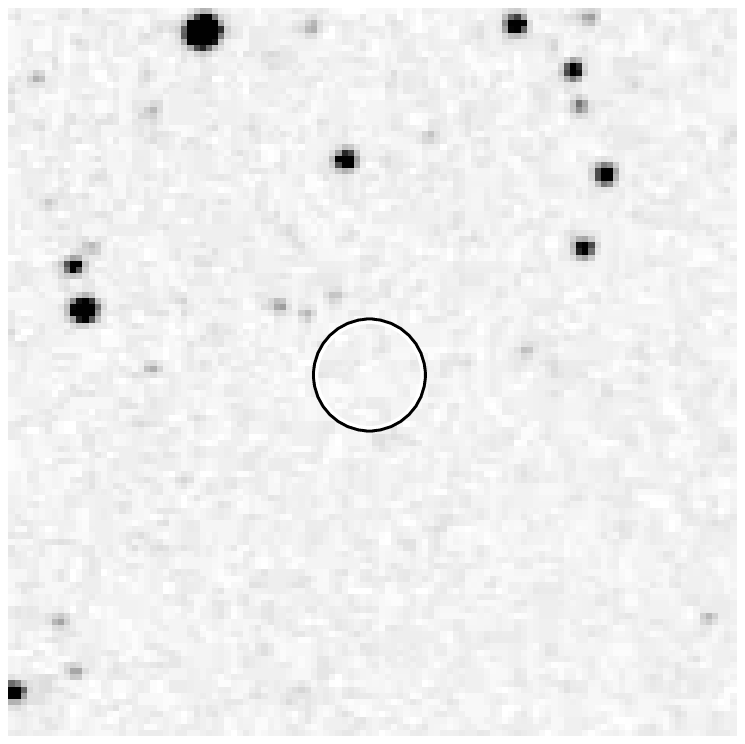}} 
\resizebox{0.32\hsize}{!}{\includegraphics{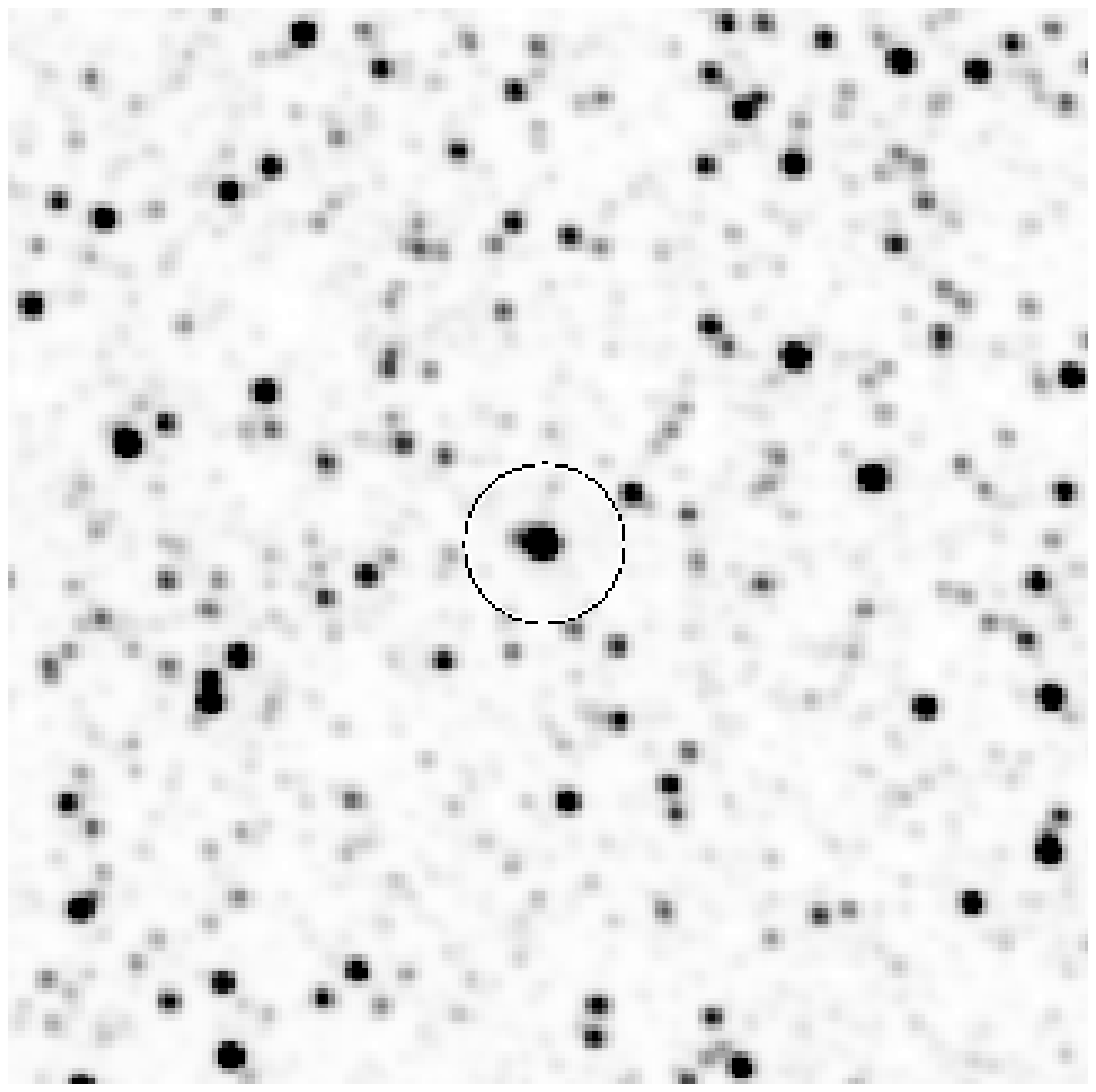}} 
\resizebox{0.32\hsize}{!}{\includegraphics{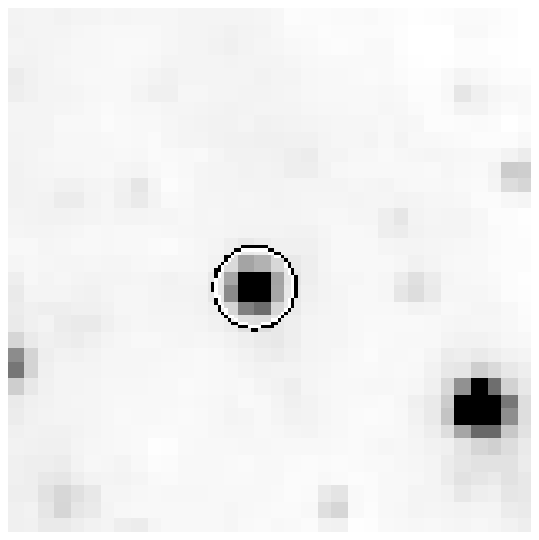}} 
\caption{\label{fig:tris.ps} Op verschillende golflengten ziet de
hemel er heel anders uit! Deze figuur laat hetzelfde gebied aan de
hemel (afmeting 1/20$^{\circ}$$\times$1/20$^{\circ}$) zien op 3
verschillende golflengten. De linker afbeelding laat zien hoe het
gebied in zichtbaar licht eruit ziet. Er zijn maar weinig sterren
zichtbaar vanwege de hoge interstellaire extinctie. Deze foto is
gemaakt met de telescoop op Mt. Palomar. De middelste afbeelding laat
hetzelfde gebied zien in het nabij-infrarood; deze afbeelding is
afkomstig uit de 2MASS survey. Rechts staat een ISOGAL afbeelding in
het mid-infrarode deel van het spectrum. Hierin kunnen we alleen
sterren zien met een circumstellaire mantel. De omcirkelde ster hebben
wij gedetecteerd in onze zoektocht naar SiO maser straling (het is
nr. 12 uit onze catalogus). Uit de SiO maser lijn afkomstig van deze
ster hebben wij afgeleid dat deze ster zich met een snelheid van -193
km per seconde langs de gezichtslijn beweegt.}
\end{center} \hfill
\end{figure}

\vskip 0.5cm
\noindent {\bf Ge{\"e}volueerde sterren en de massaverdeling en kinematica van de Melkweg }
\vskip 0.2cm

\begin{figure}[t]
\begin{center}
\resizebox{0.65\hsize}{!}{\includegraphics{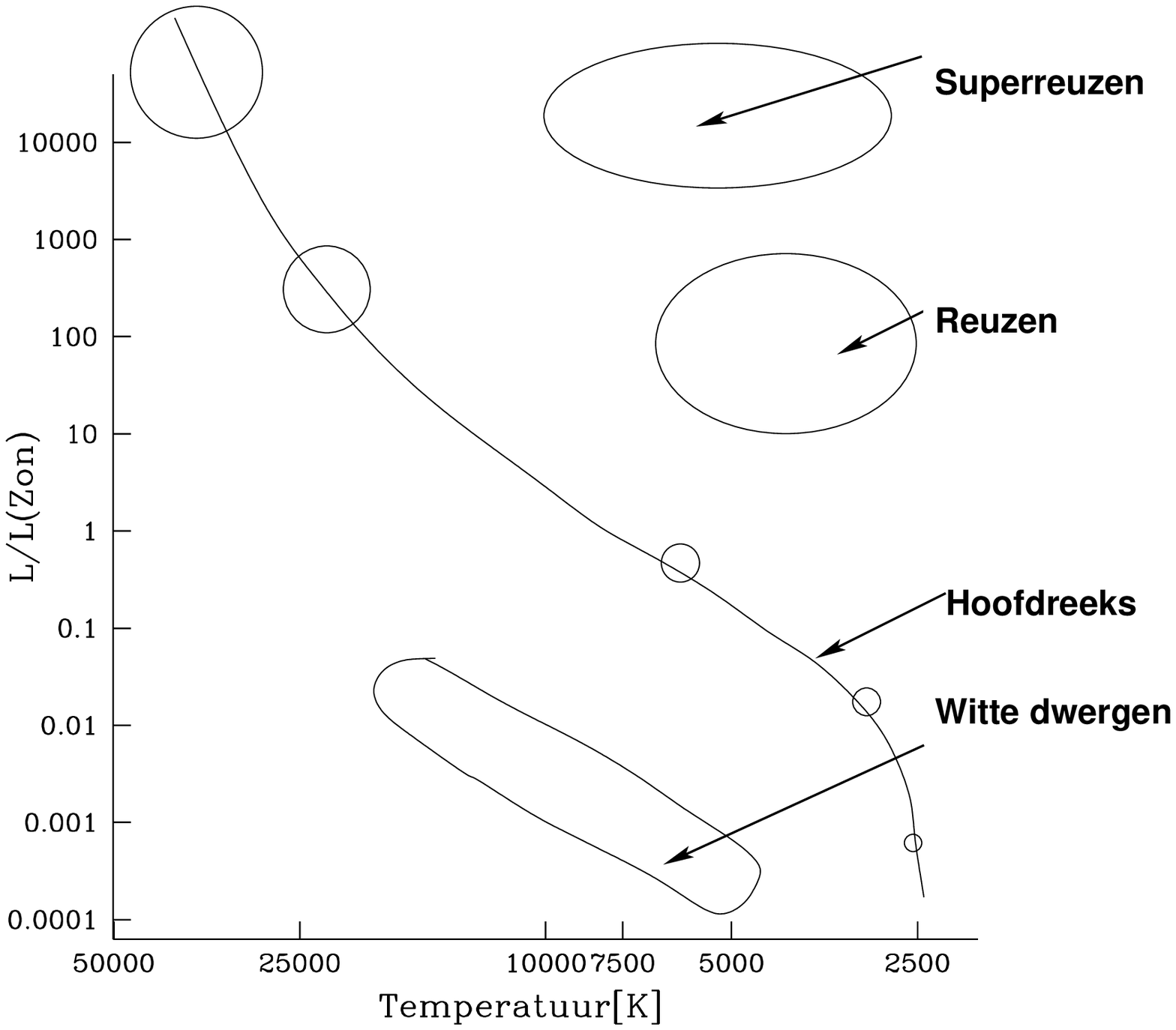}} 
\caption{\label{fig:hr.ps}Grafiek van de helderheid van een ster (de
'magnitude') versus de temperatuur van de ster (de 'kleur'): dit type
figuur heet een Hertzsprung-Russell (of HR) diagram, of ook wel een
kleur-magnitude diagram. Astronomen gebruiken 2 criteria om sterren in
deze figuur te classificeren. Het eerste criterium gebruikt het
spectrum van de ster, de kleur van het licht dat de ster uitzendt. De
kleur hangt af van de temperatuur van de ster. Zo ziet een hete ster
(zoals Sirius) er blauw uit, terwijl een koele ster (zoals Betelgeuze)
er rood uitziet. De zon is bijvoorbeeld geel. Het tweede criterium
gebruikt de helderheid van de ster, dus eigenlijk de energie die per
seconde door de ster wordt uitgezonden. Sterren liggen niet kris-kras
verspreid in een HR diagram, maar ze liggen op bepaalde banden
(reeksen). Elk van deze banden correspondeert met een zekere fase in
het leven van de ster, i.e. een zekere manier waarop kernfusie
plaatsvindt in het centrum van de ster.}
\end{center} \hfill
\end{figure}

De meeste sterren die we kunnen zien op infrarode golflengten zijn
ge{\"e}volueerde reuzesterren, koude (rode) sterren die aan hun
oppervlakte typisch zo'n 3000 K warm zijn (ter vergelijking: onze zon
heeft een oppervlaktemperatuur van zo'n 6000 K!). Wij noemen ze
reuzesterren vanwege hun grote omvang: als de zon een reuzester is
geworden zal haar straal groter zijn dan de aardbaan. De sterren waar
we het hier over hebben zijn ongeveer even zwaar of wat zwaarder dan
onze eigen zon (tot zo'n 6 keer zo zwaar). Ofschoon deze sterren
relatief koud zijn voor sterbegrippen zijn ze toch heel helder:
typisch 3000 keer helderder dan onze zon. Rode reuzen verbranden
waterstof en/of helium in een schil rond hun kern. Hun leeftijd ligt
tussen de 1 en de 15 miljard jaar; ze zijn bijna aan het eind van hun
leven aangekomen.\\ 

Ge{\"e}volueerde sterren hebben sterke winden die de ster omhullen met
een mantel van gas en stof. De aanwezigheid van stof is vastgesteld
uit een overschot aan infrarood licht dat van deze sterren
afkomt. Stof absorbeert namelijk licht afkomstig van de ster en zendt
dit weer uit op (voornamelijk) infrarode golflengten. We weten dat
zich ook gas in de mantel rond deze sterren moet bevinden omdat we
straling meten die wordt uitgezonden door sommige moleculen (zoals CO)
op golflengten van een millimeter of iets minder dan een
millimeter. Deze circumstellaire mantels zenden bovendien vaak
maserstraling uit (het equivalent van laserstraling maar dan in het
millimeter deel van het spectrum), afkomstig van moleculen zoals OH,
SiO en H$_2$O (water). We kunnen deze maserstraling opvangen uit alle
delen van de Melkweg. De golflengte van deze maserlijnen lijkt in de
waarnemingen verschoven ten gevolge van het Doppler effect, hetzelfde
effect waardoor een naderende ambulance een hogere toon lijkt te
hebben dan {\'e}{\'e}n die weg rijdt. Daarom kunnen we door heel
nauwkeurig de golflengte van de maser-emissie te meten de snelheid van
deze sterren langs de gezichtslijn bepalen.

Ge{\"e}volueerde reuzesterren die helder zijn op infrarode golflengten
zijn vanwege bovenstaande redenen uitermate geschikt om de
geschiedenis, de structuur en de kinematica (de beweging van
o.m. sterren) van de Melkweg te achterhalen. Deze sterren zijn gevormd
op verschillende tijdstippen in de geschiedenis van de Melkweg, zoals
blijkt uit hun onderlinge leeftijdsverschillen, waardoor zij ons in
staat stellen om meer te weten te komen over stervorming in
verschillende perioden van het bestaan van de Melkweg. Hun ruimtelijke
verdeling wordt bepaald door het zwaartekrachtsveld van de Melkweg,
daarom kunnen wij hen ook gebruiken om de massaverdeling die dit
zwaartekrachtsveld in de Melkweg veroorzaakt te bepalen. Tenslotte kan
de maser-emissie in deze sterren worden gebruikt om nauwkeurig de
snelheden van deze sterren te achterhalen, waardoor ze heel goed
bruikbaar zijn om de kinematica van de Melkweg te bestuderen.

\vskip 0.5cm 
\noindent {\bf Onze survey van SiO masers} 
\vskip 0.2cm

In dit promotie-onderzoek hebben wij gegevens geanalyseerd uit
verschillende surveys in het infrarood, en wij hebben ook SiO
maserlijnen waargenomen rond ge{\"e}volueerde reuzesterren. De twee
hoofddoelen van dit onderzoek zijn ten eerste het bepalen van de
snelheid langs de gezichtslijn van de waargenomen sterren en ten
tweede het bepalen van de massaverdeling van sterren in de binnenste
delen van onze Melkweg. Eigenschappen van ge{\"e}volueerde sterren,
zoals hun helderheid en massaverlies, kunnen ons verder iets vertellen
over de voorwaarden waaronder masers in dit type sterren voorkomen.

\vskip 0.5cm 
\noindent {\bf Overzicht van dit proefschrift} 
\vskip 0.2cm

We hebben gezocht naar SiO maserstraling afkomstig van sterren die in
het infrarode deel van het spectrum helder zijn. Het resultaat was dat
we het aantal van deze sterren (en daarmee ook hun snelheid langs de
gezichtslijn) dat bekend was in de binnenste delen van de Melkweg
bijna hebben kunnen verdubbelen. In {\bf hoofdstuk 2} tonen we ons
onderzoek dat is uitgevoerd met de IRAM 30 meter telescoop, die zich
op de Pico Veleta in Spanje bevindt. De sterren die we selecteerden
hebben dezelfde kleur als ge{\"e}volueerde sterren en bovendien
verandert hun helderheid op een periodieke manier. We hebben deze
criteria gebruikt omdat al bekend was dat maser-emissie vaker voorkomt
bij dergelijke sterren. We vonden SiO masers in 271 van de onderzochte
sterren, waarmee het aantal snelheden dat is bepaald voor sterren in
de binnenste delen van de Melkweg met behulp van maser lijnen is
verdubbeld.\\

De verdeling van energie van deze sterren over de verschillende
golflengtegebieden (afgeleid uit de helderheid van de sterren in het
nabije- en mid-infrarode deel van het spectrum) kan worden gebruikt om
de helderheid en het massaverlies van deze sterren te bepalen. In {\bf
hoofdstuk 3} presenteren wij de helderheid van de 441 sterren die we
hebben onderzocht op maserstraling op golflengten tussen 1\um en 25\um
(ter vergelijking: het licht dat we met onze ogen zien heeft
golflengten tussen de 0.4\um en 0.8\um) . Metingen gedaan op
verschillende tijdstippen laten zien dat de meeste van deze bronnen in
helderheid vari{\"e}ren.  Hun kleur lijkt op die van veranderlijke
ge{\"e}volueerde sterren in de zonsomgeving waarvan we weten dat ze
worden omgeven door een dunne mantel.\\

Om de intrinsieke helderheid van een ster te kunnen bepalen moeten we
corrigeren voor interstellaire extinctie. We doen dit door de
verdeling in het kleur-magnitude diagram (fig. 3) van de door ons
waargenomen ge{\"e}volueerde sterren te vergelijken met de verdeling
in het kleur-magnitude diagram van ge{\"e}volueerde sterren waarvan we
de extinctie kennen. In {\bf hoofdstuk 4} voeren we deze analyse uit,
om daarmee de extinctie voor al onze SiO maserbronnen in de
verschillende delen van de Melkweg te bepalen. Uit onze analyse volgen
ook nieuwe, interessante waarden voor de interstellaire extinctie op
verschillende golflengten.\\

In {\bf hoofdstuk 5} berekenen we de intrinsieke helderheid van onze
SiO sterren. Als we aannemen dat alle sterren die minder dan
5$^{\circ}$ van het Melkwegcentrum af staan dezelfde afstand tot de
zon hebben, dan vinden we dat de piek in de helderheidsverdeling van
onze sterren samenvalt met de piek in de verdeling van
ge{\"e}volueerde sterren met OH maser emissie die zich in de buurt van
het Melkwegcentrum bevinden (deze piek ligt bij ongeveer 8000 keer de
helderheid van de zon). Onze data laat zien dat het belangrijkste
verschil tussen sterren met OH masers en sterren met SiO masers is dat
sterren met OH masers meer massa verliezen per jaar dan sterren met
SiO masers. Sterren met SiO masers zijn echter makkelijker waar te
nemen op golflengten in het nabij-infrarood dan sterren met OH masers,
waardoor het makkelijker is om sterren met SiO masers te gebruiken om
meer te weten te komen over de sterren in het midden van de
circumstellaire mantel.\\

Een voorlopige bespreking van de banen van onze SiO maser sterren en
een opzet voor onderzoek dat we in de toekomst willen doen wordt
besproken in {\bf hoofdstuk 6}. De beweging van de SiO maser sterren
bevestigt het bestaan van een balk in het centrale deel van onze
Melkweg. We zijn begonnen met een volledige analyse van de dynamica
van onze sterren met als doel het verfijnen van de parameters die de
balk beschrijven.\\

Tenslotte beschrijven we in {\bf hoofdstuk 7} in het kort de ISOGAL
survey, een survey in het mid-infrarood van het vlak van de Melkweg,
maar voornamelijk in de richting van het Galactisch centrum. Deze
survey is uitgevoerd om de structuur van de Melkweg te bepalen, in het
bijzonder in gebieden in het centrum van de Melkweg en in gebieden
waar de interstellaire extinctie groot is. Het is erg lastig om van
afzonderlijke sterren de positie en helderheid te bepalen om daarmee
een betrouwbare catalogus samen te stellen. In dit hoofdstuk
beschrijven wij simulaties die gebruikt kunnen worden om de
nauwkeurigheid van helderheden afgeleid uit de ISOGAL afbeeldingen te
bepalen.

\end{nedervat}

\chaptermark{Biography} \thecv I am the daughter of Lucia and Leonardo
Messineo; sister of Celestina and Saro. I was born at 2 o'clock on a starry
night in a place called Salinella, high in the Sicilian mountains. From there
in the night you almost always see many stars, but my grandmother told me not
to count them because one could never finish. I attended elementary school in
Petralia Soprana and intermediate and high school in Bagheria. I began studying
Astronomy at the University of Bologna in the fall of 1989. I graduated on
March 21st 1997 cum laude with a thesis entitled  "the mixing problem in
stellar evolution and the primordial He abundance determination from population
ratios in globular clusters", which I did under the supervision of Prof.
Francesco Ferraro from Bologna University and Prof. Flavio Fusi Pecci from the
Bologna Observatory.  In 1998 I went to Baltimore for six months as a summer
student to work at the Space Telescope Science Institute with Dr. Antonella
Nota on ISO data of LBV stars. In 1999 I worked at the Bologna Observatory on
HST photometry of the globular cluster NGC 288 in collaboration with Dr.
Michele Bellazzini.  On December 1st 1999, I started my PhD research in Leiden
under the supervision of Prof. Harm Habing. For the first year I worked on the
ISOGAL catalogue, after which my research turned to late-type stars in the
inner Galaxy. This work has been done in collaboration mainly with Prof. Harm
Habing, Prof. Karl Menten from the Max-Planck-Institut fuer Radioastronomie in
Bonn, Prof. Alain Omont from the Institut d'Astrophysique de Paris and Dr.
Lorant Sjouwerman from the National Radio Astronomy Observatory in Socorro.  During
the past four years I have enjoyed observing with the IRAM 30m telescope in
Spain four times as well as with the Heinrich Hertz Telescope in Arizona, with
the ESO 3.6m and the CTIO 4m telescopes in Chile.  I spent some weeks visiting
the Institut d'Astrophysique de Paris, ESA in Villafranca del Castillo, Spain,
and the CTIO in La Serena, Chile.  I attended the YERAC 2000 school in Granada
(Spain),  the NOVA 2000 school in Dwingeloo (The Netherlands), and conferences
in Angra dos Reis (Brazil), Sendai (Japan) and Kona (Hawaii). I gave invited
presentations at the Max-Planck-Institut fuer Radioastronomie in Bonn 
(Germany), and at the Stichting Astronomisch Onderzoek in Nederland (ASTRON) 
in  Dwingeloo. In September 2004 I will start working as a fellow at ESO  in
Garching (Germany).

\begin{thepostface}

The list of people I have interacted with, learnt from and spent
time with during my PhD at Leiden Observatory is very long, and I
apologise if I do not mention you all here. 

During my first year Wing Fai was still in Leiden, always willing to
answer my technical questions.  Maria Rosa, Wouter, Inga, Garrelt,
Pedro, Petra, Dominic, Yvonne and I were Harm's group and we used to
have a group meeting every Friday, but someone would always complain
because the two Italian girls would speak too long!! 
I spent a lot of time with Roderik, Dominic and Jes, being
my officemates and I thank you for being so patient with  me.
  
I used to visit the office of the computer group, which was next to
mine, at least once per day reporting obscure behaviours of my
computer and I am grateful to David, Tycho, Erik and Aart for always
being very helpful. And the management assistants were always 
great! in particular I had the pleasure to work closely with Kirsten
in organising the AGB workshop.

I had the privilege to work with wonderful people and great
scientists, among them Karl, Alain, Lorant and Maartje: thanks for
your support.  

In the past months it has been a great pleasure to start working
with Glenn on stellar dynamics and to create the ROTBAAN consortium!

I am indebted to Kirsten, Melanie and Frank for their careful reading
of my manuscripts and to Dominic for translating the summary of this
document into Dutch. 

My numerous changes of apartments  brought me  to very close
friendships with Maria Rosa, Petra, Emanuela and Elena, with whom I
have shared so many cups of good coffee and tea.

Infine vorrei ringraziare particolarmente la mia famiglia per avermi
incoraggiato nei miei studi e per aver sempre accettato tutte le mie
scelte e la mia lontananza.

Grazie tanto!

Bedankt iedereen!

\end{thepostface}

\printindex
\end{document}